\documentclass[reprint,superscriptaddress,amsmath,amssymb,aps,pra,floatfix]{revtex4-1} 
\pdfoutput=1

\usepackage{graphicx}
\usepackage{dcolumn}
\usepackage{bm}
\usepackage{braket}
\usepackage{color}
\usepackage{hyperref}
\usepackage{multirow}
\usepackage{url}
\usepackage{threeparttable}
\usepackage{comment}
\usepackage{ulem}
\makeatletter
\newcommand{\printfnsymbol}[1]{%
  \textsuperscript{\@fnsymbol{#1}}%
}
\makeatother

\begin{document}

\title{Ab initio extended Hubbard model of short polyenes for efficient quantum computing}

\author{Yuichiro Yoshida}
\email{yoshida.yuichiro.qiqb@osaka-u.ac.jp}
\affiliation{%
Center for Quantum Information and Quantum Biology,
Osaka University, 1-2 Machikaneyama, Toyonaka, Osaka 560-0043, Japan
}%
\author{Nayuta Takemori}
\affiliation{%
Center for Quantum Information and Quantum Biology,
Osaka University, 1-2 Machikaneyama, Toyonaka, Osaka 560-0043, Japan
}%
\affiliation{Center for Emergent Matter Science, RIKEN, Wako, Saitama 351-0198, Japan}
\author{Wataru Mizukami}
\affiliation{%
Center for Quantum Information and Quantum Biology,
Osaka University, 1-2 Machikaneyama, Toyonaka, Osaka 560-0043, Japan
}%
\affiliation{%
Graduate School of Engineering Science, Osaka University, 1-3 Machikaneyama, Toyonaka, Osaka 560-8531, Japan
}%

\date{\today} 

\begin{abstract}
We propose introducing an extended Hubbard Hamiltonian derived via the \textit{ab initio} downfolding method, which was originally formulated for periodic materials, towards efficient quantum computing of molecular electronic structure calculations. 
By utilizing this method, the first-principles Hamiltonian of chemical systems can be coarse-grained by eliminating the electronic degrees of freedom in higher energy space and reducing the number of terms of electron repulsion integral from $\mathcal{O}(N^4)$ to $\mathcal{O}(N^2)$.
Our approach is validated numerically on the vertical excitation energies and excitation characters of ethylene, butadiene, and hexatriene.
The dynamical electron correlation is incorporated within the framework of the constrained random phase approximation in advance of quantum computations, and the constructed models capture the trend of experimental and high-level quantum chemical calculation results.
As expected, the $L^1$-norm of the fermion-to-qubit mapped model Hamiltonians is significantly lower than that of conventional \textit{ab initio} Hamiltonians, suggesting improved scalability of quantum computing. 
Those numerical outcomes and the results of the simulation of excited-state sampling demonstrate that the \textit{ab~initio} extended Hubbard Hamiltonian may hold significant potential for quantum chemical calculations using quantum computers.
\end{abstract}

\maketitle

\section{Introduction} \label{sec:intro}

Quantum computers are expected to solve electronic structure problems of chemistry that are potentially valuable to humanity and beyond the reach of classical computers~\cite{Motta2022emerging,Dalzell023quantum}.
Quantum phase estimation is a well-known algorithm that uses a quantum computer to estimate the eigenvalues of the chemistry Hamiltonians~\cite{Abrams1999quantum,Aspuru-Guzik2005simulated}.
Through estimating the quantum computational cost of the phase estimation algorithms, the potential applications of fault-tolerant quantum computers have been explored to address global challenges in chemistry, such as nitrogen fixation~\cite{Reiher2017elucidating,Lee2021even}, carbon dioxide reduction catalysis~\cite{von-Burg2021quantum}, materials research for batteries~\cite{Rubin2023fault}, and drug discovery~\cite{Blunt2022perspective}.

One of the traditional and most attractive themes in quantum chemistry is the computation of electronic excited states~\cite{Serrano-Andres2005quantum}.
The evaluation of excited states is still challenging for classical computation, partly because these states are often described by a linear combination of a larger number of Slater determinants, in contrast to the ground state, which is often well-described by a single Slater determinant. This highlights the suitability of quantum computational approaches leveraging the superposition nature of a quantum state. Indeed, various quantum algorithms have recently been proposed for evaluating excited states of molecules~\cite{Higgott2019variationalquantum,Jones2019,McClean2017,Nakanishi2019,Parrish2019quantum,Ollitrault2020quantum,Asthana2023quantum,Yalouz2021state,Omiya2022analytical,Yoshikura2023quantum,Yeter-Aydeniz2020practical,Tsuchimochi2023improved,Tsuchimochi2023multi,Bauman2021toward,Stair2020multireference,Cortes2022quantum}.

In principle, quantitative quantum chemical calculations on a quantum computer require a large number of quantum bits (qubits).
The increase in the number of qubits is directly related to the increase in the number of molecular orbitals: in Jordan--Wigner (JW) mapping, a typical fermion-to-qubit mapping, the number of qubits representing the mapped Hamiltonian is equal to the number of spin orbitals.
The use of many orbitals can provide a quantitative description of the electron correlation, especially the dynamical electron correlation, but more qubits require more quantum computational cost~\cite{Gonthier2022measurements}.

The complexity of the molecular electronic structure Hamiltonian, elaborately modeled using many orbitals, triggers the fatal problem of the required number of quantum gates to implement it becoming exceedingly large. The second-quantized electronic structure Hamiltonian of a chemical system has an $\mathcal{O}(N^4)$ electron repulsion integral tensor, where $N$ is the number of spin orbitals.
It induces a single Trotter step of the time-evolution operator to become $\mathcal{O}(N^4)$ circuit depth with the naive implementation~\cite{Wecker2014gate}.
Such an inherent complexity poses challenges for fault-tolerant quantum computations as well as quantum simulations using near-term quantum devices; inevitably, the Coulomb operator of electronic structure Hamiltonians is factorized or sparsified to reduce the non-Clifford gate counts~\cite{berry2019qubitization,von-Burg2021quantum,Lee2021even,Motta2021low,Matsuzawa2020jastrow,Rubin2022compressing,Dario2024reducing}.
Simplification of electronic structure Hamiltonians in chemistry is crucial to avoid the excessively complicated quantum circuit operations.

A prospective way to save the number of qubits is to construct an effective Hamiltonian of the active space, consisting of chemically essential orbitals, prior to quantum computation.
Such approaches to effectively reducing the number of electronic degrees of freedom are known as `downfolding' approaches, and several downfolding methods for quantum computation have been proposed recently~\cite{bauman2019downfolding,Metcalf2020,Nicholas2021,Huang2023leveraging,Bauman2022coupled,Le2023,Motta2020quantum,McArdle2020Improving,Vorwerk2022quantum}.
Downfolding approaches are very powerful because they incorporate the dynamical electron correlation related to the huge exterior space into the effective Hamiltonian in a relatively small active space. However, the $\mathcal{O}(N^4)$ complexity of \textit{ab initio} electronic Hamiltonians still remains in the previous studies.

Considering those above, it is desired to construct an effective subspace model that possesses sufficient capability for discussing the electronic properties of molecules while reducing the complexity of the electron-electron interaction operators.
In the context of condensed matter physics, the \textit{ab initio} downfolding method was developed to make a low-energy model for periodic materials~\cite{Aryasetiawan2004frequency,RESPACK1}.
This method can parameterize an extended Hubbard model by incorporating the contributions from the higher energy degrees of freedom into the electronic interactions of the lower energy degrees of freedom near the Fermi level based on constrained random phase approximation (cRPA).
The \textit{ab initio} downfolding approach has been widely applied to strongly correlated electronic phenomena in materials, such as the superconductivity of iron-based materials~\cite{Misawa2014superconductivity} and the quantum spin liquid behavior of a molecular solid~\cite{Misawa2020electronic,Yoshimi2021Abinitio}.
Extending this approach to quantum computing, the quasi-one-dimensional CuBr$_2$ material as an analog of cuprate was modeled to test~\cite{amsler2023quantumenhanced}.
Nonetheless, there has been little quantitative discussion about the advantage of introducing such an \textit{ab initio} extended Hubbard model in quantum computing.

In this paper, we propose to use an extended Hubbard model of an isolated chemical system based on the \textit{ab initio} downfolding method for efficient quantum computing.
We numerically test this approach by constructing models of polyenes of short conjugation length: ethylene, butadiene, and hexatriene.
Such polyene molecules, especially their excited states, have been traditionally studied as benchmarks for quantum chemistry methods~\cite{Luis1993towards,Nakayama1998theoretical,Kurashige2004pi,Schreiber2008benchmarks,Watson2012excited,Daday2012full,Chien2018excited,Manna2020taming} as well as spectral experiments~\cite{Mulliken1977excited,doering1980electron,gavin1973spectroscopic,Flicker1977low,Fujii1985two}. The $L^1$-norm of our model Hamiltonians after the fermion-to-qubit mapping is analyzed to show that the low-dimensionality and sparsity of the effective Hamiltonian can contribute to efficient quantum computing.
Finally, we simulate the sampling calculations of the excited states of our models using quantum circuits optimized via the variational quantum deflation (VQD) algorithm~\cite{Higgott2019variationalquantum}.

Moreover, we examine how the number of bands of the referential first-principles electronic structure calculation affects the quality of our models.
The construction of \textit{ab initio} downfolded models for isolated chemical systems is still largely unexplored, and it is vital to investigate how to construct a reasonable model.
The constructed models are validated by comparing the excitation energies with experimental and several quantum chemical calculation results.

The rest of this paper is organized as follows. In Sec.~\ref{sec:theo}, we briefly review the extended Hubbard model Hamiltonian and its construction via the \textit{ab initio} downfolding method. 
Next, we explain the proposed approach that leverages the model Hamiltonian for quantum computation and briefly summarize the relationship between the $L^1$-norm of the fermion-to-qubit mapped Hamiltonian and quantum algorithmic scaling.
The computational details are explained in Sec.~\ref{sec:comput}. 
In Section~\ref{sec:result}, we present a comprehensive analysis of our models. This includes examining the dependency of the model Hamiltonian on the number of bands, validating the models through excitation energy comparisons, $L^1$-norm analysis, and discussing the implications of classical simulations for quantum computation.
Finally, the conclusion of this study is given in Sec.~\ref{sec:conclusion}.

\section{Computational Methods} \label{sec:theo}

\subsection{Ab initio extended Hubbard Hamiltonian for molecules}

We briefly review the extended Hubbard Hamiltonian of the \textit{ab initio} downfolding method~\cite{Aryasetiawan2004frequency,RESPACK1}.
For explanation, the effective Hamiltonian for molecules is explicitly described, with minor modifications not including the summations of inter-unit cell interactions.

We employ an extended Hubbard Hamiltonian to describe isolated molecules, which is given by,
\begin{align}
\mathcal{H}_{\textrm{ex-Hub}} = \mathcal{H}_{0}  
+ \mathcal{H}_{\rm int}. \label{eq:exH}
\end{align}
Here, $\mathcal{H}_0$ and  $\mathcal{H}_{\rm int}$ consists of one-body and  two-body operators, respectively.
The one-body term is expressed as:
\begin{align}
\mathcal{H}_0  
= \sum_{i,j,\sigma} ( t_{ij} - t_{ij}^{\rm DC} ) c^\dagger_{i\sigma} c_{j\sigma},
\end{align}
where $c^{(\dagger)}_{i\sigma}$ denote the annihilation (creation) operators of the $i$-th orbital with spin $\sigma$.
Here, $t_{ij}$ is a matrix element of the effective one-electron operator defined by
\begin{align}
    t_{ij} = \int \textrm{d}{\bm r} \phi_i^*({\bm r}) \mathcal{H}_\textrm{KS} \phi_j({\bm r}),
\end{align}
where $\phi_i({\bm r})$ is the $i$-th Wannier orbital.
This integral is executed over the crystal volume, and $\mathcal{H}_{\rm KS}$ is the Kohn-Sham (KS) Hamiltonian.

The term $t_{ij}^{\textrm{DC}}$ is introduced to prevent double counting (DC) of two-electron integrals from the one-electron integral $t_{ij}$. It can be defined as
\begin{align}
t_{ij}^{\rm DC} &\equiv
    \begin{cases}
        \alpha U_{ii} D_{ii}
+ \sum_{k \neq i} U_{ik} D_{kk} & (i = j) \\
        0 & (i \neq j),
    \end{cases}
\end{align}
following the approach in Ref.~\cite{HPhi1}.
$\alpha$ is a parameter ranging $0 \leq \alpha \leq 1$ for tuning the correction.
The DC of the electron-electron interactions arises from the exchange-correlation functional of KS--density functional theory (DFT), and it should be noted that it cannot be eliminated completely.
The one-body density matrix $D_{ij}$ can be defined in the KS orbital basis as:
\begin{align}
    D_{ij} &\equiv \sum_{\sigma}
    \left\langle c_{i \sigma}^{\dagger} c_{j \sigma}\right\rangle_{\rm KS}.
\end{align}
$U_{ij}$ denotes a screened Coulomb integral represented as:
\begin{align}
    U_{ij}(\omega) &= \int \textrm{d}{\bm r} \int \textrm{d}{\bm r}' \phi_i^*({\bm r}) \phi_i({\bm r}) W({\bm r}, {\bm r}', \omega) \phi_j^*({\bm r}') \phi_j({\bm r}').
\end{align}
In this formulation, the inverse operator $1/|{\bm r} - {\bm r}'|$ of the bare Coulomb interaction is replaced with the frequency-dependent screened Coulomb interaction $W({\bm r}, {\bm r}', \omega)$, where $\omega$ is frequency. The $U_{ij}(\omega)$ value can be derived from cRPA~\cite{Aryasetiawan2004frequency,RESPACK1}.
For our purposes, we utilize the static limit of the frequency-dependent Coulomb integrals:
\begin{align}
    U_{ij} &= \lim_{\omega \rightarrow 0} U_{ij}(\omega).
\end{align}

For the two-body electron-electron interaction component, $\mathcal{H}_{\rm int}$, we compare the following terms: $\mathcal{H}_{\rm int}^{(1)}$ and $\mathcal{H}_{\rm int}^{(2)}$. 
These terms are expressed as follows:
\begin{align}
    {\cal H}_{\rm int}^{(1)} &= \frac{1}{2} \sum_{i,j} \sum_{\sigma, \rho} \left \{ U_{ij} c_{i\sigma}^\dagger c_{j\rho}^\dagger c_{j\rho} c_{i\sigma} \notag \right. \\
    &\quad+ \left. J_{ij} \left ( c_{i\sigma}^\dagger c_{j\rho}^\dagger c_{i\rho} c_{j\sigma} 
    + c_{i\sigma}^\dagger c_{i\rho}^\dagger c_{j\rho} c_{j\sigma} \right ) \right \}, 
     \label{eq:Hint2p} \\
    {\cal H}_{\rm int}^{(2)} &= \frac{1}{2} \sum_{i,j} \sum_{\sigma, \rho} U_{ij} c_{i\sigma}^\dagger c_{j\rho}^\dagger c_{j\rho} c_{i\sigma}
     \label{eq:Hint2p_2}.
\end{align}
Here, $J_{ij}$ is a screened exchange integral represented as:
\begin{align}
    J_{ij}(\omega) &=\int \textrm{d}{\bm r} \int \textrm{d}{\bm r}' \phi_i^*({\bm r}) \phi_j({\bm r}) W({\bm r}, {\bm r}', \omega) \phi_j^*({\bm r}') \phi_i({\bm r}'),
\end{align}
where we also utilize the static limit for exchange integrals and refer to it as $J_{ij}$.
The $\mathcal{H}_{\rm int}^{(2)}$ term discards the exchange interactions of the $\mathcal{H}_{\rm int}^{(1)}$ term.

Note that the \textit{ab initio} downfolding method has scarcely been applied to isolated chemical systems; very recently, Chang \textit{et al.} reported to apply to vanadocene~\cite{Yueqing2023downfolding}.
Other related studies include the assessment of screened Coulomb interaction based on random phase approximation (RPA) for Nb$_x$Co ($1 \leq x \leq 9$) clusters~\cite{Peters2018abinitio} and benzene~\cite{vanLoon2021random}.
Scott \textit{et al.} recently developed the moment-constrained RPA to evaluate static effective interaction rather than relying on the static limit of cRPA~\cite{Scott2024rigorous}.

\subsection{Ab initio extended Hubbard model approach for quantum computation}

\begin{figure*}[ht]
    \centering
    \includegraphics[bb=0 0 946 427, width=0.90\hsize]{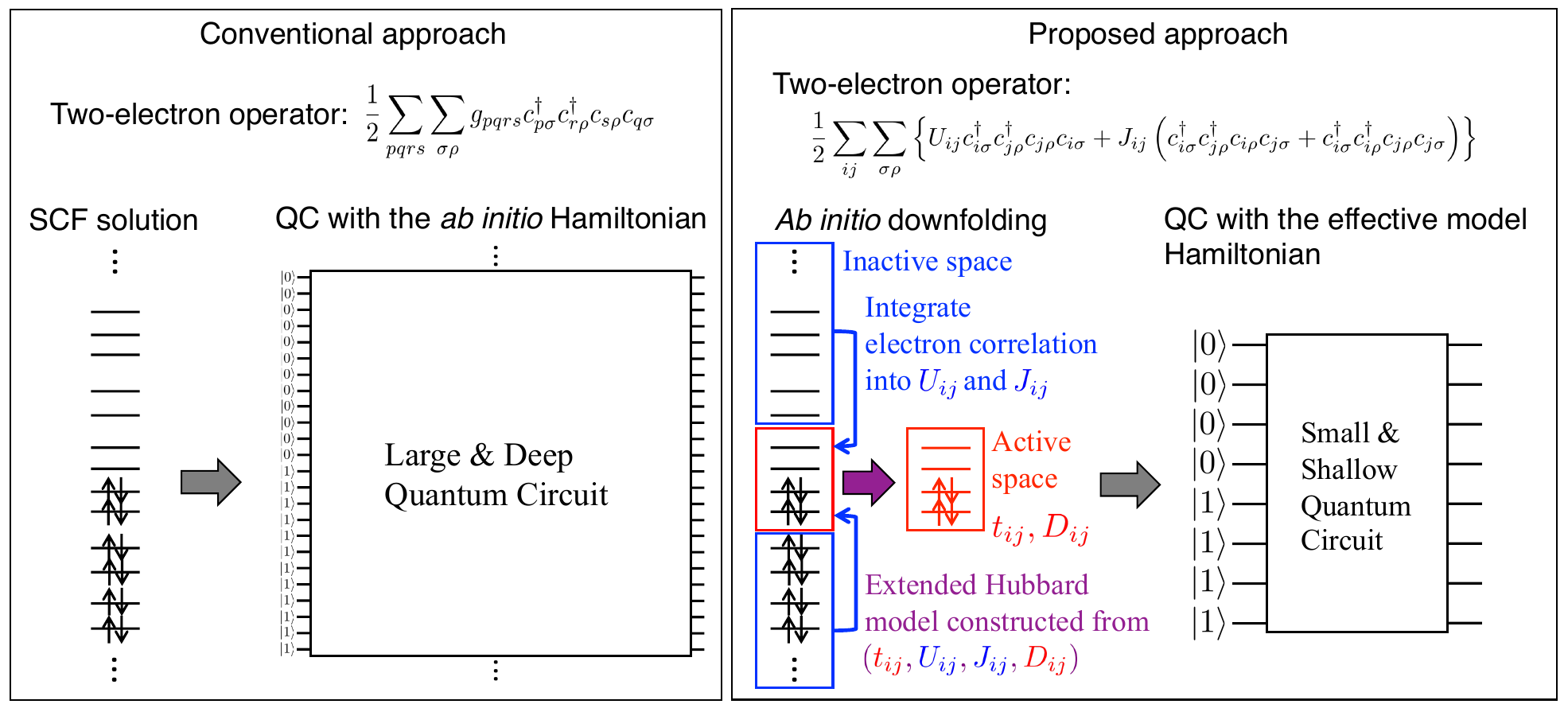}
    \caption{Comparative illustration of the conventional approach with an \textit{ab initio} Hamiltonian (left) versus the {\it ab initio} extended Hubbard model approach (right).
    }
    \label{fig:scheme}
\end{figure*}

Our approach utilizes the extended Hubbard model for an isolated molecule constructed via the \textit{ab initio} downfolding method in quantum computation. The conceptual figure comparing our approach with the conventional one is shown in Figure~\ref{fig:scheme}.
In this approach, the simplified model Hamiltonian is expected to perform quantum computations with a relatively small and shallow quantum circuit. 
The two-electron operator of the extended Hubbard model Hamiltonian consists of the second-order tensors ${\bm U}$ and ${\bm J}$, whose indices belong to a smaller but physically essential space. It becomes a much sparse and compact representation compared to the \textit{ab initio} Hamiltonian, the fourth-order tensor ${\bm g}$, whose indices belong to the entire space of orbitals.

It should be noted that our research direction is closely related to that of dynamical self-energy mapping (DSEM)~\cite{Dhawan2021dynamical,Daniel2021sparse}. The DSEM procedure parametrizes a sparse Hamiltonian to reproduce the dynamical self-energy of the original molecular Hamiltonian. The sparse Hamiltonian contains at most $\mathcal{O}(N^2)$ two-body interaction terms, and it makes quantum circuits shallower and increases the feasibility of quantum computation for molecular systems.

\subsection{\texorpdfstring{$L^1$}{}-norm of fermion-to-qubit mapped Hamiltonian and computational costs in quantum computing}

A second quantized Hamiltonian can be transformed by fermion-to-qubit mappings, such as Jordan-Wigner mapping, into the linear combination of the Pauli operators as
\begin{align}
    \mathcal{H} = \sum_{l=1}^{N_{\rm term}} h_l P_l, \label{eq:transformedH}
\end{align}
where $P_l$ is the $l$-th Pauli operator and $h_l$ is its coefficient.
The summation runs over the number of the Pauli operators, denoted as $N_{\rm term}$.

$L^1$-norm of the coefficient vector of the fermion-to-qubit mapped Hamiltonian $\mathcal{H}$ is defined as the sum of the absolute values of the coefficients
\begin{align}
\label{eq:l1}
    \lambda = \sum_{l=1}^{N_{\rm term}} |h_l|.
\end{align}

Given the crucial role that the parameter $\lambda$ of a given Hamiltonian plays in determining the scalability of various quantum algorithms~\cite{Lee2021even,Koridon2021}, evaluating the value of $\lambda$ is essential for demonstrating the feasibility of quantum computing.
For example, the scaling of the phase estimation algorithms based on the qubitization method using the single factorization and the tensor hypercontraction of the Coulomb operator is $\tilde{\mathcal{O}}(N^{3/2}\lambda / \varepsilon)$~\cite{berry2019qubitization} and $\tilde{\mathcal{O}}(N\lambda / \varepsilon)$~\cite{Lee2021even}, respectively. Here, $N$ is the number of spin orbitals, and $\varepsilon$ is the target precision.
The qDRIFT algorithm~\cite{campbell2019random}, a randomized compiler for Hamiltonian simulation, has the $\mathcal{O}(\lambda^2/\varepsilon^2)$ scaling.

Measurement is one of the most troublesome processes on variational quantum eigensolver (VQE)~\cite{Peruzzo2014variational,Tilly2022variational}, and the number of measurements of VQE also depends on the $L^1$-norm $\lambda$ of the Hamiltonian~\cite{wecker2015progress,Rubin2018}. 
The total number of measurements $M$ is given by the sum of each number of measurements $M_l$ for each Pauli operator $P_l$ of the Hamiltonian as 
\begin{align}
    M = \sum_{l=1}^{N_{\rm term}}M_l.
\end{align} 
The optimal $M_l$ is proportional to $|h_l|$, which is derived through the method of Lagrange multipliers~\cite{Rubin2018}.
The optimal number of measurements is represented as
\begin{align}
M = \frac{1}{\varepsilon^2} \left (
\sum_{l=1}^{N_{\rm term}} |h_l| \sigma_l
\right )^2
\leq \frac{\lambda^2}{\varepsilon^2}, \label{eq:Meas}
\end{align}
where $\sigma_l$ is the intrinsic standard deviation of the Pauli operator $P_l$.
Here, $\varepsilon$ is the sampling error for the expectation value of the Hamiltonian.
The right inequality of Eq.~(\ref{eq:Meas}) is derived from the condition of the intrinsic variance $\sigma_l^2 \leq 1$.
Hence, the total number of measurements of VQE is bounded using the $L^1$-norm $\lambda$.

\section{Computational details} \label{sec:comput}

Here, we explain the computational details in a form that aids understanding of our calculation process.

In preparation for model construction, the first-principles electronic structure calculations were performed by Quantum ESPRESSO version 7.2~\cite{Espresso1,Espresso2,Espresso3,PP}.
We employed a simple cubic lattice with a lattice constant $a=17.0~\textrm{\AA}$, and set the plane-wave cutoff of the wave function $E_{\rm cut}^\psi=30.0~\textrm{Ry}$.
We used the optimized norm-conserving Vanderbilt pseudopotential~\cite{ONCV}.
For our usage, calculations were confined only to the $\Gamma$ point. 
The number of bands $N_{\rm band}$ computed in the band structure calculations is an essential parameter involving the following computation of the screened Coulomb and exchange integrals, and a sufficiently large number of $N_{\rm band}$ is necessary.
In this study, we construct models varying $N_{\rm band}$ and confirm that the sufficiently large value of $N_{\rm band}$.
The details of increasing $N_{\rm band}$ and the maximum number of bands $N_{\rm band}^{\rm max}$ for each molecule are discussed in Sec.~\ref{sec:hyper}.
The molecular structure of the polyenes were all-trans isomers and optimized by DFT calculations at the B3LYP functional~\cite{B3LYP1,B3LYP2} and 6-31G$*$ basis-set level using Gaussian~16 software Revision C~\cite{g16}.

The extended Hubbard models of the polyene molecules were constructed with RESPACK-20200113~\cite{RESPACK1,Fujiwara2003generalization,Nohara2009electronic,Nakamura2008abinitio,Nakamura2009abinitio,Nakamura2016abinitio}.
The maximally localized Wannier functions and the effective one-electron integrals were obtained to reproduce the target band energies of the $\pi$-orbitals perpendicular to the carbon plane.
The screened Coulomb and exchange integrals are evaluated via cRPA.
The plane-wave cutoff of the polarization function $E_{\rm cut}^\varepsilon$ was set to one-tenth of $E_{\rm cut}^\psi$.
The parameter $\alpha$ was set to 1.
We used VESTA version 3~\cite{VESTA} for visualization of the Wannier functions.

The constructed models are handled using PySCF version 2.4.0~\cite{PySCF,PySCF2} and OpenFermion version 1.6.0~\cite{McClean2020} or their earlier versions.
OpenFermion-PySCF~\cite{PySCF,PySCF2,McClean2020} was also used.
Jordan-Wigner mapping was employed for the fermion-to-qubit mapping, and the fermion-to-qubit mapped Hamiltonians were diagonalized to obtain their eigenstates if otherwise specified.
When characterizing the eigenstates, the basis of the Hamiltonian was rotated by self-consistent field (SCF) calculation using PySCF, discussed further in Appendix~\ref{sec:appendix}.
Quantum chemical calculations for comparison were performed using the following program packages:
The $e^T$ program version 1.5.11 ~\cite{Folkestad2020eT} was used to perform CC3~\cite{Paul2021new} calculations.
The OpenMolcas program version 22.06~\cite{OpenMolcas1,OpenMolcas2,OpenMolcas3} was used to perform CASCI and CASPT2 calculations.
The Gaussian~16 software Revision C~\cite{g16} was used to perform time-dependent density functional theory (TD-DFT) calculations, and the functional was B3LYP.
Edmiston-Ruedenberg (ER) localization was performed using \texttt{fcidump\_rotation.f90} in the NECI program package~\cite{guther2020neci}.

VQD calculations and the following sampling estimations were classically simulated by employing Qiskit version 0.43.2, Qiskit-aer version 0.12.1~\cite{Qiskit}, and Qiskit-Nature version 0.6.2~\cite{the-qiskit-nature-developers-and-contrib-2023-7828768}.
For the ansatz, we used a quantum circuit arranged with the particle-conserving A-gates~\cite{Gard2020} in a brick-wall pattern.

\section{Results and discussion} \label{sec:result}

\subsection{Extended Hubbard model of short polyenes}

\begin{figure*}[ht]
    \centering
    \includegraphics[bb=0 0 1272 512, width=0.85\hsize]{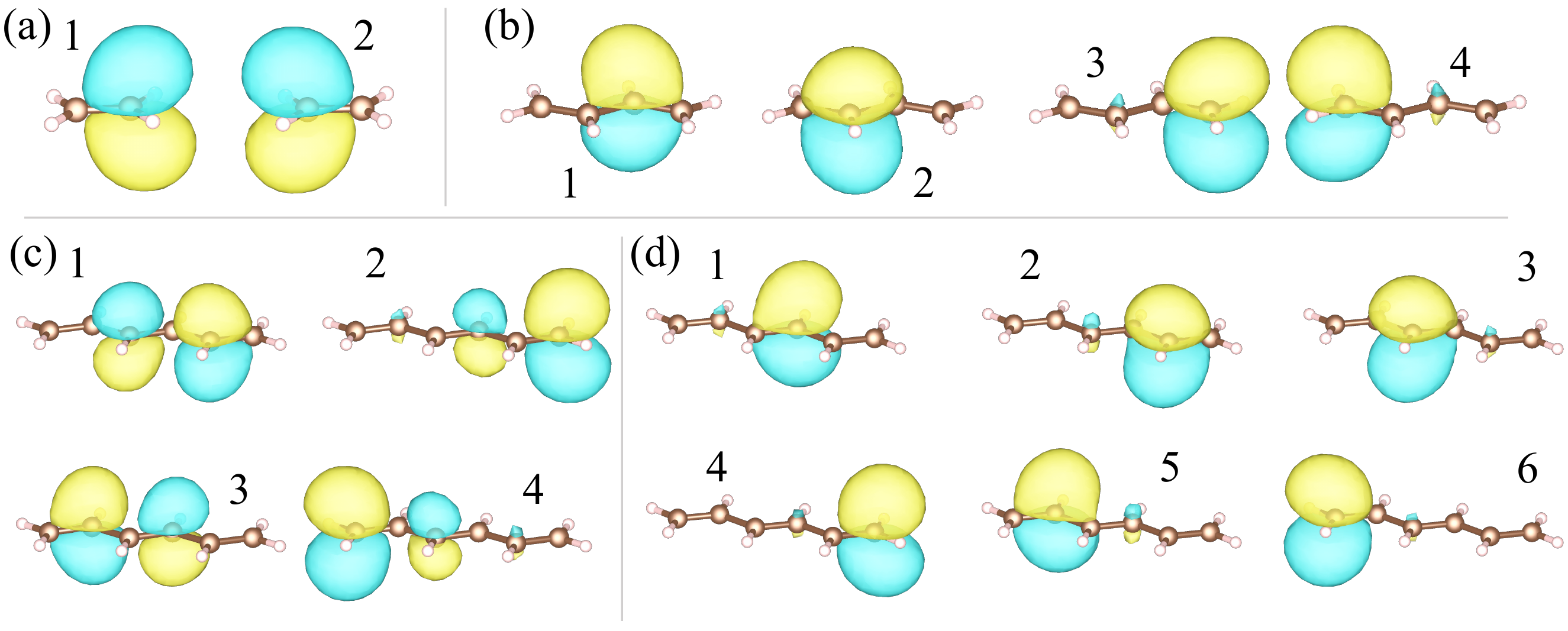}
    \caption{Wannier functions $\{\phi_i \}$ of (a) ethylene (2e, 2o), (b) butadiene (4e, 4o), (c) hexatriene (4e, 4o), and (d) hexatriene (6e, 6o) models. The index $i$ and the shape of each Wannier function $\phi_i$ are specified. The isosurface value is set to 0.03.}
    \label{fig:wannier}
\end{figure*}

The Wannier functions $\{ \phi_i \}$ of our models are shown in Figure~\ref{fig:wannier}.
The modeled active space is represented as ($m$e, $n$o), where $(m, n)$ is the number of electrons and orbitals.
The shape of these Wannier functions is reasonable because they have the $\pi$-orbital character, which is a key to understanding the lower electron excitations of the polyenes.
In the (4e, 4o) and (6e, 6o) cases of hexatriene, the shape of the Wannier functions is reasonable but different.
The difference is that the former uses two fewer $\pi$-type KS orbitals for localization.

In particular, the Wannier functions of ethylene (2e, 2o), butadiene (4e, 4o), and hexatriene (6e, 6o) have shapes reminiscent of a $p_z$-orbital.
This characteristic suggests that the model Hamiltonian with such a basis closely resembles the second-quantized Pariser-Parr-Pople (PPP) Hamiltonian~\cite{Schulten1972origin,Tavan1986low}, which is also the extended Hubbard-type (analogous to the model with $\mathcal{H}_{\rm int} = \mathcal{H}_{\rm int}^{(2)}$) and composed of the set of the orthonormal $p_z$-orbital basis.

Note that the PPP model is a semi-empirical molecular orbital method for $\pi$-conjugated systems~\cite{PPP1,PPP2,PPP3}. 
The PPP-multireference double excitation configuration interaction (PPP-MRD-CI) calculation, a configuration interaction (CI) calculation based on PPP, had achieved qualitative success for polyenes of relatively long conjugation length, even though the molecular integrals were given empirically.
The PPP-MRD-CI calculations provided a qualitatively reasonable explanation for spectroscopic findings for the energetic order of the optically allowed single excitation and the forbidden double excitation~\cite{Hudson1972,Hudson1984}, although the evaluated state order of butadiene is now known to be reversed~\cite{Watson2012excited}.

The major difference between the second-quantized PPP and our extended Hubbard models is how the molecular integrals are determined.
Whereas the former is given empirically, the latter is derived by the \textit{ab initio} downfolding approach.
The integral parameters of our models are summarized in Table~\ref{tab:params}.
\begin{table*}[ht] 
\centering
\caption{Model parameters in ethylene (2e, 2o), butadiene (4e, 4o), hexatriene (4e, 4o), and hexatriene (6e, 6o) models. The indices of the Wannier functions $(i, j)$ are specified in Figure~\ref{fig:wannier}.}    
\label{tab:params}
\begin{minipage}[]{0.95 \columnwidth}
\centering
\resizebox{1 \linewidth}{!}{
\begin{tabular}{lll rrrr} \hline \hline
Molecule   & Active space & $(i, j)$ & $t_{ij}$/eV & $U_{ij}$/eV & $J_{ij}$/eV & $D_{ij}$   \\ \hline 
Ethylene   & (2e, 2o) & $(1, 1)$ & $-$3.820 & 10.442 &        &    1.000 \\
           &          & $(1, 2)$ & $-$2.874 &  6.376 &  0.161 &    0.948 \\
           &          & $(2, 2)$ & $-$3.820 & 10.442 &        &    1.000 \\ \hline 
Butadiene  & (4e, 4o) & $(1, 1)$ & $-$3.663 &  8.298 &        &    0.965 \\
           &          & $(1, 2)$ & $-$2.423 &  5.651 &  0.240 &    0.459 \\
           &          & $(1, 3)$ & $-$2.749 &  5.817 &  0.211 &    0.862 \\
           &          & $(1, 4)$ & $-$0.129 &  4.440 &  0.066 &    0.056 \\
           &          & $(2, 2)$ & $-$3.663 &  8.298 &        &    0.965 \\
           &          & $(2, 3)$ & $-$0.129 &  4.440 &  0.066 &    0.056 \\
           &          & $(2, 4)$ & $-$2.749 &  5.817 &  0.211 &    0.862 \\
           &          & $(3, 3)$ & $-$3.924 &  9.248 &        &    1.035 \\
           &          & $(3, 4)$ &    0.282 &  3.610 &  0.023 & $-$0.387 \\
           &          & $(4, 4)$ & $-$3.924 &  9.248 &        &    1.035 \\ \hline
Hexatriene & (4e, 4o) & $(1, 1)$ & $-$3.707 &  6.977 &        &    1.011 \\
           &          & $(1, 2)$ & $-$2.354 &  5.466 &  0.115 &    0.887 \\
           &          & $(1, 3)$ &    1.887 &  4.738 &  0.074 & $-$0.388 \\
           &          & $(1, 4)$ & $-$0.067 &  3.889 &  0.173 &    0.055 \\
           &          & $(2, 2)$ & $-$3.583 &  8.281 &        &    0.989 \\
           &          & $(2, 3)$ & $-$0.067 &  3.889 &  0.173 &    0.055 \\
           &          & $(2, 4)$ &    0.076 &  2.720 &  0.015 &    0.292 \\
           &          & $(3, 3)$ & $-$3.707 &  6.977 &        &    1.011 \\
           &          & $(3, 4)$ & $-$2.354 &  5.466 &  0.115 &    0.887 \\
           &          & $(4, 4)$ & $-$3.583 &  8.281 &        &    0.989 \\ \hline \hline
\end{tabular}
}
\end{minipage}
\begin{minipage}[]{0.95 \columnwidth}
\centering
\vspace{-6.8mm}
\resizebox{1 \linewidth}{!}{
\begin{tabular}{lll rrrr} \hline \hline
Molecule   & Active space & $(i, j)$ & $t_{ij}$/eV & $U_{ij}$/eV & $J_{ij}$/eV & $D_{ij}$   \\ \hline 
Hexatriene & (6e, 6o) & $(1, 1)$ & $-$3.535 &  7.658 &        &    0.944 \\
           &          & $(1, 2)$ & $-$2.340 &  5.528 &  0.216 &    0.473 \\
           &          & $(1, 3)$ & $-$2.726 &  5.310 &  0.312 &    0.754 \\
           &          & $(1, 4)$ &    0.046 &  4.267 &  0.059 &    0.026 \\
           &          & $(1, 5)$ & $-$0.322 &  4.314 &  0.077 &    0.082 \\
           &          & $(1, 6)$ &    0.270 &  3.410 &  0.026 & $-$0.292 \\
           &          & $(2, 2)$ & $-$4.001 &  8.766 &        &    1.081 \\
           &          & $(2, 3)$ & $-$0.322 &  4.314 &  0.077 &    0.082 \\
           &          & $(2, 4)$ & $-$2.747 &  5.951 &  0.179 &    0.851 \\
           &          & $(2, 5)$ &    0.268 &  3.475 &  0.029 & $-$0.097 \\
           &          & $(2, 6)$ & $-$0.127 &  2.852 &  0.008 & $-$0.060 \\
           &          & $(3, 3)$ & $-$3.535 &  7.658 &        &    0.944 \\
           &          & $(3, 4)$ &    0.270 &  3.410 &  0.026 & $-$0.292 \\
           &          & $(3, 5)$ & $-$2.340 &  5.528 &  0.216 &    0.473 \\
           &          & $(3, 6)$ &    0.046 &  4.267 &  0.059 &    0.026 \\
           &          & $(4, 4)$ & $-$3.746 &  9.281 &        &    0.975 \\ 
           &          & $(4, 5)$ & $-$0.127 &  2.852 &  0.008 & $-$0.060 \\
           &          & $(4, 6)$ &    0.052 &  2.485 &  0.002 &    0.178 \\
           &          & $(5, 5)$ & $-$4.001 &  8.766 &        &    1.081 \\
           &          & $(5, 6)$ & $-$2.747 &  5.951 &  0.179 &    0.851 \\
           &          & $(6, 6)$ & $-$3.746 &  9.281 &        &    0.975 \\ \hline \hline
\end{tabular}
}
\end{minipage}
\end{table*}

These $U_{ij}$ and $J_{ij}$ parameters include the electron correlation effects from outside of the active space.
In the next section, we examine the incorporation of the electron correlation effects by increasing the number of bands and extrapolating the continuum limit.

\subsection{Number of bands dependency \label{sec:hyper}}

We investigate the excitation energies of the models varying the number of bands $N_{\rm band}$.
The values of excitation energy $\Delta E$ are extrapolated to confirm the reliability of our models via estimating the continuum limit.
For extrapolation, the following equation is used:
\begin{align}
    f(N_{\rm band}) = \Delta E_\infty + b \times {\rm exp}(-N_{\rm band}/c), \label{eq:extrpl}
\end{align}
where $\Delta E_\infty$, $b$, and $c$ are real fitting parameters. $\Delta E_\infty$ corresponds to the extrapolated value of the excitation energy.
Fitting is performed based on the eigenvalues of the model Hamiltonian with $\mathcal{H}_{\rm int} = \mathcal{H}_{\rm int}^{(1)}$ obtained by H$\Phi$~\cite{HPhi1,HPhi2}.
In this section, we focus on the Hamiltonian employing $\mathcal{H}_{\rm int}^{(1)}$ rather than both $\mathcal{H}_{\rm int}^{(1)}$ and $\mathcal{H}_{\rm int}^{(2)}$. As shown in Table~\ref{tab:params}, the magnitude of $J_{ij}$ is smaller than $U_{ij}$, and it is considered that the Hamiltonian with $\mathcal{H}_{\rm int}^{(2)}$ exhibits a similar trend to that with $\mathcal{H}_{\rm int}^{(1)}$.

For further confirmation of the convergence behavior, we also vary the cutoff parameter set by comparing the $(a, E_\textrm{cut}^\psi, E_\textrm{cut}^\varepsilon) = (17.0, 30.0, 3.0)$ condition with the $(13.0, 50.0, 5.0)$ condition, where the units of the lattice constant and cutoff parameters are $\textrm{\AA}$ and Ry, respectively.
Due to our computational resources, it is difficult to increase the value of cutoff parameters while maintaining the lattice constant $a$ at 17.0~$\textrm{\AA}$. However, it is important to note that this lattice constant is sufficiently large to treat the molecule as isolated, ensuring the relevance and accuracy of our results within this computational framework.

\begin{figure*}[ht]
    \centering
    \includegraphics[bb=0 0 881 639, width=0.87\hsize]{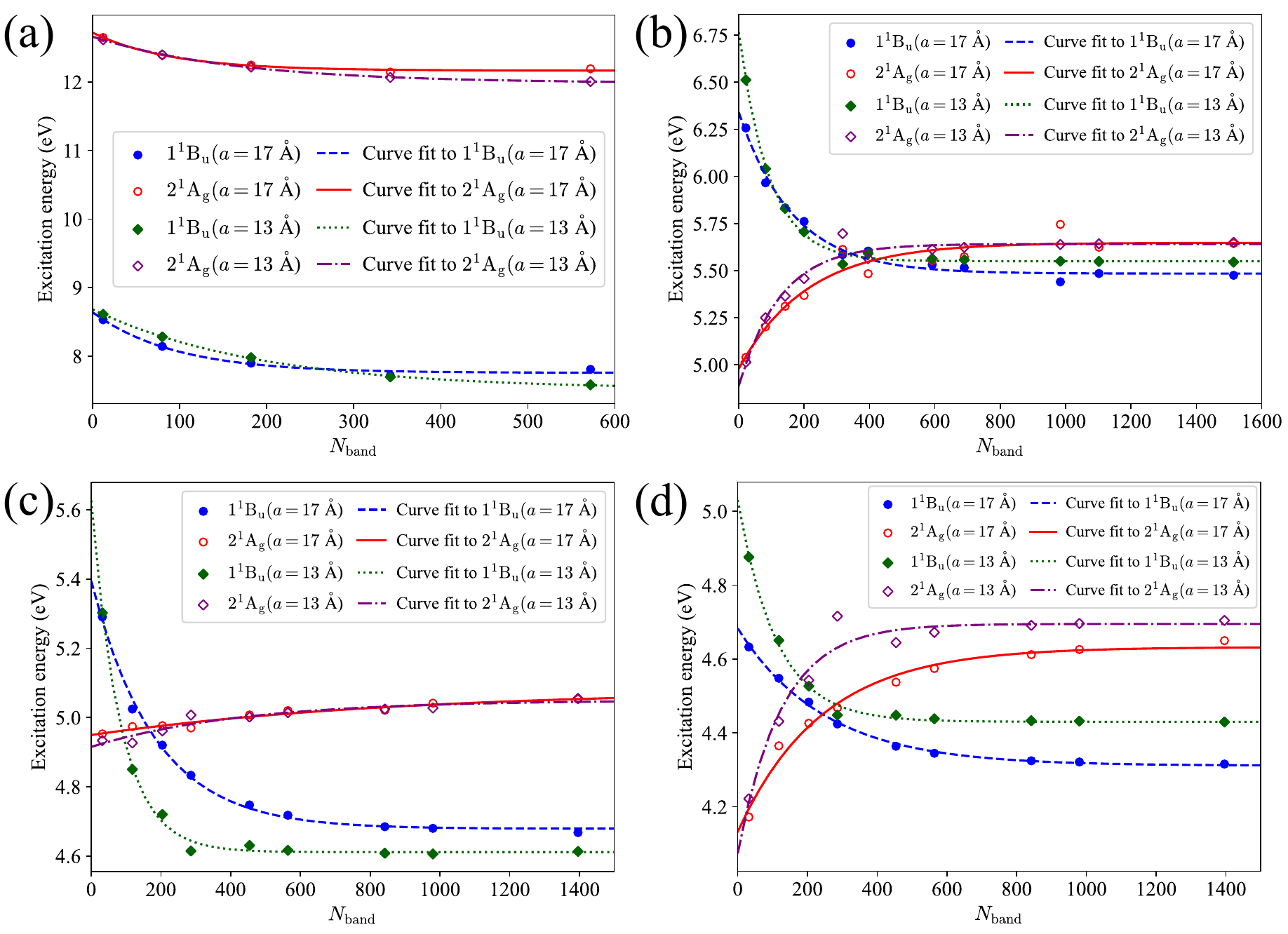}
    \caption{Dependence of the excitation energies on $N_{\rm band}$ in the cases of (a) ethylene (2e, 2o), (b) butadiene (4e, 4o), (c) hexatriene (4e, 4o), and (d) hexatriene (6e, 6o) models. The conditions of $(a, E_\textrm{cut}^\psi, E_\textrm{cut}^\varepsilon) = (17.0, 30.0, 3.0)$ and $(13.0, 50.0, 5.0)$ are investigated.}
    \label{fig:nband}
\end{figure*}

Figure \ref{fig:nband} shows the $N_{\rm band}$ dependency of our systems and the result of extrapolation.
The state characterization is conducted in the way shown in Appendix~\ref{sec:appendix}.
As a result, the excitation energy data varying on $N_{\rm band}$ are successfully fitted using Eq.~(\ref{eq:extrpl}), as shown in Figure~\ref{fig:nband}.
Comparing the two $(a, E_\textrm{cut}^\psi, E_\textrm{cut}^\varepsilon)$ conditions, the shapes, the intersection point, and the convergence behavior to the large $N_\textrm{band}$ limit of the curves fitted to the $1{}^1{\rm B_u}$ and $2{}^1{\rm A_g}$ states are similar.
It suggests that our excitation energy results do not vary significantly with the setting of these hyperparameters.

It is found that there is a qualitative difference in the $N_{\rm band}$ dependence among the molecules.
In the ethylene (2e, 2o) case, as shown in Figure~\ref{fig:nband} (a), the excitation energies to the $1{}^1{\rm B_u}$ and $2{}^1{\rm A_g}$ states decrease in a similar way as $N_{\rm band}$ increases.
Conversely, in the other cases, the excitation energy to the $1{}^1{\rm B_u}$ state decreases, but that to the $2{}^1{\rm A_g}$ state increases as $N_{\rm band}$ increases, as shown in Figure~\ref{fig:nband} (b)--(d).
In the small $N_{\rm band}$ region, lacking the dynamical correlation, the $2{}^1{\rm A_g}$ state is lower than the $1{}^1{\rm B_u}$ state.
As $N_{\rm band}$ increases, the order is reversed, and the excitation energies become saturated eventually.
It is crucial to take a sufficiently large $N_{\rm band}$ to reasonably discuss the energetic order of the $1{}^1{\rm B_u}$ and $2{}^1{\rm A_g}$ states and its gap width within the framework of cRPA.

In Table~\ref{tab:extrp}, we compare the extrapolated $\Delta E_\infty$ values with the $\Delta E$ values of the models, which employ $\mathcal{H}_{\rm int}=\mathcal{H}_{\rm int}^{(1)}$ and $N_\textrm{band}= N_\textrm{band}^{\textrm{max}}$.
\begin{table}[ht]
    \centering
    \caption{Comparison of $\Delta E$ and $\Delta E_{\infty}$ in eV. The $\Delta E$ values correspond to those in the maximum number of bands $N_{\rm band}^{\rm max}$ in Figure~\ref{fig:nband} (a)--(d) in the condition of $a=17.0$~$\textrm{\AA}$,
    $E_\textrm{cut}^\psi = 30.0$~Ry, and $E_\textrm{cut}^\varepsilon = 3.0$~Ry.}
    \begin{tabular}{llccrr} \hline \hline
    Molecule   & Active space & $N_{\rm band}^{\rm max}$ & State & $\Delta E$ & $\Delta E_\infty$ \\ \hline
    Ethylene   & (2e, 2o) & 572  & $1{}^1{\rm B_u}$ & 7.81 & 7.76 \\
               &          &      & $2{}^1{\rm A_g}$ & 12.19 & 12.17 \\ \hline
    Butadiene  & (4e, 4o) & 1514 & $1{}^1{\rm B_u}$ & 5.47 & 5.48
    \\
               &          &      & $2{}^1{\rm A_g}$ & 5.64 & 5.65 \\ \hline
    Hexatriene & (4e, 4o) & 1396 & $1{}^1{\rm B_u}$ & 4.67 & 4.68 \\
               &          &      & $2{}^1{\rm A_g}$ & 5.05 & 5.08 \\ \cline{2-6}
               & (6e, 6o) & 1396 & $1{}^1{\rm B_u}$ & 4.32 & 4.31 \\
               &          &      & $2{}^1{\rm A_g}$ & 4.65 & 4.63 \\
    \hline \hline
    \end{tabular}
    \label{tab:extrp}
\end{table}
These values are similar in each molecule, and it is considered that the models at the $N_{\rm band}^{\rm max}$ points are constructed by incorporating the electron correlation effects almost the same as the effects in the continuum limit.
The values of $U_{ij}$ and $J_{ij}$ at the $N_{\rm band}^{\rm max}$ points are shown in Table~\ref{tab:params}.

\subsection{Vertical excitation energies\label{sec:vert}}

We compare the excitation energies of our models with the results of other quantum chemical calculations and experiments shown in Table~\ref{tab:exc}.
\begin{table*}[!htpb]
    \centering
    \caption{Excitation energies (eV) of short polyenes. The values without and with parentheses for hexatriene correspond to (4e, 4o) and (6e, 6o) active spaces, respectively. 
    The results of CC3, TD-DFT, CASCI, and CASPT2 are obtained using the same molecular structure as our models.} \label{tab:exc}
    \begin{threeparttable}
    \centering
    \begin{tabular}{l w{r}{2.8em} w{r}{2.8em} w{r}{2.8em} w{r}{2.8em} w{r}{4.65em} w{r}{4.65em}} \hline \hline
    \multirow{2}{*}{Model or methods} & \multicolumn{2}{c}{Ethylene} & \multicolumn{2}{c}{Butadiene} & \multicolumn{2}{c}{Hexatriene} \\ 
    & \multicolumn{1}{c}{$1{}^1{\rm B_u}$} & \multicolumn{1}{c}{$2{}^1{\rm A_g}$} &  \multicolumn{1}{c}{$1{}^1{\rm B_u}$} & \multicolumn{1}{c}{$2{}^1{\rm A_g}$} & \multicolumn{1}{c}{$1{}^1{\rm B_u}$} & \multicolumn{1}{c}{$2{}^1{\rm A_g}$} \\ \hline  
    Model 1 & 7.81 & 12.19 & 5.47 & 5.64 & 4.67 (4.32) & 5.05 (4.65) \\  
    Model 2 & 8.13 & 12.19 & 5.82 & 5.98 & 4.81 (4.65) & 5.12 (4.96) \\
    CC3/aug-cc-pVDZ & 7.95 &     & 6.25 & 6.68 & \multicolumn{1}{c}{5.35} & \multicolumn{1}{c}{5.73} \\ 
    TD-DFT/aug-cc-pVDZ & 7.42 &   & 5.59 & 6.53 & \multicolumn{1}{c}{4.61} & \multicolumn{1}{c}{5.64} \\
    CASCI/cc-pVDZ  & 10.17 & 15.82 & 8.15 & 7.06 & 6.72 (6.93) & 6.88 (5.84) \\ 
    CASPT2/cc-pVDZ &  8.33 & 14.10 & 6.26 & 6.57 & 5.09 (5.76) & 4.74 (5.24) \\ \hline 
    Experiment     & 7.66\tnote{a}  &   & 5.92\tnote{b} &      & \multicolumn{1}{c}{4.93\tnote{c}, 4.95\tnote{d}} & \multicolumn{1}{c}{5.2\tnote{e}}     \\
    PPP-CI\tnote{f} &              &   & 5.83 & 5.34 & (5.05) & (4.36) \\
    SHCI with Extrp.\tnote{g} & 8.05 &       & 6.45 & 6.58 & \multicolumn{1}{r}{(5.59)} & \multicolumn{1}{r}{(5.58)} \\ \hline \hline  
    \end{tabular}
    \begin{tablenotes}
    \item[a] Band maximum from Ref.~\cite{Mulliken1977excited}.
    \item[b] Spectral peak from Ref.~\cite{doering1980electron}.
    \item[c] Band center from Ref.~\cite{gavin1973spectroscopic}.
    \item[d] A band peak reported in Ref.~\cite{Flicker1977low}, while the intensity maximum is an adjacent peak at 5.13 eV.
    \item[e] This value was obtained using WebPlotDigitizer~\cite{Rohatgi2022} from the intensity maximum of the spectral data in Ref.~\cite{Fujii1985two} and consistent with the well-known literature value~\cite{Luis1993towards,Chien2018excited}.
    \item[f] Ref.~\cite{Tavan1986low}.
    \item[g] Ref.~\cite{Chien2018excited}. The excitation energy of ethylene was obtained without extrapolation.
    \end{tablenotes}
    \end{threeparttable}
\end{table*}
Hereafter, the models that employ $\mathcal{H}_{\rm int}^{(1)}$ and $\mathcal{H}_{\rm int}^{(2)}$ together with the parameters in Table~\ref{tab:params} are referred to as Model~1 and Model~2, respectively.

Overall, Model~2 practically yields a better result than Model~1.
Model~1 tends to underestimate the excitation energies compared to Model~2, which is considered to be due to the exchange interactions in Model~1.

The excitation energies of Models 1 and 2 become smaller as the conjugation length becomes longer. This tendency is similar to the results of quantum chemical calculations and experiments.
In particular, the excitation energies of Model~2 are in good agreement with the experimental values in either molecule.
We employ the (4e, 4o) and (6e, 6o) active spaces for hexatriene, and the former model has the excitation energies closer to the experimental values.
This suggests that a larger active space does not necessarily lead to a better result in the construction of the \textit{ab initio} downfolding model.

The magnitude of the excitation energies to the lower excited states of polyenes has been an important research topic in traditional quantum chemistry~\cite{Schulten1972origin,Tavan1986low,Nakayama1998theoretical,Kurashige2004pi}.
In butadiene and hexatriene, it is known that the singly excited $1{}^1{\rm B_u}$ and doubly excited $2{}^1{\rm A_g}$ states are energetically close to each other.
It can be said that the gap of our models is small --- the excitation energy gaps of Model~2 of butadiene and hexatriene are 0.16~eV and 0.31~eV, respectively.
These narrow gaps are attributed to the inclusion of dynamical electron correlation from outside of the active space.
Comparison between CASCI and CASPT2 results also indicates that the dynamical electron correlation reduces the energy gap between $1{}^1{\rm B_u}$ and $2{}^1{\rm A_g}$.

We compare the results of our models with those of CC3 and CASPT2 calculations.
In our CC3 calculations, the excitation energies get close to the models' and experimental results in each molecule.
The states of CC3 are characterized using the oscillator strengths and the molecular orbital shapes related to the major single and double amplitudes.
The excitation energy gaps of butadiene and hexatriene are 0.43 eV and 0.38 eV, comparable to those in Model 2.
In our CASPT2 calculations, the excitation energies to the $1{}^1{\rm B_u}$ state capture the trend of the results of Model~2, experiment, and CC3.
The excitation energy gaps of butadiene and hexatriene are 0.31 eV and $-$0.35 eV in CASPT2(4e, 4o) --- The excitation energy to the $2{}^1{\rm A_g}$ state is lower than that of the $1{}^1{\rm B_u}$ state in hexatriene.

Highly precise quantum chemical calculations have provided insight into the energetic order of the $1{}^1{\rm B_u}$ and $2{}^1{\rm A_g}$ states. It has been known that the $1{}^1{\rm B_u}$ state is lower in butadiene~\cite{Watson2012excited,Chien2018excited}.
Our models of butadiene reproduce the order of the $1{}^1{\rm B_u}$ and $2{}^1{\rm A_g}$ states with a reasonably small energy gap.

In hexatriene, high-level quantum chemical calculations show that the $2{}^1{\rm A_g}$ state is much closer to $1{}^1{\rm B_u}$ in energy and slightly lower~\cite{Nakayama1998theoretical,Kurashige2004pi}.
For instance, the result of the semi-stochastic heat-bath CI~(SHCI) reported by Chien \textit{et al.}~\cite{Chien2018excited} is shown in Table~\ref{tab:exc}, and the $2{}^1{\rm A_g}$ state is slightly lower than and accidentally degenerates to the $1{}^1{\rm B_u}$ state in hexatriene.
The state order of our models is reversed to the SHCI result, yet the energy gap remains narrow.
The order of the CC3 hexatriene result is also reversed in the order of these states.
Note that the state order from experiments is similar to the results of our models and CC3. There seems to be room for a more detailed discussion about the energy gap in hexatriene.
For example, the geometry optimization calculation of the ground state at a much higher level might refine our discussion.

We also compare the result of our models with that of the CI calculation using the PPP model, termed PPP-CI~\cite{Tavan1986low}.
The single excitation energy of PPP-CI for butadiene closely matches the experimental value, whereas the double excitation energy is smaller than the single excitation energy.
On the other hand, our models reproduce the relationship between $1{}^1{\rm B_u}$ and $2{}^1{\rm A_g}$ in butadiene well.
For hexatriene, PPP-CI qualitatively reproduced the order that the $2{}^1{\rm A_g}$ state is lower than the $1{}^1{\rm B_u}$, but the gap seems large ($-$0.69~eV).
In our models, the $1{}^1{\rm B_u}$ state is lower than the $2{}^1{\rm A_g}$ state, albeit with a small gap.

Finally, we discuss the TD-DFT result.
TD-DFT reproduces the excitation energy to the $1{}^1{\rm B_u}$ state well, but in the butadiene and hexatriene cases, the excitation energy to the $2{}^1{\rm A_g}$ state is about 1~eV larger than that to the $1{}^1{\rm B_u}$ state.
This feature differs from our models and other established quantum chemical calculations.
Conventional TD-DFT is known to encounter challenges in describing double electron excitations. While it can reasonably predict the single excitation to the $1{}^1{\rm B_u}$ state, TD-DFT struggles with representing the double excitation to the $2{}^1{\rm A_g}$ state, as highlighted in Ref.~\cite{CAVE2004}.

\subsection{\texorpdfstring{$L^1$}{}-norm values of Hamiltonians}

Here, we analyze the extent to which the extended Hubbard model Hamiltonian can contribute to efficient quantum computation.
The $L^1$-norm $\lambda$ and the number of terms $N_{\rm term}$ of the fermion-to-qubit mapped Hamiltonians are shown in Table~\ref{tab:norm}.
\begin{table}[htpb]
    \centering
    \caption{$L^1$-norm value $\lambda$ (in a.u.) and the number of terms $N_{\rm term}$ of fermion-to-qubit mapped Hamiltonians. The $\lambda$ values do not include the coefficient of the constant term if not specified. Full space and active space Hamiltonians are prepared using the canonical molecular orbitals (CMOs) at the RHF/aug-cc-pVDZ level as a basis.} \label{tab:norm}
    \resizebox{\linewidth}{!}{
    \begin{threeparttable}
    \begin{tabular}{llllrrr} \hline \hline
        Molecule   & Space & Hamiltonian & Basis   & $\lambda$ & $N_{\textrm{term}}$ \\ \hline
        Ethylene   & (16e, 82o)   & Full space & CMO     & 7076      & $8.8 \times 10^6$              \\
                   & (2e, 2o)     & Active space & CMO     & 0.8         & 15                   \\
                   & (2e, 2o)     & Model 1 & Wannier & 1.3         & 19                   \\
                   & (2e, 2o)     & Model 2 & Wannier & 1.3         & 15  \\
                   \hline
        Butadiene  & (30e, 146o)  & Full space & CMO     & 40392\tnote{a}   & -                    \\
                   & (4e, 4o)     & Active space & CMO     & 3.1       & 185                  \\
                   & (4e, 4o)     & Active space & ER      & 3.6       & 361        \\
                   & (4e, 4o)     & Model 1 & Wannier & 3.3       & 85                   \\
                   & (4e, 4o)     & Model 2 & Wannier & 3.2       & 61  \\
                   \hline
        Hexatriene & (44e, 210o)  & Full space & CMO     & 101802\tnote{a}   & -                      \\
                   & (4e, 4o)     & Active space & CMO     & 2.7       & 185                  \\
                   & (4e, 4o)     & Active space & ER      & 3.1       & 361                  \\
                   & (4e, 4o)     & Model 1 & Wannier & 2.9       & 85                   \\
                   & (4e, 4o)     & Model 2 & Wannier & 2.8       & 61    \\
                   & (6e, 6o)     & Active space & CMO     & 7.1       & 919                  \\
                   & (6e, 6o)     & Active space & ER      & 7.0       & 1819                 \\
                   & (6e, 6o)     & Model 1 & Wannier & 5.8       & 199                  \\
                   & (6e, 6o)     & Model 2 & Wannier & 5.7       & 139     \\
                   \hline \hline
    \end{tabular}
    \begin{tablenotes}
    \item[a] Values with constant term evaluated using Ref.~\cite{Emieeel,Koridon2021}.
    \end{tablenotes}
    \end{threeparttable}}
\end{table}

Our model Hamiltonians have the $L^1$-norm value comparable to the active space Hamiltonian in each molecule.
Since Model~2 does not include the exchange interaction terms, the $\lambda$ values of Model~2 are reduced compared to Model~1, yet their differences are negligible.
On the contrary, the first-principles Hamiltonians of the full space have a huge $L^1$-norm value.
This contrast clearly shows that the quantum computing cost of our extended Hubbard Hamiltonian is similar to that of the active space Hamiltonian and significantly cheaper than that of the full space first-principles Hamiltonian.
Recalling that our models include dynamical electron correlations from outside of the active space, as discussed in Section \ref{sec:vert}, they have a significant advantage in the efficiency of quantum computation compared to the active space Hamiltonians, which do not include such correlations.

Since the required active space is very small for these targeted molecules, we could not show numerically that the extended Hubbard models have the advantage regarding the $L^1$-norm value against the corresponding active space Hamiltonian. 
The $L^1$-norm values are, however, expected to decrease in much larger active space cases because the extended Hubbard and active space Hamiltonians have the square and quartic number of the electron-electron interaction terms, respectively.
This effect may be seen in the hexatriene (6e, 6o) case and is expected to be more remarkable in a chemical system requiring a larger active space, {\it e.g.}, larger $\pi$-conjugated systems or transition metal complexes.

Table~\ref{tab:norm} also shows the results of ER localization of active space Hamiltonians in the butadiene and hexatriene cases.
The maximally localized Wannier function is localized orbital rather than canonical molecular orbital (CMO).
Compared to the canonical orbital cases, the $L^1$-norm values in the ER-localized basis have slightly increased.
Koridon {\it et al.} have previously reported that localized orbitals can decrease the $L^1$-norm values~\cite{Koridon2021}, but our numerical investigation does not clearly show this trend. It can be just because the necessary active space for our system is very small.

Regarding $N_\textrm{term}$, Model~2 is the smallest within the Hamiltonians of the same active space in either molecule.
The effect of neglecting the exchange terms appears in comparing Model~1 and Model~2, even though our systems are small.
The maximally localized Wannier functions do not reflect the high symmetry of the molecules.
However, the extended Hubbard model Hamiltonians have a significant advantage in $N_\textrm{term}$, whose second-quantized representation does not have the electron-electron interaction terms of the three and four kinds of distinct indices.
It is expected to facilitate the execution of quantum computations.

On the contrary, in the cases of active space Hamiltonians in an ER-localized basis, the number of terms $N_{\rm term}$ has almost doubled compared to that in a CMO basis.
Since the all-trans polyenes are highly symmetrized, the fermion-to-qubit mapped active space Hamiltonian in a CMO basis is sparse, and the number of terms is reduced.
Orbital localization leads to the lower symmetry of the orbitals, and it can be considered that the number of the terms increases.

The small number of terms in our models is considered to have advantages in quantum computations.
For example, it can be robust against errors from the noises of the noisy intermediate-scale quantum devices~\cite{Preskill2018quantumcomputingin}  because the number of measurements to evaluate the expectation values of Pauli operators is expected to become smaller.
For quantum algorithms such as quantum phase estimation, the quantum circuits become smaller and shallower.

\subsection{Application: Estimation of excitation energies from VQD calculation and sampling}

We further apply our extended Hubbard models to the classical simulation of quantum computations.
We estimate the expectation values of the models and the ER-localized active space Hamiltonians by sampling.
The quantum circuits were prepared for the ground and excited states using VQD calculations.

The VQD algorithm is an extension of VQE for calculating excited states, proposed by Higgott \textit{et al.}~\cite{Higgott2019variationalquantum}. VQD’s cost function $F({\bm \theta_k})$ is defined as 
\begin{align}
    F({\bm \theta_k}) := \braket{\Psi({\bm \theta_k}) | \mathcal{H} | \Psi({\bm \theta_k}) } + \sum_{i=0}^{k-1} \beta_i |\braket{\Psi({\bm \theta_k}) | \Psi({\bm \theta_i}) }|^2,
\end{align}
where $|\Psi({\bm \theta_k})\rangle$ is the $k$-th target excited state that can be obtained by optimizing the quantum circuit parameters ${\bm \theta_k}$.
The first term of the cost function corresponds to the expectation value of the energy of the $k$-th excited state. The second term is the penalty term that is introduced as the sum of the overlap between the $k$-th state and $i$-th ansatz states ($i=0, \cdots, k-1$). The parameter $\beta_i$ is the weight factor of each overlap term.

Table~\ref{tab:smpl} shows the estimated energy differences and the errors of the (4e, 4o) models and ER-localized active space Hamiltonians of butadiene and hexatriene.
\begin{table*}[htpb] 
    \centering
    \caption{Difference between the estimated energy from sampling and the exact ground state energy in the case of the extended Hubbard models and the ER-localized active space Hamiltonians. The sampling using $10^4$ shots is repeated $10^4$ times, and the sampling error is also shown here. The states are characterized based on the energy values of the particle-number conserving states. The exact diagonalization results are also included. The energy unit is eV.}
    \resizebox{2.0 \columnwidth}{!}{
    \begin{tabular}{lcrrrrrrrrr} \hline \hline
\multirow{2}{*}{Molecule} & \multirow{2}{*}{State} & \multicolumn{3}{c}{Model 1}  & \multicolumn{3}{c}{Model 2} & \multicolumn{3}{c}{ER-localized active space Hamiltonian}  \\
                   &                  & \multicolumn{1}{c}{Abelian grouping} & \multicolumn{1}{c}{No grouping} & \multicolumn{1}{c}{Exact} & \multicolumn{1}{c}{Abelian grouping} & \multicolumn{1}{c}{No grouping} & \multicolumn{1}{c}{Exact} & \multicolumn{1}{c}{Abelian grouping} & \multicolumn{1}{c}{No grouping} & \multicolumn{1}{c}{Exact} \\ \hline
        Butadiene  & S$_0$ & $0.01 \pm 0.13$ & $0.00 \pm 0.14$ & 0.00 & $0.05 \pm 0.13$ & $0.05 \pm 0.14$ & 0.00 & $0.00 \pm 0.05$ & $0.00 \pm 0.07$ & 0.00 \\
                   & S$_1$ & $5.47 \pm 0.16$ & $5.47 \pm 0.14$ & 5.47 & $5.79 \pm 0.15$ & $5.78 \pm 0.14$ & 5.82 & $6.93 \pm 0.09$ & $6.93 \pm 0.11$ & 6.96 \\
                   & S$_2$ & $5.73 \pm 0.10$ & $5.73 \pm 0.14$ & 5.64 & $6.02 \pm 0.12$ & $6.02 \pm 0.14$ & 5.98 & $7.61 \pm 0.09$ & $7.61 \pm 0.13$ & 7.71 \\ \hline
        Hexatriene & S$_0$ & $0.01 \pm 0.11$ & $0.02 \pm 0.13$ & 0.00 & $0.02 \pm 0.11$ & $0.02 \pm 0.12$ & 0.00 & $0.00 \pm 0.08$ & $0.00 \pm 0.10$ & 0.00 \\
                   & S$_1$ & $4.71 \pm 0.16$ & $4.71 \pm 0.13$ & 4.67 & $4.90 \pm 0.14$ & $4.90 \pm 0.13$ & 4.81 & $6.09 \pm 0.08$ & $6.09 \pm 0.11$ & 6.08 \\
                   & S$_2$ & $5.08 \pm 0.09$ & $5.08 \pm 0.13$ & 5.05 & $5.22 \pm 0.09$ & $5.22 \pm 0.13$ & 5.12 & $7.06 \pm 0.08$ & $7.06 \pm 0.11$ & 7.05 \\
        \hline \hline
    \label{tab:smpl}
    \end{tabular}
    }
\end{table*}
The energy difference is between the energy estimated by sampling and the exact ground-state energy.
The energy is estimated by repeating the $10^4$ shots energy calculation $10^4$ times.

The estimated energy differences using our models are in good agreement with those of exact diagonalization.
Those using the ER-localized active space Hamiltonians are similarly estimated well but overestimate the excitation energies due to the absence of electron correlation related to the outside of the active space.
The errors are similar for both our models and ER-localized active space Hamiltonians.
Hence, we validate the applicability of the \textit{ab initio} downfolded model to compute excited states using quantum devices through numerical calculations using our models.

\section{Conclusions} \label{sec:conclusion}

We propose introducing extended Hubbard models for isolated molecules based on the \textit{ab initio} downfolding method~\cite{Aryasetiawan2004frequency,RESPACK1} toward efficient quantum computing.
To demonstrate the effectiveness of this approach, we constructed models for short polyenes: ethylene, butadiene, and hexatriene. 
The constructed models are intended to describe the lower-lying excited states of these conjugated systems.
They exhibit two key features: 
First, screened Coulomb and exchange integrals effectively incorporate the electron-correlation effect from outside of the active space through cRPA.
Second, the extended Hubbard model has fewer terms than the usual active space Hamiltonian because the electron repulsion integral tensor is expressed as second-order rather than fourth-order.
In summary, the proposed approach depicted in Figure~\ref{fig:scheme} efficiently uses the \textit{ab initio} downfolding method to construct an extended Hubbard model for a molecule, reducing the need for numerous qubits and quantum operations compared to the conventional approach.

It found that our models, particularly Model~2, demonstrate reasonable excitation energies compared to experimental values and high-level quantum chemical calculation results, offering practicality under limited quantum resources. 
The impact of introducing an approximation such as cRPA has been quantitatively analyzed: The incorporation of the dynamical electron correlation effect from outside of the active space is examined through the comparison with the continuum limit.
Besides, one may say that this study has redetermined the parameters of the second-quantized PPP model Hamiltonian for short polyenes in an \textit{ab initio} manner.

We also contribute to showing that our models maintain a small $L^1$-norm and a reduced number of terms, making them advantageous in quantum computing. 
In simulations of sampling estimation, our models successfully estimate the excitation energies in molecules like butadiene and hexatriene, and they include doubly excited states, which are challenging for conventional TD-DFT. It suggests a promising direction for quantum chemical computation on quantum computers. 

It is worthwhile to evaluate the performance of the \textit{ab initio} extended Hubbard models for molecules with a real quantum computer.
This approach can be beneficial for fault-tolerant quantum computers as well as near-term quantum devices with restricted numbers of reliable qubits and quantum gate operations~\cite{yoshioka2022hunting,ichikawa2023comprehensive}.

Updating the framework of \textit{ab initio} downfolding also seems a meaningful research direction. For example, cRPA is known to overestimate the screening effect on two-body interactions~\cite{Shinaoka2015accuracy,Carsten2018limitations}, and Scott \textit{et al.} address this issue~\cite{Scott2024rigorous}.
Considering the future application to larger molecular systems, the fast evaluation of the effective two-body integrals, whose bottleneck is the parallelized calculation of the polarization function, as Nakamura \textit{et al.} reported~\cite{RESPACK1}, is also considered important.

\section*{Acknowledgements} \label{sec:acknowledgements}

We would like to thank Hideaki Hakoshima, Hiroshi Shinaoka, and Takahiro Misawa for fruitful discussion.
This project has been supported by funding from the MEXT Quantum Leap Flagship Program (MEXTQLEAP) through Grant No. JPMXS0120319794, and the JST COI-NEXT Program through Grant No. JPMJPF2014. 
YY wishes to thank JSPS KAKENHI Grant No. JP21K20536.
NT wishes to thank JSPS KAKENHI Grant Nos. JP19H05817, JP19H05820, JST PRESTO Grant No. JPMJPR23F6.
WM wishes to thank JSPS KAKENHI Grant Nos. JP23H03819 and JP21K18933. 
We thank the Supercomputer Center, the Institute for Solid State Physics, the University of Tokyo for the use of the facilities.
YY thank Research Center for Computational Science, Okazaki, Japan (Project: 23-IMS-C183).

\bibliography{exHub}

\begin{thebibliography}{116}%
\makeatletter
\providecommand \@ifxundefined [1]{%
 \@ifx{#1\undefined}
}%
\providecommand \@ifnum [1]{%
 \ifnum #1\expandafter \@firstoftwo
 \else \expandafter \@secondoftwo
 \fi
}%
\providecommand \@ifx [1]{%
 \ifx #1\expandafter \@firstoftwo
 \else \expandafter \@secondoftwo
 \fi
}%
\providecommand \natexlab [1]{#1}%
\providecommand \enquote  [1]{``#1''}%
\providecommand \bibnamefont  [1]{#1}%
\providecommand \bibfnamefont [1]{#1}%
\providecommand \citenamefont [1]{#1}%
\providecommand \href@noop [0]{\@secondoftwo}%
\providecommand \href [0]{\begingroup \@sanitize@url \@href}%
\providecommand \@href[1]{\@@startlink{#1}\@@href}%
\providecommand \@@href[1]{\endgroup#1\@@endlink}%
\providecommand \@sanitize@url [0]{\catcode `\\12\catcode `\$12\catcode
  `\&12\catcode `\#12\catcode `\^12\catcode `\_12\catcode `\%12\relax}%
\providecommand \@@startlink[1]{}%
\providecommand \@@endlink[0]{}%
\providecommand \url  [0]{\begingroup\@sanitize@url \@url }%
\providecommand \@url [1]{\endgroup\@href {#1}{\urlprefix }}%
\providecommand \urlprefix  [0]{URL }%
\providecommand \Eprint [0]{\href }%
\providecommand \doibase [0]{http://dx.doi.org/}%
\providecommand \selectlanguage [0]{\@gobble}%
\providecommand \bibinfo  [0]{\@secondoftwo}%
\providecommand \bibfield  [0]{\@secondoftwo}%
\providecommand \translation [1]{[#1]}%
\providecommand \BibitemOpen [0]{}%
\providecommand \bibitemStop [0]{}%
\providecommand \bibitemNoStop [0]{.\EOS\space}%
\providecommand \EOS [0]{\spacefactor3000\relax}%
\providecommand \BibitemShut  [1]{\csname bibitem#1\endcsname}%
\let\auto@bib@innerbib\@empty
\bibitem [{\citenamefont {Motta}\ and\ \citenamefont
  {Rice}(2022)}]{Motta2022emerging}%
  \BibitemOpen
  \bibfield  {author} {\bibinfo {author} {\bibfnamefont {M.}~\bibnamefont
  {Motta}}\ and\ \bibinfo {author} {\bibfnamefont {J.~E.}\ \bibnamefont
  {Rice}},\ }\href {\doibase https://doi.org/10.1002/wcms.1580} {\bibfield
  {journal} {\bibinfo  {journal} {WIREs Comput. Mol. Sci.}\ }\textbf {\bibinfo
  {volume} {12}},\ \bibinfo {pages} {e1580} (\bibinfo {year}
  {2022})}\BibitemShut {NoStop}%
\bibitem [{\citenamefont {{Dalzell}}\ \emph {et~al.}(2023)\citenamefont
  {{Dalzell}}, \citenamefont {{McArdle}}, \citenamefont {{Berta}},
  \citenamefont {{Bienias}}, \citenamefont {{Chen}}, \citenamefont
  {{Gily{\'e}n}}, \citenamefont {{Hann}}, \citenamefont {{Kastoryano}},
  \citenamefont {{Khabiboulline}}, \citenamefont {{Kubica}}, \citenamefont
  {{Salton}}, \citenamefont {{Wang}},\ and\ \citenamefont
  {{Brand{\~a}o}}}]{Dalzell023quantum}%
  \BibitemOpen
  \bibfield  {author} {\bibinfo {author} {\bibfnamefont {A.~M.}\ \bibnamefont
  {{Dalzell}}}, \bibinfo {author} {\bibfnamefont {S.}~\bibnamefont
  {{McArdle}}}, \bibinfo {author} {\bibfnamefont {M.}~\bibnamefont {{Berta}}},
  \bibinfo {author} {\bibfnamefont {P.}~\bibnamefont {{Bienias}}}, \bibinfo
  {author} {\bibfnamefont {C.-F.}\ \bibnamefont {{Chen}}}, \bibinfo {author}
  {\bibfnamefont {A.}~\bibnamefont {{Gily{\'e}n}}}, \bibinfo {author}
  {\bibfnamefont {C.~T.}\ \bibnamefont {{Hann}}}, \bibinfo {author}
  {\bibfnamefont {M.~J.}\ \bibnamefont {{Kastoryano}}}, \bibinfo {author}
  {\bibfnamefont {E.~T.}\ \bibnamefont {{Khabiboulline}}}, \bibinfo {author}
  {\bibfnamefont {A.}~\bibnamefont {{Kubica}}}, \bibinfo {author}
  {\bibfnamefont {G.}~\bibnamefont {{Salton}}}, \bibinfo {author}
  {\bibfnamefont {S.}~\bibnamefont {{Wang}}}, \ and\ \bibinfo {author}
  {\bibfnamefont {F.~G.~S.~L.}\ \bibnamefont {{Brand{\~a}o}}},\ }\href@noop {}
  {\bibfield  {journal} {\bibinfo  {journal} {arXiv}\ } (\bibinfo {year}
  {2023})},\ \Eprint {http://arxiv.org/abs/2310.03011} {arXiv:2310.03011
  [quant-ph]} \BibitemShut {NoStop}%
\bibitem [{\citenamefont {Abrams}\ and\ \citenamefont
  {Lloyd}(1999)}]{Abrams1999quantum}%
  \BibitemOpen
  \bibfield  {author} {\bibinfo {author} {\bibfnamefont {D.~S.}\ \bibnamefont
  {Abrams}}\ and\ \bibinfo {author} {\bibfnamefont {S.}~\bibnamefont {Lloyd}},\
  }\href {\doibase 10.1103/PhysRevLett.83.5162} {\bibfield  {journal} {\bibinfo
   {journal} {Phys. Rev. Lett.}\ }\textbf {\bibinfo {volume} {83}},\ \bibinfo
  {pages} {5162} (\bibinfo {year} {1999})}\BibitemShut {NoStop}%
\bibitem [{\citenamefont {Aspuru-Guzik}\ \emph {et~al.}(2005)\citenamefont
  {Aspuru-Guzik}, \citenamefont {Dutoi}, \citenamefont {Love},\ and\
  \citenamefont {Head-Gordon}}]{Aspuru-Guzik2005simulated}%
  \BibitemOpen
  \bibfield  {author} {\bibinfo {author} {\bibfnamefont {A.}~\bibnamefont
  {Aspuru-Guzik}}, \bibinfo {author} {\bibfnamefont {A.~D.}\ \bibnamefont
  {Dutoi}}, \bibinfo {author} {\bibfnamefont {P.~J.}\ \bibnamefont {Love}}, \
  and\ \bibinfo {author} {\bibfnamefont {M.}~\bibnamefont {Head-Gordon}},\
  }\href {\doibase 10.1126/science.1113479} {\bibfield  {journal} {\bibinfo
  {journal} {Science}\ }\textbf {\bibinfo {volume} {309}},\ \bibinfo {pages}
  {1704} (\bibinfo {year} {2005})}\BibitemShut {NoStop}%
\bibitem [{\citenamefont {Reiher}\ \emph {et~al.}(2017)\citenamefont {Reiher},
  \citenamefont {Wiebe}, \citenamefont {Svore}, \citenamefont {Wecker},\ and\
  \citenamefont {Troyer}}]{Reiher2017elucidating}%
  \BibitemOpen
  \bibfield  {author} {\bibinfo {author} {\bibfnamefont {M.}~\bibnamefont
  {Reiher}}, \bibinfo {author} {\bibfnamefont {N.}~\bibnamefont {Wiebe}},
  \bibinfo {author} {\bibfnamefont {K.~M.}\ \bibnamefont {Svore}}, \bibinfo
  {author} {\bibfnamefont {D.}~\bibnamefont {Wecker}}, \ and\ \bibinfo {author}
  {\bibfnamefont {M.}~\bibnamefont {Troyer}},\ }\href {\doibase
  10.1073/pnas.1619152114} {\bibfield  {journal} {\bibinfo  {journal} {Proc.
  Natl. Acad. Sci. U.S.A.}\ }\textbf {\bibinfo {volume} {114}},\ \bibinfo
  {pages} {7555} (\bibinfo {year} {2017})}\BibitemShut {NoStop}%
\bibitem [{\citenamefont {Lee}\ \emph {et~al.}(2021)\citenamefont {Lee},
  \citenamefont {Berry}, \citenamefont {Gidney}, \citenamefont {Huggins},
  \citenamefont {McClean}, \citenamefont {Wiebe},\ and\ \citenamefont
  {Babbush}}]{Lee2021even}%
  \BibitemOpen
  \bibfield  {author} {\bibinfo {author} {\bibfnamefont {J.}~\bibnamefont
  {Lee}}, \bibinfo {author} {\bibfnamefont {D.~W.}\ \bibnamefont {Berry}},
  \bibinfo {author} {\bibfnamefont {C.}~\bibnamefont {Gidney}}, \bibinfo
  {author} {\bibfnamefont {W.~J.}\ \bibnamefont {Huggins}}, \bibinfo {author}
  {\bibfnamefont {J.~R.}\ \bibnamefont {McClean}}, \bibinfo {author}
  {\bibfnamefont {N.}~\bibnamefont {Wiebe}}, \ and\ \bibinfo {author}
  {\bibfnamefont {R.}~\bibnamefont {Babbush}},\ }\href {\doibase
  10.1103/PRXQuantum.2.030305} {\bibfield  {journal} {\bibinfo  {journal} {PRX
  Quantum}\ }\textbf {\bibinfo {volume} {2}},\ \bibinfo {pages} {030305}
  (\bibinfo {year} {2021})}\BibitemShut {NoStop}%
\bibitem [{\citenamefont {von Burg}\ \emph {et~al.}(2021)\citenamefont {von
  Burg}, \citenamefont {Low}, \citenamefont {H\"aner}, \citenamefont {Steiger},
  \citenamefont {Reiher}, \citenamefont {Roetteler},\ and\ \citenamefont
  {Troyer}}]{von-Burg2021quantum}%
  \BibitemOpen
  \bibfield  {author} {\bibinfo {author} {\bibfnamefont {V.}~\bibnamefont {von
  Burg}}, \bibinfo {author} {\bibfnamefont {G.~H.}\ \bibnamefont {Low}},
  \bibinfo {author} {\bibfnamefont {T.}~\bibnamefont {H\"aner}}, \bibinfo
  {author} {\bibfnamefont {D.~S.}\ \bibnamefont {Steiger}}, \bibinfo {author}
  {\bibfnamefont {M.}~\bibnamefont {Reiher}}, \bibinfo {author} {\bibfnamefont
  {M.}~\bibnamefont {Roetteler}}, \ and\ \bibinfo {author} {\bibfnamefont
  {M.}~\bibnamefont {Troyer}},\ }\href {\doibase
  10.1103/PhysRevResearch.3.033055} {\bibfield  {journal} {\bibinfo  {journal}
  {Phys. Rev. Res.}\ }\textbf {\bibinfo {volume} {3}},\ \bibinfo {pages}
  {033055} (\bibinfo {year} {2021})}\BibitemShut {NoStop}%
\bibitem [{\citenamefont {Rubin}\ \emph {et~al.}(2023)\citenamefont {Rubin},
  \citenamefont {Berry}, \citenamefont {Malone}, \citenamefont {White},
  \citenamefont {Khattar}, \citenamefont {DePrince}, \citenamefont {Sicolo},
  \citenamefont {K\"uehn}, \citenamefont {Kaicher}, \citenamefont {Lee},\ and\
  \citenamefont {Babbush}}]{Rubin2023fault}%
  \BibitemOpen
  \bibfield  {author} {\bibinfo {author} {\bibfnamefont {N.~C.}\ \bibnamefont
  {Rubin}}, \bibinfo {author} {\bibfnamefont {D.~W.}\ \bibnamefont {Berry}},
  \bibinfo {author} {\bibfnamefont {F.~D.}\ \bibnamefont {Malone}}, \bibinfo
  {author} {\bibfnamefont {A.~F.}\ \bibnamefont {White}}, \bibinfo {author}
  {\bibfnamefont {T.}~\bibnamefont {Khattar}}, \bibinfo {author} {\bibfnamefont
  {A.~E.}\ \bibnamefont {DePrince}}, \bibinfo {author} {\bibfnamefont
  {S.}~\bibnamefont {Sicolo}}, \bibinfo {author} {\bibfnamefont
  {M.}~\bibnamefont {K\"uehn}}, \bibinfo {author} {\bibfnamefont
  {M.}~\bibnamefont {Kaicher}}, \bibinfo {author} {\bibfnamefont
  {J.}~\bibnamefont {Lee}}, \ and\ \bibinfo {author} {\bibfnamefont
  {R.}~\bibnamefont {Babbush}},\ }\href {\doibase 10.1103/PRXQuantum.4.040303}
  {\bibfield  {journal} {\bibinfo  {journal} {PRX Quantum}\ }\textbf {\bibinfo
  {volume} {4}},\ \bibinfo {pages} {040303} (\bibinfo {year}
  {2023})}\BibitemShut {NoStop}%
\bibitem [{\citenamefont {Blunt}\ \emph {et~al.}(2022)\citenamefont {Blunt},
  \citenamefont {Camps}, \citenamefont {Crawford}, \citenamefont {Izs\'ak},
  \citenamefont {Leontica}, \citenamefont {Mirani}, \citenamefont {Moylett},
  \citenamefont {Scivier}, \citenamefont {S\"underhauf}, \citenamefont
  {Schopf}, \citenamefont {Taylor},\ and\ \citenamefont
  {Holzmann}}]{Blunt2022perspective}%
  \BibitemOpen
  \bibfield  {author} {\bibinfo {author} {\bibfnamefont {N.~S.}\ \bibnamefont
  {Blunt}}, \bibinfo {author} {\bibfnamefont {J.}~\bibnamefont {Camps}},
  \bibinfo {author} {\bibfnamefont {O.}~\bibnamefont {Crawford}}, \bibinfo
  {author} {\bibfnamefont {R.}~\bibnamefont {Izs\'ak}}, \bibinfo {author}
  {\bibfnamefont {S.}~\bibnamefont {Leontica}}, \bibinfo {author}
  {\bibfnamefont {A.}~\bibnamefont {Mirani}}, \bibinfo {author} {\bibfnamefont
  {A.~E.}\ \bibnamefont {Moylett}}, \bibinfo {author} {\bibfnamefont {S.~A.}\
  \bibnamefont {Scivier}}, \bibinfo {author} {\bibfnamefont {C.}~\bibnamefont
  {S\"underhauf}}, \bibinfo {author} {\bibfnamefont {P.}~\bibnamefont
  {Schopf}}, \bibinfo {author} {\bibfnamefont {J.~M.}\ \bibnamefont {Taylor}},
  \ and\ \bibinfo {author} {\bibfnamefont {N.}~\bibnamefont {Holzmann}},\
  }\href {\doibase 10.1021/acs.jctc.2c00574} {\bibfield  {journal} {\bibinfo
  {journal} {J. Chem. Theory Comput.}\ }\textbf {\bibinfo {volume} {18}},\
  \bibinfo {pages} {7001} (\bibinfo {year} {2022})}\BibitemShut {NoStop}%
\bibitem [{\citenamefont {Serrano-Andr\'{e}s}\ and\ \citenamefont
  {Merch\'{a}n}(2005)}]{Serrano-Andres2005quantum}%
  \BibitemOpen
  \bibfield  {author} {\bibinfo {author} {\bibfnamefont {L.}~\bibnamefont
  {Serrano-Andr\'{e}s}}\ and\ \bibinfo {author} {\bibfnamefont
  {M.}~\bibnamefont {Merch\'{a}n}},\ }\href {\doibase
  https://doi.org/10.1016/j.theochem.2005.03.020} {\bibfield  {journal}
  {\bibinfo  {journal} {J. Mol. Struct.: THEOCHEM}\ }\textbf {\bibinfo {volume}
  {729}},\ \bibinfo {pages} {99} (\bibinfo {year} {2005})}\BibitemShut
  {NoStop}%
\bibitem [{\citenamefont {Higgott}\ \emph {et~al.}(2019)\citenamefont
  {Higgott}, \citenamefont {Wang},\ and\ \citenamefont
  {Brierley}}]{Higgott2019variationalquantum}%
  \BibitemOpen
  \bibfield  {author} {\bibinfo {author} {\bibfnamefont {O.}~\bibnamefont
  {Higgott}}, \bibinfo {author} {\bibfnamefont {D.}~\bibnamefont {Wang}}, \
  and\ \bibinfo {author} {\bibfnamefont {S.}~\bibnamefont {Brierley}},\ }\href
  {\doibase 10.22331/q-2019-07-01-156} {\bibfield  {journal} {\bibinfo
  {journal} {{Quantum}}\ }\textbf {\bibinfo {volume} {3}},\ \bibinfo {pages}
  {156} (\bibinfo {year} {2019})}\BibitemShut {NoStop}%
\bibitem [{\citenamefont {Jones}\ \emph {et~al.}(2019)\citenamefont {Jones},
  \citenamefont {Endo}, \citenamefont {McArdle}, \citenamefont {Yuan},\ and\
  \citenamefont {Benjamin}}]{Jones2019}%
  \BibitemOpen
  \bibfield  {author} {\bibinfo {author} {\bibfnamefont {T.}~\bibnamefont
  {Jones}}, \bibinfo {author} {\bibfnamefont {S.}~\bibnamefont {Endo}},
  \bibinfo {author} {\bibfnamefont {S.}~\bibnamefont {McArdle}}, \bibinfo
  {author} {\bibfnamefont {X.}~\bibnamefont {Yuan}}, \ and\ \bibinfo {author}
  {\bibfnamefont {S.~C.}\ \bibnamefont {Benjamin}},\ }\href {\doibase
  10.1103/PhysRevA.99.062304} {\bibfield  {journal} {\bibinfo  {journal} {Phys.
  Rev. A}\ }\textbf {\bibinfo {volume} {99}},\ \bibinfo {pages} {062304}
  (\bibinfo {year} {2019})}\BibitemShut {NoStop}%
\bibitem [{\citenamefont {McClean}\ \emph {et~al.}(2017)\citenamefont
  {McClean}, \citenamefont {Kimchi-Schwartz}, \citenamefont {Carter},\ and\
  \citenamefont {de~Jong}}]{McClean2017}%
  \BibitemOpen
  \bibfield  {author} {\bibinfo {author} {\bibfnamefont {J.~R.}\ \bibnamefont
  {McClean}}, \bibinfo {author} {\bibfnamefont {M.~E.}\ \bibnamefont
  {Kimchi-Schwartz}}, \bibinfo {author} {\bibfnamefont {J.}~\bibnamefont
  {Carter}}, \ and\ \bibinfo {author} {\bibfnamefont {W.~A.}\ \bibnamefont
  {de~Jong}},\ }\href {\doibase 10.1103/PhysRevA.95.042308} {\bibfield
  {journal} {\bibinfo  {journal} {Phys. Rev. A}\ }\textbf {\bibinfo {volume}
  {95}},\ \bibinfo {pages} {042308} (\bibinfo {year} {2017})}\BibitemShut
  {NoStop}%
\bibitem [{\citenamefont {Nakanishi}\ \emph {et~al.}(2019)\citenamefont
  {Nakanishi}, \citenamefont {Mitarai},\ and\ \citenamefont
  {Fujii}}]{Nakanishi2019}%
  \BibitemOpen
  \bibfield  {author} {\bibinfo {author} {\bibfnamefont {K.~M.}\ \bibnamefont
  {Nakanishi}}, \bibinfo {author} {\bibfnamefont {K.}~\bibnamefont {Mitarai}},
  \ and\ \bibinfo {author} {\bibfnamefont {K.}~\bibnamefont {Fujii}},\ }\href
  {\doibase 10.1103/PhysRevResearch.1.033062} {\bibfield  {journal} {\bibinfo
  {journal} {Phys. Rev. Res.}\ }\textbf {\bibinfo {volume} {1}},\ \bibinfo
  {pages} {033062} (\bibinfo {year} {2019})}\BibitemShut {NoStop}%
\bibitem [{\citenamefont {Parrish}\ \emph {et~al.}(2019)\citenamefont
  {Parrish}, \citenamefont {Hohenstein}, \citenamefont {McMahon},\ and\
  \citenamefont {Mart\'{\i}nez}}]{Parrish2019quantum}%
  \BibitemOpen
  \bibfield  {author} {\bibinfo {author} {\bibfnamefont {R.~M.}\ \bibnamefont
  {Parrish}}, \bibinfo {author} {\bibfnamefont {E.~G.}\ \bibnamefont
  {Hohenstein}}, \bibinfo {author} {\bibfnamefont {P.~L.}\ \bibnamefont
  {McMahon}}, \ and\ \bibinfo {author} {\bibfnamefont {T.~J.}\ \bibnamefont
  {Mart\'{\i}nez}},\ }\href {\doibase 10.1103/PhysRevLett.122.230401}
  {\bibfield  {journal} {\bibinfo  {journal} {Phys. Rev. Lett.}\ }\textbf
  {\bibinfo {volume} {122}},\ \bibinfo {pages} {230401} (\bibinfo {year}
  {2019})}\BibitemShut {NoStop}%
\bibitem [{\citenamefont {Ollitrault}\ \emph {et~al.}(2020)\citenamefont
  {Ollitrault}, \citenamefont {Kandala}, \citenamefont {Chen}, \citenamefont
  {Barkoutsos}, \citenamefont {Mezzacapo}, \citenamefont {Pistoia},
  \citenamefont {Sheldon}, \citenamefont {Woerner}, \citenamefont {Gambetta},\
  and\ \citenamefont {Tavernelli}}]{Ollitrault2020quantum}%
  \BibitemOpen
  \bibfield  {author} {\bibinfo {author} {\bibfnamefont {P.~J.}\ \bibnamefont
  {Ollitrault}}, \bibinfo {author} {\bibfnamefont {A.}~\bibnamefont {Kandala}},
  \bibinfo {author} {\bibfnamefont {C.-F.}\ \bibnamefont {Chen}}, \bibinfo
  {author} {\bibfnamefont {P.~K.}\ \bibnamefont {Barkoutsos}}, \bibinfo
  {author} {\bibfnamefont {A.}~\bibnamefont {Mezzacapo}}, \bibinfo {author}
  {\bibfnamefont {M.}~\bibnamefont {Pistoia}}, \bibinfo {author} {\bibfnamefont
  {S.}~\bibnamefont {Sheldon}}, \bibinfo {author} {\bibfnamefont
  {S.}~\bibnamefont {Woerner}}, \bibinfo {author} {\bibfnamefont {J.~M.}\
  \bibnamefont {Gambetta}}, \ and\ \bibinfo {author} {\bibfnamefont
  {I.}~\bibnamefont {Tavernelli}},\ }\href {\doibase
  10.1103/PhysRevResearch.2.043140} {\bibfield  {journal} {\bibinfo  {journal}
  {Phys. Rev. Res.}\ }\textbf {\bibinfo {volume} {2}},\ \bibinfo {pages}
  {043140} (\bibinfo {year} {2020})}\BibitemShut {NoStop}%
\bibitem [{\citenamefont {Asthana}\ \emph {et~al.}(2023)\citenamefont
  {Asthana}, \citenamefont {Kumar}, \citenamefont {Abraham}, \citenamefont
  {Grimsley}, \citenamefont {Zhang}, \citenamefont {Cincio}, \citenamefont
  {Tretiak}, \citenamefont {Dub}, \citenamefont {Economou}, \citenamefont
  {Barnes},\ and\ \citenamefont {Mayhall}}]{Asthana2023quantum}%
  \BibitemOpen
  \bibfield  {author} {\bibinfo {author} {\bibfnamefont {A.}~\bibnamefont
  {Asthana}}, \bibinfo {author} {\bibfnamefont {A.}~\bibnamefont {Kumar}},
  \bibinfo {author} {\bibfnamefont {V.}~\bibnamefont {Abraham}}, \bibinfo
  {author} {\bibfnamefont {H.}~\bibnamefont {Grimsley}}, \bibinfo {author}
  {\bibfnamefont {Y.}~\bibnamefont {Zhang}}, \bibinfo {author} {\bibfnamefont
  {L.}~\bibnamefont {Cincio}}, \bibinfo {author} {\bibfnamefont
  {S.}~\bibnamefont {Tretiak}}, \bibinfo {author} {\bibfnamefont {P.~A.}\
  \bibnamefont {Dub}}, \bibinfo {author} {\bibfnamefont {S.~E.}\ \bibnamefont
  {Economou}}, \bibinfo {author} {\bibfnamefont {E.}~\bibnamefont {Barnes}}, \
  and\ \bibinfo {author} {\bibfnamefont {N.~J.}\ \bibnamefont {Mayhall}},\
  }\href {\doibase 10.1039/D2SC05371C} {\bibfield  {journal} {\bibinfo
  {journal} {Chem. Sci.}\ }\textbf {\bibinfo {volume} {14}},\ \bibinfo {pages}
  {2405} (\bibinfo {year} {2023})}\BibitemShut {NoStop}%
\bibitem [{\citenamefont {Yalouz}\ \emph {et~al.}(2021)\citenamefont {Yalouz},
  \citenamefont {Senjean}, \citenamefont {G\"{u}nther}, \citenamefont {Buda},
  \citenamefont {O’Brien},\ and\ \citenamefont {Visscher}}]{Yalouz2021state}%
  \BibitemOpen
  \bibfield  {author} {\bibinfo {author} {\bibfnamefont {S.}~\bibnamefont
  {Yalouz}}, \bibinfo {author} {\bibfnamefont {B.}~\bibnamefont {Senjean}},
  \bibinfo {author} {\bibfnamefont {J.}~\bibnamefont {G\"{u}nther}}, \bibinfo
  {author} {\bibfnamefont {F.}~\bibnamefont {Buda}}, \bibinfo {author}
  {\bibfnamefont {T.~E.}\ \bibnamefont {O’Brien}}, \ and\ \bibinfo {author}
  {\bibfnamefont {L.}~\bibnamefont {Visscher}},\ }\href {\doibase
  10.1088/2058-9565/abd334} {\bibfield  {journal} {\bibinfo  {journal} {Quantum
  Sci. Technol.}\ }\textbf {\bibinfo {volume} {6}},\ \bibinfo {pages} {024004}
  (\bibinfo {year} {2021})}\BibitemShut {NoStop}%
\bibitem [{\citenamefont {Omiya}\ \emph {et~al.}(2022)\citenamefont {Omiya},
  \citenamefont {Nakagawa}, \citenamefont {Koh}, \citenamefont {Mizukami},
  \citenamefont {Gao},\ and\ \citenamefont {Kobayashi}}]{Omiya2022analytical}%
  \BibitemOpen
  \bibfield  {author} {\bibinfo {author} {\bibfnamefont {K.}~\bibnamefont
  {Omiya}}, \bibinfo {author} {\bibfnamefont {Y.~O.}\ \bibnamefont {Nakagawa}},
  \bibinfo {author} {\bibfnamefont {S.}~\bibnamefont {Koh}}, \bibinfo {author}
  {\bibfnamefont {W.}~\bibnamefont {Mizukami}}, \bibinfo {author}
  {\bibfnamefont {Q.}~\bibnamefont {Gao}}, \ and\ \bibinfo {author}
  {\bibfnamefont {T.}~\bibnamefont {Kobayashi}},\ }\href {\doibase
  10.1021/acs.jctc.1c00877} {\bibfield  {journal} {\bibinfo  {journal} {J.
  Chem. Theory Comput.}\ }\textbf {\bibinfo {volume} {18}},\ \bibinfo {pages}
  {741} (\bibinfo {year} {2022})}\BibitemShut {NoStop}%
\bibitem [{\citenamefont {Yoshikura}\ \emph {et~al.}(2023)\citenamefont
  {Yoshikura}, \citenamefont {Ten-no},\ and\ \citenamefont
  {Tsuchimochi}}]{Yoshikura2023quantum}%
  \BibitemOpen
  \bibfield  {author} {\bibinfo {author} {\bibfnamefont {T.}~\bibnamefont
  {Yoshikura}}, \bibinfo {author} {\bibfnamefont {S.~L.}\ \bibnamefont
  {Ten-no}}, \ and\ \bibinfo {author} {\bibfnamefont {T.}~\bibnamefont
  {Tsuchimochi}},\ }\href {\doibase 10.1021/acs.jpca.3c02800} {\bibfield
  {journal} {\bibinfo  {journal} {J. Phys. Chem. A}\ }\textbf {\bibinfo
  {volume} {127}},\ \bibinfo {pages} {6577} (\bibinfo {year}
  {2023})}\BibitemShut {NoStop}%
\bibitem [{\citenamefont {Yeter-Aydeniz}\ \emph {et~al.}(2020)\citenamefont
  {Yeter-Aydeniz}, \citenamefont {Pooser},\ and\ \citenamefont
  {Siopsis}}]{Yeter-Aydeniz2020practical}%
  \BibitemOpen
  \bibfield  {author} {\bibinfo {author} {\bibfnamefont {K.}~\bibnamefont
  {Yeter-Aydeniz}}, \bibinfo {author} {\bibfnamefont {R.~C.}\ \bibnamefont
  {Pooser}}, \ and\ \bibinfo {author} {\bibfnamefont {G.}~\bibnamefont
  {Siopsis}},\ }\href {\doibase 10.1038/s41534-020-00290-1} {\bibfield
  {journal} {\bibinfo  {journal} {Npj Quantum Inf.}\ }\textbf {\bibinfo
  {volume} {6}},\ \bibinfo {pages} {63} (\bibinfo {year} {2020})}\BibitemShut
  {NoStop}%
\bibitem [{\citenamefont {Tsuchimochi}\ \emph
  {et~al.}(2023{\natexlab{a}})\citenamefont {Tsuchimochi}, \citenamefont {Ryo},
  \citenamefont {Ten-no},\ and\ \citenamefont
  {Sasasako}}]{Tsuchimochi2023improved}%
  \BibitemOpen
  \bibfield  {author} {\bibinfo {author} {\bibfnamefont {T.}~\bibnamefont
  {Tsuchimochi}}, \bibinfo {author} {\bibfnamefont {Y.}~\bibnamefont {Ryo}},
  \bibinfo {author} {\bibfnamefont {S.~L.}\ \bibnamefont {Ten-no}}, \ and\
  \bibinfo {author} {\bibfnamefont {K.}~\bibnamefont {Sasasako}},\ }\href
  {\doibase 10.1021/acs.jctc.2c00906} {\bibfield  {journal} {\bibinfo
  {journal} {J. Chem. Theory Comput.}\ }\textbf {\bibinfo {volume} {19}},\
  \bibinfo {pages} {503} (\bibinfo {year} {2023}{\natexlab{a}})}\BibitemShut
  {NoStop}%
\bibitem [{\citenamefont {Tsuchimochi}\ \emph
  {et~al.}(2023{\natexlab{b}})\citenamefont {Tsuchimochi}, \citenamefont {Ryo},
  \citenamefont {Tsang},\ and\ \citenamefont {Ten-no}}]{Tsuchimochi2023multi}%
  \BibitemOpen
  \bibfield  {author} {\bibinfo {author} {\bibfnamefont {T.}~\bibnamefont
  {Tsuchimochi}}, \bibinfo {author} {\bibfnamefont {Y.}~\bibnamefont {Ryo}},
  \bibinfo {author} {\bibfnamefont {S.~C.}\ \bibnamefont {Tsang}}, \ and\
  \bibinfo {author} {\bibfnamefont {S.~L.}\ \bibnamefont {Ten-no}},\ }\href
  {\doibase 10.1038/s41534-023-00780-y} {\bibfield  {journal} {\bibinfo
  {journal} {Npj Quantum Inf.}\ }\textbf {\bibinfo {volume} {9}},\ \bibinfo
  {pages} {113} (\bibinfo {year} {2023}{\natexlab{b}})}\BibitemShut {NoStop}%
\bibitem [{\citenamefont {Bauman}\ \emph
  {et~al.}(2021{\natexlab{a}})\citenamefont {Bauman}, \citenamefont {Liu},
  \citenamefont {Bylaska}, \citenamefont {Krishnamoorthy}, \citenamefont {Low},
  \citenamefont {Granade}, \citenamefont {Wiebe}, \citenamefont {Baker},
  \citenamefont {Peng}, \citenamefont {Roetteler}, \citenamefont {Troyer},\
  and\ \citenamefont {Kowalski}}]{Bauman2021toward}%
  \BibitemOpen
  \bibfield  {author} {\bibinfo {author} {\bibfnamefont {N.~P.}\ \bibnamefont
  {Bauman}}, \bibinfo {author} {\bibfnamefont {H.}~\bibnamefont {Liu}},
  \bibinfo {author} {\bibfnamefont {E.~J.}\ \bibnamefont {Bylaska}}, \bibinfo
  {author} {\bibfnamefont {S.}~\bibnamefont {Krishnamoorthy}}, \bibinfo
  {author} {\bibfnamefont {G.~H.}\ \bibnamefont {Low}}, \bibinfo {author}
  {\bibfnamefont {C.~E.}\ \bibnamefont {Granade}}, \bibinfo {author}
  {\bibfnamefont {N.}~\bibnamefont {Wiebe}}, \bibinfo {author} {\bibfnamefont
  {N.~A.}\ \bibnamefont {Baker}}, \bibinfo {author} {\bibfnamefont
  {B.}~\bibnamefont {Peng}}, \bibinfo {author} {\bibfnamefont {M.}~\bibnamefont
  {Roetteler}}, \bibinfo {author} {\bibfnamefont {M.}~\bibnamefont {Troyer}}, \
  and\ \bibinfo {author} {\bibfnamefont {K.}~\bibnamefont {Kowalski}},\ }\href
  {\doibase 10.1021/acs.jctc.0c00909} {\bibfield  {journal} {\bibinfo
  {journal} {J. Chem. Theory Comput.}\ }\textbf {\bibinfo {volume} {17}},\
  \bibinfo {pages} {201} (\bibinfo {year} {2021}{\natexlab{a}})}\BibitemShut
  {NoStop}%
\bibitem [{\citenamefont {Stair}\ \emph {et~al.}(2020)\citenamefont {Stair},
  \citenamefont {Huang},\ and\ \citenamefont
  {Evangelista}}]{Stair2020multireference}%
  \BibitemOpen
  \bibfield  {author} {\bibinfo {author} {\bibfnamefont {N.~H.}\ \bibnamefont
  {Stair}}, \bibinfo {author} {\bibfnamefont {R.}~\bibnamefont {Huang}}, \ and\
  \bibinfo {author} {\bibfnamefont {F.~A.}\ \bibnamefont {Evangelista}},\
  }\href {\doibase 10.1021/acs.jctc.9b01125} {\bibfield  {journal} {\bibinfo
  {journal} {J. Chem. Theory Comput.}\ }\textbf {\bibinfo {volume} {16}},\
  \bibinfo {pages} {2236} (\bibinfo {year} {2020})}\BibitemShut {NoStop}%
\bibitem [{\citenamefont {Cortes}\ and\ \citenamefont
  {Gray}(2022)}]{Cortes2022quantum}%
  \BibitemOpen
  \bibfield  {author} {\bibinfo {author} {\bibfnamefont {C.~L.}\ \bibnamefont
  {Cortes}}\ and\ \bibinfo {author} {\bibfnamefont {S.~K.}\ \bibnamefont
  {Gray}},\ }\href {\doibase 10.1103/PhysRevA.105.022417} {\bibfield  {journal}
  {\bibinfo  {journal} {Phys. Rev. A}\ }\textbf {\bibinfo {volume} {105}},\
  \bibinfo {pages} {022417} (\bibinfo {year} {2022})}\BibitemShut {NoStop}%
\bibitem [{\citenamefont {Gonthier}\ \emph {et~al.}(2022)\citenamefont
  {Gonthier}, \citenamefont {Radin}, \citenamefont {Buda}, \citenamefont
  {Doskocil}, \citenamefont {Abuan},\ and\ \citenamefont
  {Romero}}]{Gonthier2022measurements}%
  \BibitemOpen
  \bibfield  {author} {\bibinfo {author} {\bibfnamefont {J.~F.}\ \bibnamefont
  {Gonthier}}, \bibinfo {author} {\bibfnamefont {M.~D.}\ \bibnamefont {Radin}},
  \bibinfo {author} {\bibfnamefont {C.}~\bibnamefont {Buda}}, \bibinfo {author}
  {\bibfnamefont {E.~J.}\ \bibnamefont {Doskocil}}, \bibinfo {author}
  {\bibfnamefont {C.~M.}\ \bibnamefont {Abuan}}, \ and\ \bibinfo {author}
  {\bibfnamefont {J.}~\bibnamefont {Romero}},\ }\href {\doibase
  10.1103/PhysRevResearch.4.033154} {\bibfield  {journal} {\bibinfo  {journal}
  {Phys. Rev. Res.}\ }\textbf {\bibinfo {volume} {4}},\ \bibinfo {pages}
  {033154} (\bibinfo {year} {2022})}\BibitemShut {NoStop}%
\bibitem [{\citenamefont {Wecker}\ \emph {et~al.}(2014)\citenamefont {Wecker},
  \citenamefont {Bauer}, \citenamefont {Clark}, \citenamefont {Hastings},\ and\
  \citenamefont {Troyer}}]{Wecker2014gate}%
  \BibitemOpen
  \bibfield  {author} {\bibinfo {author} {\bibfnamefont {D.}~\bibnamefont
  {Wecker}}, \bibinfo {author} {\bibfnamefont {B.}~\bibnamefont {Bauer}},
  \bibinfo {author} {\bibfnamefont {B.~K.}\ \bibnamefont {Clark}}, \bibinfo
  {author} {\bibfnamefont {M.~B.}\ \bibnamefont {Hastings}}, \ and\ \bibinfo
  {author} {\bibfnamefont {M.}~\bibnamefont {Troyer}},\ }\href {\doibase
  10.1103/PhysRevA.90.022305} {\bibfield  {journal} {\bibinfo  {journal} {Phys.
  Rev. A}\ }\textbf {\bibinfo {volume} {90}},\ \bibinfo {pages} {022305}
  (\bibinfo {year} {2014})}\BibitemShut {NoStop}%
\bibitem [{\citenamefont {Berry}\ \emph {et~al.}(2019)\citenamefont {Berry},
  \citenamefont {Gidney}, \citenamefont {Motta}, \citenamefont {McClean},\ and\
  \citenamefont {Babbush}}]{berry2019qubitization}%
  \BibitemOpen
  \bibfield  {author} {\bibinfo {author} {\bibfnamefont {D.~W.}\ \bibnamefont
  {Berry}}, \bibinfo {author} {\bibfnamefont {C.}~\bibnamefont {Gidney}},
  \bibinfo {author} {\bibfnamefont {M.}~\bibnamefont {Motta}}, \bibinfo
  {author} {\bibfnamefont {J.~R.}\ \bibnamefont {McClean}}, \ and\ \bibinfo
  {author} {\bibfnamefont {R.}~\bibnamefont {Babbush}},\ }\href {\doibase
  https://doi.org/10.22331/q-2019-12-02-208} {\bibfield  {journal} {\bibinfo
  {journal} {Quantum}\ }\textbf {\bibinfo {volume} {3}},\ \bibinfo {pages}
  {208} (\bibinfo {year} {2019})}\BibitemShut {NoStop}%
\bibitem [{\citenamefont {Motta}\ \emph {et~al.}(2021)\citenamefont {Motta},
  \citenamefont {Ye}, \citenamefont {McClean}, \citenamefont {Li},
  \citenamefont {Minnich}, \citenamefont {Babbush},\ and\ \citenamefont
  {Chan}}]{Motta2021low}%
  \BibitemOpen
  \bibfield  {author} {\bibinfo {author} {\bibfnamefont {M.}~\bibnamefont
  {Motta}}, \bibinfo {author} {\bibfnamefont {E.}~\bibnamefont {Ye}}, \bibinfo
  {author} {\bibfnamefont {J.~R.}\ \bibnamefont {McClean}}, \bibinfo {author}
  {\bibfnamefont {Z.}~\bibnamefont {Li}}, \bibinfo {author} {\bibfnamefont
  {A.~J.}\ \bibnamefont {Minnich}}, \bibinfo {author} {\bibfnamefont
  {R.}~\bibnamefont {Babbush}}, \ and\ \bibinfo {author} {\bibfnamefont
  {G.~K.-L.}\ \bibnamefont {Chan}},\ }\href {\doibase
  10.1038/s41534-021-00416-z} {\bibfield  {journal} {\bibinfo  {journal} {npj
  Quantum Information}\ }\textbf {\bibinfo {volume} {7}},\ \bibinfo {pages}
  {83} (\bibinfo {year} {2021})}\BibitemShut {NoStop}%
\bibitem [{\citenamefont {Matsuzawa}\ and\ \citenamefont
  {Kurashige}(2020)}]{Matsuzawa2020jastrow}%
  \BibitemOpen
  \bibfield  {author} {\bibinfo {author} {\bibfnamefont {Y.}~\bibnamefont
  {Matsuzawa}}\ and\ \bibinfo {author} {\bibfnamefont {Y.}~\bibnamefont
  {Kurashige}},\ }\href {\doibase 10.1021/acs.jctc.9b00963} {\bibfield
  {journal} {\bibinfo  {journal} {J. Chem. Theory Comput.}\ }\textbf {\bibinfo
  {volume} {16}},\ \bibinfo {pages} {944} (\bibinfo {year} {2020})}\BibitemShut
  {NoStop}%
\bibitem [{\citenamefont {Rubin}\ \emph {et~al.}(2022)\citenamefont {Rubin},
  \citenamefont {Lee},\ and\ \citenamefont {Babbush}}]{Rubin2022compressing}%
  \BibitemOpen
  \bibfield  {author} {\bibinfo {author} {\bibfnamefont {N.~C.}\ \bibnamefont
  {Rubin}}, \bibinfo {author} {\bibfnamefont {J.}~\bibnamefont {Lee}}, \ and\
  \bibinfo {author} {\bibfnamefont {R.}~\bibnamefont {Babbush}},\ }\href
  {\doibase 10.1021/acs.jctc.1c00912} {\bibfield  {journal} {\bibinfo
  {journal} {J. Chem. Theory Comput.}\ }\textbf {\bibinfo {volume} {18}},\
  \bibinfo {pages} {1480} (\bibinfo {year} {2022})}\BibitemShut {NoStop}%
\bibitem [{\citenamefont {{Rocca}}\ \emph {et~al.}(2024)\citenamefont
  {{Rocca}}, \citenamefont {{Cortes}}, \citenamefont {{Gonthier}},
  \citenamefont {{Ollitrault}}, \citenamefont {{Parrish}}, \citenamefont
  {{Anselmetti}}, \citenamefont {{Degroote}}, \citenamefont {{Moll}},
  \citenamefont {{Santagati}},\ and\ \citenamefont
  {{Streif}}}]{Dario2024reducing}%
  \BibitemOpen
  \bibfield  {author} {\bibinfo {author} {\bibfnamefont {D.}~\bibnamefont
  {{Rocca}}}, \bibinfo {author} {\bibfnamefont {C.~L.}\ \bibnamefont
  {{Cortes}}}, \bibinfo {author} {\bibfnamefont {J.}~\bibnamefont
  {{Gonthier}}}, \bibinfo {author} {\bibfnamefont {P.~J.}\ \bibnamefont
  {{Ollitrault}}}, \bibinfo {author} {\bibfnamefont {R.~M.}\ \bibnamefont
  {{Parrish}}}, \bibinfo {author} {\bibfnamefont {G.-L.}\ \bibnamefont
  {{Anselmetti}}}, \bibinfo {author} {\bibfnamefont {M.}~\bibnamefont
  {{Degroote}}}, \bibinfo {author} {\bibfnamefont {N.}~\bibnamefont {{Moll}}},
  \bibinfo {author} {\bibfnamefont {R.}~\bibnamefont {{Santagati}}}, \ and\
  \bibinfo {author} {\bibfnamefont {M.}~\bibnamefont {{Streif}}},\ }\href@noop
  {} {\bibfield  {journal} {\bibinfo  {journal} {arXiv}\ } (\bibinfo {year}
  {2024})},\ \Eprint {http://arxiv.org/abs/2403.03502} {arXiv:2403.03502
  [quant-ph]} \BibitemShut {NoStop}%
\bibitem [{\citenamefont {Bauman}\ \emph {et~al.}(2019)\citenamefont {Bauman},
  \citenamefont {Bylaska}, \citenamefont {Krishnamoorthy}, \citenamefont {Low},
  \citenamefont {Wiebe}, \citenamefont {Granade}, \citenamefont {Roetteler},
  \citenamefont {Troyer},\ and\ \citenamefont
  {Kowalski}}]{bauman2019downfolding}%
  \BibitemOpen
  \bibfield  {author} {\bibinfo {author} {\bibfnamefont {N.~P.}\ \bibnamefont
  {Bauman}}, \bibinfo {author} {\bibfnamefont {E.~J.}\ \bibnamefont {Bylaska}},
  \bibinfo {author} {\bibfnamefont {S.}~\bibnamefont {Krishnamoorthy}},
  \bibinfo {author} {\bibfnamefont {G.~H.}\ \bibnamefont {Low}}, \bibinfo
  {author} {\bibfnamefont {N.}~\bibnamefont {Wiebe}}, \bibinfo {author}
  {\bibfnamefont {C.~E.}\ \bibnamefont {Granade}}, \bibinfo {author}
  {\bibfnamefont {M.}~\bibnamefont {Roetteler}}, \bibinfo {author}
  {\bibfnamefont {M.}~\bibnamefont {Troyer}}, \ and\ \bibinfo {author}
  {\bibfnamefont {K.}~\bibnamefont {Kowalski}},\ }\href {\doibase
  10.1063/1.5094643} {\bibfield  {journal} {\bibinfo  {journal} {J. Chem.
  Phys.}\ }\textbf {\bibinfo {volume} {151}},\ \bibinfo {pages} {014107}
  (\bibinfo {year} {2019})}\BibitemShut {NoStop}%
\bibitem [{\citenamefont {Metcalf}\ \emph {et~al.}(2020)\citenamefont
  {Metcalf}, \citenamefont {Bauman}, \citenamefont {Kowalski},\ and\
  \citenamefont {de~Jong}}]{Metcalf2020}%
  \BibitemOpen
  \bibfield  {author} {\bibinfo {author} {\bibfnamefont {M.}~\bibnamefont
  {Metcalf}}, \bibinfo {author} {\bibfnamefont {N.~P.}\ \bibnamefont {Bauman}},
  \bibinfo {author} {\bibfnamefont {K.}~\bibnamefont {Kowalski}}, \ and\
  \bibinfo {author} {\bibfnamefont {W.~A.}\ \bibnamefont {de~Jong}},\ }\href
  {\doibase 10.1021/acs.jctc.0c00421} {\bibfield  {journal} {\bibinfo
  {journal} {J. Chem. Theory Comput.}\ }\textbf {\bibinfo {volume} {16}},\
  \bibinfo {pages} {6165} (\bibinfo {year} {2020})}\BibitemShut {NoStop}%
\bibitem [{\citenamefont {Bauman}\ \emph
  {et~al.}(2021{\natexlab{b}})\citenamefont {Bauman}, \citenamefont {Jaroslav},
  \citenamefont {Libor}, \citenamefont {Ji\v{r}\'{i}},\ and\ \citenamefont
  {Kowalski}}]{Nicholas2021}%
  \BibitemOpen
  \bibfield  {author} {\bibinfo {author} {\bibfnamefont {N.~P.}\ \bibnamefont
  {Bauman}}, \bibinfo {author} {\bibfnamefont {C.}~\bibnamefont {Jaroslav}},
  \bibinfo {author} {\bibfnamefont {V.}~\bibnamefont {Libor}}, \bibinfo
  {author} {\bibfnamefont {P.}~\bibnamefont {Ji\v{r}\'{i}}}, \ and\ \bibinfo
  {author} {\bibfnamefont {K.}~\bibnamefont {Kowalski}},\ }\href {\doibase
  10.1088/2058-9565/abf602} {\bibfield  {journal} {\bibinfo  {journal} {Quantum
  Sci. Technol.}\ }\textbf {\bibinfo {volume} {6}},\ \bibinfo {pages} {034008}
  (\bibinfo {year} {2021}{\natexlab{b}})}\BibitemShut {NoStop}%
\bibitem [{\citenamefont {Huang}\ \emph {et~al.}(2023)\citenamefont {Huang},
  \citenamefont {Li},\ and\ \citenamefont {Evangelista}}]{Huang2023leveraging}%
  \BibitemOpen
  \bibfield  {author} {\bibinfo {author} {\bibfnamefont {R.}~\bibnamefont
  {Huang}}, \bibinfo {author} {\bibfnamefont {C.}~\bibnamefont {Li}}, \ and\
  \bibinfo {author} {\bibfnamefont {F.~A.}\ \bibnamefont {Evangelista}},\
  }\href {\doibase 10.1103/PRXQuantum.4.020313} {\bibfield  {journal} {\bibinfo
   {journal} {PRX Quantum}\ }\textbf {\bibinfo {volume} {4}},\ \bibinfo {pages}
  {020313} (\bibinfo {year} {2023})}\BibitemShut {NoStop}%
\bibitem [{\citenamefont {Bauman}\ and\ \citenamefont
  {Kowalski}(2022)}]{Bauman2022coupled}%
  \BibitemOpen
  \bibfield  {author} {\bibinfo {author} {\bibfnamefont {N.~P.}\ \bibnamefont
  {Bauman}}\ and\ \bibinfo {author} {\bibfnamefont {K.}~\bibnamefont
  {Kowalski}},\ }\href {\doibase 10.1186/s41313-022-00046-8} {\bibfield
  {journal} {\bibinfo  {journal} {Mater. Theory}\ }\textbf {\bibinfo {volume}
  {6}},\ \bibinfo {pages} {17} (\bibinfo {year} {2022})}\BibitemShut {NoStop}%
\bibitem [{\citenamefont {Le}\ and\ \citenamefont {Tran}(2023)}]{Le2023}%
  \BibitemOpen
  \bibfield  {author} {\bibinfo {author} {\bibfnamefont {N.~T.}\ \bibnamefont
  {Le}}\ and\ \bibinfo {author} {\bibfnamefont {L.~N.}\ \bibnamefont {Tran}},\
  }\href {\doibase 10.1021/acs.jpca.3c00993} {\bibfield  {journal} {\bibinfo
  {journal} {J. Phys. Chem. A}\ }\textbf {\bibinfo {volume} {127}},\ \bibinfo
  {pages} {5222} (\bibinfo {year} {2023})}\BibitemShut {NoStop}%
\bibitem [{\citenamefont {Motta}\ \emph {et~al.}(2020)\citenamefont {Motta},
  \citenamefont {Gujarati}, \citenamefont {Rice}, \citenamefont {Kumar},
  \citenamefont {Masteran}, \citenamefont {Latone}, \citenamefont {Lee},
  \citenamefont {Valeev},\ and\ \citenamefont {Takeshita}}]{Motta2020quantum}%
  \BibitemOpen
  \bibfield  {author} {\bibinfo {author} {\bibfnamefont {M.}~\bibnamefont
  {Motta}}, \bibinfo {author} {\bibfnamefont {T.~P.}\ \bibnamefont {Gujarati}},
  \bibinfo {author} {\bibfnamefont {J.~E.}\ \bibnamefont {Rice}}, \bibinfo
  {author} {\bibfnamefont {A.}~\bibnamefont {Kumar}}, \bibinfo {author}
  {\bibfnamefont {C.}~\bibnamefont {Masteran}}, \bibinfo {author}
  {\bibfnamefont {J.~A.}\ \bibnamefont {Latone}}, \bibinfo {author}
  {\bibfnamefont {E.}~\bibnamefont {Lee}}, \bibinfo {author} {\bibfnamefont
  {E.~F.}\ \bibnamefont {Valeev}}, \ and\ \bibinfo {author} {\bibfnamefont
  {T.~Y.}\ \bibnamefont {Takeshita}},\ }\href {\doibase 10.1039/D0CP04106H}
  {\bibfield  {journal} {\bibinfo  {journal} {Phys. Chem. Chem. Phys.}\
  }\textbf {\bibinfo {volume} {22}},\ \bibinfo {pages} {24270} (\bibinfo {year}
  {2020})}\BibitemShut {NoStop}%
\bibitem [{\citenamefont {{McArdle}}\ and\ \citenamefont
  {{Tew}}(2020)}]{McArdle2020Improving}%
  \BibitemOpen
  \bibfield  {author} {\bibinfo {author} {\bibfnamefont {S.}~\bibnamefont
  {{McArdle}}}\ and\ \bibinfo {author} {\bibfnamefont {D.~P.}\ \bibnamefont
  {{Tew}}},\ }\href@noop {} {\bibfield  {journal} {\bibinfo  {journal} {arXiv}\
  } (\bibinfo {year} {2020})},\ \Eprint {http://arxiv.org/abs/2006.11181}
  {arXiv:2006.11181 [quant-ph]} \BibitemShut {NoStop}%
\bibitem [{\citenamefont {Vorwerk}\ \emph {et~al.}(2022)\citenamefont
  {Vorwerk}, \citenamefont {Sheng}, \citenamefont {Govoni}, \citenamefont
  {Huang},\ and\ \citenamefont {Galli}}]{Vorwerk2022quantum}%
  \BibitemOpen
  \bibfield  {author} {\bibinfo {author} {\bibfnamefont {C.}~\bibnamefont
  {Vorwerk}}, \bibinfo {author} {\bibfnamefont {N.}~\bibnamefont {Sheng}},
  \bibinfo {author} {\bibfnamefont {M.}~\bibnamefont {Govoni}}, \bibinfo
  {author} {\bibfnamefont {B.}~\bibnamefont {Huang}}, \ and\ \bibinfo {author}
  {\bibfnamefont {G.}~\bibnamefont {Galli}},\ }\href {\doibase
  10.1038/s43588-022-00279-0} {\bibfield  {journal} {\bibinfo  {journal} {Nat.
  Comput. Sci.}\ }\textbf {\bibinfo {volume} {2}},\ \bibinfo {pages} {424}
  (\bibinfo {year} {2022})}\BibitemShut {NoStop}%
\bibitem [{\citenamefont {Aryasetiawan}\ \emph {et~al.}(2004)\citenamefont
  {Aryasetiawan}, \citenamefont {Imada}, \citenamefont {Georges}, \citenamefont
  {Kotliar}, \citenamefont {Biermann},\ and\ \citenamefont
  {Lichtenstein}}]{Aryasetiawan2004frequency}%
  \BibitemOpen
  \bibfield  {author} {\bibinfo {author} {\bibfnamefont {F.}~\bibnamefont
  {Aryasetiawan}}, \bibinfo {author} {\bibfnamefont {M.}~\bibnamefont {Imada}},
  \bibinfo {author} {\bibfnamefont {A.}~\bibnamefont {Georges}}, \bibinfo
  {author} {\bibfnamefont {G.}~\bibnamefont {Kotliar}}, \bibinfo {author}
  {\bibfnamefont {S.}~\bibnamefont {Biermann}}, \ and\ \bibinfo {author}
  {\bibfnamefont {A.~I.}\ \bibnamefont {Lichtenstein}},\ }\href {\doibase
  10.1103/PhysRevB.70.195104} {\bibfield  {journal} {\bibinfo  {journal} {Phys.
  Rev. B}\ }\textbf {\bibinfo {volume} {70}},\ \bibinfo {pages} {195104}
  (\bibinfo {year} {2004})}\BibitemShut {NoStop}%
\bibitem [{\citenamefont {Nakamura}\ \emph {et~al.}(2021)\citenamefont
  {Nakamura}, \citenamefont {Yoshimoto}, \citenamefont {Nomura}, \citenamefont
  {Tadano}, \citenamefont {Kawamura}, \citenamefont {Kosugi}, \citenamefont
  {Yoshimi}, \citenamefont {Misawa},\ and\ \citenamefont
  {Motoyama}}]{RESPACK1}%
  \BibitemOpen
  \bibfield  {author} {\bibinfo {author} {\bibfnamefont {K.}~\bibnamefont
  {Nakamura}}, \bibinfo {author} {\bibfnamefont {Y.}~\bibnamefont {Yoshimoto}},
  \bibinfo {author} {\bibfnamefont {Y.}~\bibnamefont {Nomura}}, \bibinfo
  {author} {\bibfnamefont {T.}~\bibnamefont {Tadano}}, \bibinfo {author}
  {\bibfnamefont {M.}~\bibnamefont {Kawamura}}, \bibinfo {author}
  {\bibfnamefont {T.}~\bibnamefont {Kosugi}}, \bibinfo {author} {\bibfnamefont
  {K.}~\bibnamefont {Yoshimi}}, \bibinfo {author} {\bibfnamefont
  {T.}~\bibnamefont {Misawa}}, \ and\ \bibinfo {author} {\bibfnamefont
  {Y.}~\bibnamefont {Motoyama}},\ }\href {\doibase
  https://doi.org/10.1016/j.cpc.2020.107781} {\bibfield  {journal} {\bibinfo
  {journal} {Comput. Phys. Commun.}\ }\textbf {\bibinfo {volume} {261}},\
  \bibinfo {pages} {107781} (\bibinfo {year} {2021})}\BibitemShut {NoStop}%
\bibitem [{\citenamefont {Misawa}\ and\ \citenamefont
  {Imada}(2014)}]{Misawa2014superconductivity}%
  \BibitemOpen
  \bibfield  {author} {\bibinfo {author} {\bibfnamefont {T.}~\bibnamefont
  {Misawa}}\ and\ \bibinfo {author} {\bibfnamefont {M.}~\bibnamefont {Imada}},\
  }\href {\doibase 10.1038/ncomms6738} {\bibfield  {journal} {\bibinfo
  {journal} {Nat. Commun.}\ }\textbf {\bibinfo {volume} {5}},\ \bibinfo {pages}
  {5738} (\bibinfo {year} {2014})}\BibitemShut {NoStop}%
\bibitem [{\citenamefont {Misawa}\ \emph {et~al.}(2020)\citenamefont {Misawa},
  \citenamefont {Yoshimi},\ and\ \citenamefont
  {Tsumuraya}}]{Misawa2020electronic}%
  \BibitemOpen
  \bibfield  {author} {\bibinfo {author} {\bibfnamefont {T.}~\bibnamefont
  {Misawa}}, \bibinfo {author} {\bibfnamefont {K.}~\bibnamefont {Yoshimi}}, \
  and\ \bibinfo {author} {\bibfnamefont {T.}~\bibnamefont {Tsumuraya}},\ }\href
  {\doibase 10.1103/PhysRevResearch.2.032072} {\bibfield  {journal} {\bibinfo
  {journal} {Phys. Rev. Res.}\ }\textbf {\bibinfo {volume} {2}},\ \bibinfo
  {pages} {032072} (\bibinfo {year} {2020})}\BibitemShut {NoStop}%
\bibitem [{\citenamefont {Yoshimi}\ \emph {et~al.}(2021)\citenamefont
  {Yoshimi}, \citenamefont {Tsumuraya},\ and\ \citenamefont
  {Misawa}}]{Yoshimi2021Abinitio}%
  \BibitemOpen
  \bibfield  {author} {\bibinfo {author} {\bibfnamefont {K.}~\bibnamefont
  {Yoshimi}}, \bibinfo {author} {\bibfnamefont {T.}~\bibnamefont {Tsumuraya}},
  \ and\ \bibinfo {author} {\bibfnamefont {T.}~\bibnamefont {Misawa}},\ }\href
  {\doibase 10.1103/PhysRevResearch.3.043224} {\bibfield  {journal} {\bibinfo
  {journal} {Phys. Rev. Res.}\ }\textbf {\bibinfo {volume} {3}},\ \bibinfo
  {pages} {043224} (\bibinfo {year} {2021})}\BibitemShut {NoStop}%
\bibitem [{\citenamefont {Amsler}\ \emph {et~al.}(2023)\citenamefont {Amsler},
  \citenamefont {Deglmann}, \citenamefont {Degroote}, \citenamefont {Kaicher},
  \citenamefont {Kiser}, \citenamefont {Kühn}, \citenamefont {Kumar},
  \citenamefont {Maier}, \citenamefont {Samsonidze}, \citenamefont {Schroeder},
  \citenamefont {Streif}, \citenamefont {Vodola},\ and\ \citenamefont
  {Wever}}]{amsler2023quantumenhanced}%
  \BibitemOpen
  \bibfield  {author} {\bibinfo {author} {\bibfnamefont {M.}~\bibnamefont
  {Amsler}}, \bibinfo {author} {\bibfnamefont {P.}~\bibnamefont {Deglmann}},
  \bibinfo {author} {\bibfnamefont {M.}~\bibnamefont {Degroote}}, \bibinfo
  {author} {\bibfnamefont {M.~P.}\ \bibnamefont {Kaicher}}, \bibinfo {author}
  {\bibfnamefont {M.}~\bibnamefont {Kiser}}, \bibinfo {author} {\bibfnamefont
  {M.}~\bibnamefont {Kühn}}, \bibinfo {author} {\bibfnamefont
  {C.}~\bibnamefont {Kumar}}, \bibinfo {author} {\bibfnamefont
  {A.}~\bibnamefont {Maier}}, \bibinfo {author} {\bibfnamefont
  {G.}~\bibnamefont {Samsonidze}}, \bibinfo {author} {\bibfnamefont
  {A.}~\bibnamefont {Schroeder}}, \bibinfo {author} {\bibfnamefont
  {M.}~\bibnamefont {Streif}}, \bibinfo {author} {\bibfnamefont
  {D.}~\bibnamefont {Vodola}}, \ and\ \bibinfo {author} {\bibfnamefont
  {C.}~\bibnamefont {Wever}},\ }\href@noop {} {\bibfield  {journal} {\bibinfo
  {journal} {arXiv}\ } (\bibinfo {year} {2023})},\ \Eprint
  {http://arxiv.org/abs/2301.11838} {arXiv:2301.11838 [quant-ph]} \BibitemShut
  {NoStop}%
\bibitem [{\citenamefont {Serrano‐Andr\'{e}s}\ \emph
  {et~al.}(1993)\citenamefont {Serrano‐Andr\'{e}s}, \citenamefont
  {Merch\'{a}n}, \citenamefont {Nebot‐Gil}, \citenamefont {Lindh},\ and\
  \citenamefont {Roos}}]{Luis1993towards}%
  \BibitemOpen
  \bibfield  {author} {\bibinfo {author} {\bibfnamefont {L.}~\bibnamefont
  {Serrano‐Andr\'{e}s}}, \bibinfo {author} {\bibfnamefont {M.}~\bibnamefont
  {Merch\'{a}n}}, \bibinfo {author} {\bibfnamefont {I.}~\bibnamefont
  {Nebot‐Gil}}, \bibinfo {author} {\bibfnamefont {R.}~\bibnamefont {Lindh}},
  \ and\ \bibinfo {author} {\bibfnamefont {B.~O.}\ \bibnamefont {Roos}},\
  }\href {\doibase 10.1063/1.465071} {\bibfield  {journal} {\bibinfo  {journal}
  {J. Chem. Phys.}\ }\textbf {\bibinfo {volume} {98}},\ \bibinfo {pages} {3151}
  (\bibinfo {year} {1993})}\BibitemShut {NoStop}%
\bibitem [{\citenamefont {Nakayama}\ \emph {et~al.}(1998)\citenamefont
  {Nakayama}, \citenamefont {Nakano},\ and\ \citenamefont
  {Hirao}}]{Nakayama1998theoretical}%
  \BibitemOpen
  \bibfield  {author} {\bibinfo {author} {\bibfnamefont {K.}~\bibnamefont
  {Nakayama}}, \bibinfo {author} {\bibfnamefont {H.}~\bibnamefont {Nakano}}, \
  and\ \bibinfo {author} {\bibfnamefont {K.}~\bibnamefont {Hirao}},\ }\href
  {\doibase
  https://doi.org/10.1002/(SICI)1097-461X(1998)66:2<157::AID-QUA7>3.0.CO;2-U}
  {\bibfield  {journal} {\bibinfo  {journal} {Int. J. Quantum Chem.}\ }\textbf
  {\bibinfo {volume} {66}},\ \bibinfo {pages} {157} (\bibinfo {year}
  {1998})}\BibitemShut {NoStop}%
\bibitem [{\citenamefont {Kurashige}\ \emph {et~al.}(2004)\citenamefont
  {Kurashige}, \citenamefont {Nakano}, \citenamefont {Nakao},\ and\
  \citenamefont {Hirao}}]{Kurashige2004pi}%
  \BibitemOpen
  \bibfield  {author} {\bibinfo {author} {\bibfnamefont {Y.}~\bibnamefont
  {Kurashige}}, \bibinfo {author} {\bibfnamefont {H.}~\bibnamefont {Nakano}},
  \bibinfo {author} {\bibfnamefont {Y.}~\bibnamefont {Nakao}}, \ and\ \bibinfo
  {author} {\bibfnamefont {K.}~\bibnamefont {Hirao}},\ }\href {\doibase
  https://doi.org/10.1016/j.cplett.2004.10.141} {\bibfield  {journal} {\bibinfo
   {journal} {Chem. Phys. Lett.}\ }\textbf {\bibinfo {volume} {400}},\ \bibinfo
  {pages} {425} (\bibinfo {year} {2004})}\BibitemShut {NoStop}%
\bibitem [{\citenamefont {Schreiber}\ \emph {et~al.}(2008)\citenamefont
  {Schreiber}, \citenamefont {Silva-Junior}, \citenamefont {Sauer},\ and\
  \citenamefont {Thiel}}]{Schreiber2008benchmarks}%
  \BibitemOpen
  \bibfield  {author} {\bibinfo {author} {\bibfnamefont {M.}~\bibnamefont
  {Schreiber}}, \bibinfo {author} {\bibfnamefont {M.~R.}\ \bibnamefont
  {Silva-Junior}}, \bibinfo {author} {\bibfnamefont {S.~P.~A.}\ \bibnamefont
  {Sauer}}, \ and\ \bibinfo {author} {\bibfnamefont {W.}~\bibnamefont
  {Thiel}},\ }\href {\doibase 10.1063/1.2889385} {\bibfield  {journal}
  {\bibinfo  {journal} {J. Chem. Phys.}\ }\textbf {\bibinfo {volume} {128}},\
  \bibinfo {pages} {134110} (\bibinfo {year} {2008})}\BibitemShut {NoStop}%
\bibitem [{\citenamefont {Watson}\ and\ \citenamefont
  {Chan}(2012)}]{Watson2012excited}%
  \BibitemOpen
  \bibfield  {author} {\bibinfo {author} {\bibfnamefont {M.~A.}\ \bibnamefont
  {Watson}}\ and\ \bibinfo {author} {\bibfnamefont {G.~K.-L.}\ \bibnamefont
  {Chan}},\ }\href {\doibase 10.1021/ct300591z} {\bibfield  {journal} {\bibinfo
   {journal} {J. Chem. Theory Comput.}\ }\textbf {\bibinfo {volume} {8}},\
  \bibinfo {pages} {4013} (\bibinfo {year} {2012})}\BibitemShut {NoStop}%
\bibitem [{\citenamefont {Daday}\ \emph {et~al.}(2012)\citenamefont {Daday},
  \citenamefont {Smart}, \citenamefont {Booth}, \citenamefont {Alavi},\ and\
  \citenamefont {Filippi}}]{Daday2012full}%
  \BibitemOpen
  \bibfield  {author} {\bibinfo {author} {\bibfnamefont {C.}~\bibnamefont
  {Daday}}, \bibinfo {author} {\bibfnamefont {S.}~\bibnamefont {Smart}},
  \bibinfo {author} {\bibfnamefont {G.~H.}\ \bibnamefont {Booth}}, \bibinfo
  {author} {\bibfnamefont {A.}~\bibnamefont {Alavi}}, \ and\ \bibinfo {author}
  {\bibfnamefont {C.}~\bibnamefont {Filippi}},\ }\href {\doibase
  10.1021/ct300486d} {\bibfield  {journal} {\bibinfo  {journal} {J. Chem.
  Theory Comput.}\ }\textbf {\bibinfo {volume} {8}},\ \bibinfo {pages} {4441}
  (\bibinfo {year} {2012})}\BibitemShut {NoStop}%
\bibitem [{\citenamefont {Chien}\ \emph {et~al.}(2018)\citenamefont {Chien},
  \citenamefont {Holmes}, \citenamefont {Otten}, \citenamefont {Umrigar},
  \citenamefont {Sharma},\ and\ \citenamefont {Zimmerman}}]{Chien2018excited}%
  \BibitemOpen
  \bibfield  {author} {\bibinfo {author} {\bibfnamefont {A.~D.}\ \bibnamefont
  {Chien}}, \bibinfo {author} {\bibfnamefont {A.~A.}\ \bibnamefont {Holmes}},
  \bibinfo {author} {\bibfnamefont {M.}~\bibnamefont {Otten}}, \bibinfo
  {author} {\bibfnamefont {C.~J.}\ \bibnamefont {Umrigar}}, \bibinfo {author}
  {\bibfnamefont {S.}~\bibnamefont {Sharma}}, \ and\ \bibinfo {author}
  {\bibfnamefont {P.~M.}\ \bibnamefont {Zimmerman}},\ }\href {\doibase
  10.1021/acs.jpca.8b01554} {\bibfield  {journal} {\bibinfo  {journal} {J.
  Phys. Chem. A}\ }\textbf {\bibinfo {volume} {122}},\ \bibinfo {pages} {2714}
  (\bibinfo {year} {2018})}\BibitemShut {NoStop}%
\bibitem [{\citenamefont {Manna}\ \emph {et~al.}(2020)\citenamefont {Manna},
  \citenamefont {Chaudhuri},\ and\ \citenamefont
  {Chattopadhyay}}]{Manna2020taming}%
  \BibitemOpen
  \bibfield  {author} {\bibinfo {author} {\bibfnamefont {S.}~\bibnamefont
  {Manna}}, \bibinfo {author} {\bibfnamefont {R.~K.}\ \bibnamefont
  {Chaudhuri}}, \ and\ \bibinfo {author} {\bibfnamefont {S.}~\bibnamefont
  {Chattopadhyay}},\ }\href {\doibase 10.1063/5.0007198} {\bibfield  {journal}
  {\bibinfo  {journal} {J. Chem. Phys.}\ }\textbf {\bibinfo {volume} {152}},\
  \bibinfo {pages} {244105} (\bibinfo {year} {2020})}\BibitemShut {NoStop}%
\bibitem [{\citenamefont {Mulliken}(1977)}]{Mulliken1977excited}%
  \BibitemOpen
  \bibfield  {author} {\bibinfo {author} {\bibfnamefont {R.~S.}\ \bibnamefont
  {Mulliken}},\ }\href {\doibase 10.1063/1.434239} {\bibfield  {journal}
  {\bibinfo  {journal} {J. Chem. Phys.}\ }\textbf {\bibinfo {volume} {66}},\
  \bibinfo {pages} {2448} (\bibinfo {year} {1977})}\BibitemShut {NoStop}%
\bibitem [{\citenamefont {Doering}\ and\ \citenamefont
  {McDiarmid}(1980)}]{doering1980electron}%
  \BibitemOpen
  \bibfield  {author} {\bibinfo {author} {\bibfnamefont {J.~P.}\ \bibnamefont
  {Doering}}\ and\ \bibinfo {author} {\bibfnamefont {R.}~\bibnamefont
  {McDiarmid}},\ }\href {\doibase 10.1063/1.440587} {\bibfield  {journal}
  {\bibinfo  {journal} {The Journal of Chemical Physics}\ }\textbf {\bibinfo
  {volume} {73}},\ \bibinfo {pages} {3617} (\bibinfo {year}
  {1980})}\BibitemShut {NoStop}%
\bibitem [{\citenamefont {{Gavin, Jr.}}\ \emph {et~al.}(1973)\citenamefont
  {{Gavin, Jr.}}, \citenamefont {Risemberg},\ and\ \citenamefont
  {Rice}}]{gavin1973spectroscopic}%
  \BibitemOpen
  \bibfield  {author} {\bibinfo {author} {\bibfnamefont {R.~M.}\ \bibnamefont
  {{Gavin, Jr.}}}, \bibinfo {author} {\bibfnamefont {S.}~\bibnamefont
  {Risemberg}}, \ and\ \bibinfo {author} {\bibfnamefont {S.~A.}\ \bibnamefont
  {Rice}},\ }\href {\doibase 10.1063/1.1679637} {\bibfield  {journal} {\bibinfo
   {journal} {J. Chem. Phys.}\ }\textbf {\bibinfo {volume} {58}},\ \bibinfo
  {pages} {3160} (\bibinfo {year} {1973})}\BibitemShut {NoStop}%
\bibitem [{\citenamefont {Flicker}\ \emph {et~al.}(1977)\citenamefont
  {Flicker}, \citenamefont {Mosher},\ and\ \citenamefont
  {Kuppermann}}]{Flicker1977low}%
  \BibitemOpen
  \bibfield  {author} {\bibinfo {author} {\bibfnamefont {W.~M.}\ \bibnamefont
  {Flicker}}, \bibinfo {author} {\bibfnamefont {O.~A.}\ \bibnamefont {Mosher}},
  \ and\ \bibinfo {author} {\bibfnamefont {A.}~\bibnamefont {Kuppermann}},\
  }\href {\doibase https://doi.org/10.1016/0009-2614(77)80073-0} {\bibfield
  {journal} {\bibinfo  {journal} {Chem. Phys. Lett.}\ }\textbf {\bibinfo
  {volume} {45}},\ \bibinfo {pages} {492} (\bibinfo {year} {1977})}\BibitemShut
  {NoStop}%
\bibitem [{\citenamefont {Fujii}\ \emph {et~al.}(1985)\citenamefont {Fujii},
  \citenamefont {Kamata}, \citenamefont {Shimizu}, \citenamefont {Adachi},\
  and\ \citenamefont {Maeda}}]{Fujii1985two}%
  \BibitemOpen
  \bibfield  {author} {\bibinfo {author} {\bibfnamefont {T.}~\bibnamefont
  {Fujii}}, \bibinfo {author} {\bibfnamefont {A.}~\bibnamefont {Kamata}},
  \bibinfo {author} {\bibfnamefont {M.}~\bibnamefont {Shimizu}}, \bibinfo
  {author} {\bibfnamefont {Y.}~\bibnamefont {Adachi}}, \ and\ \bibinfo {author}
  {\bibfnamefont {S.}~\bibnamefont {Maeda}},\ }\href {\doibase
  https://doi.org/10.1016/0009-2614(85)85150-2} {\bibfield  {journal} {\bibinfo
   {journal} {Chem. Phys. Lett.}\ }\textbf {\bibinfo {volume} {115}},\ \bibinfo
  {pages} {369} (\bibinfo {year} {1985})}\BibitemShut {NoStop}%
\bibitem [{\citenamefont {Kawamura}\ \emph {et~al.}(2017)\citenamefont
  {Kawamura}, \citenamefont {Yoshimi}, \citenamefont {Misawa}, \citenamefont
  {Yamaji}, \citenamefont {Todo},\ and\ \citenamefont {Kawashima}}]{HPhi1}%
  \BibitemOpen
  \bibfield  {author} {\bibinfo {author} {\bibfnamefont {M.}~\bibnamefont
  {Kawamura}}, \bibinfo {author} {\bibfnamefont {K.}~\bibnamefont {Yoshimi}},
  \bibinfo {author} {\bibfnamefont {T.}~\bibnamefont {Misawa}}, \bibinfo
  {author} {\bibfnamefont {Y.}~\bibnamefont {Yamaji}}, \bibinfo {author}
  {\bibfnamefont {S.}~\bibnamefont {Todo}}, \ and\ \bibinfo {author}
  {\bibfnamefont {N.}~\bibnamefont {Kawashima}},\ }\href {\doibase
  https://doi.org/10.1016/j.cpc.2017.04.006} {\bibfield  {journal} {\bibinfo
  {journal} {Comput. Phys. Commun.}\ }\textbf {\bibinfo {volume} {217}},\
  \bibinfo {pages} {180} (\bibinfo {year} {2017})}\BibitemShut {NoStop}%
\bibitem [{\citenamefont {{Chang}}\ \emph {et~al.}(2023)\citenamefont
  {{Chang}}, \citenamefont {{van Loon}}, \citenamefont {{Eskridge}},
  \citenamefont {{Busemeyer}}, \citenamefont {{Morales}}, \citenamefont
  {{Dreyer}}, \citenamefont {{Millis}}, \citenamefont {{Zhang}}, \citenamefont
  {{Wehling}}, \citenamefont {{Wagner}},\ and\ \citenamefont
  {{R{\"o}sner}}}]{Yueqing2023downfolding}%
  \BibitemOpen
  \bibfield  {author} {\bibinfo {author} {\bibfnamefont {Y.}~\bibnamefont
  {{Chang}}}, \bibinfo {author} {\bibfnamefont {E.~G.~C.~P.}\ \bibnamefont
  {{van Loon}}}, \bibinfo {author} {\bibfnamefont {B.}~\bibnamefont
  {{Eskridge}}}, \bibinfo {author} {\bibfnamefont {B.}~\bibnamefont
  {{Busemeyer}}}, \bibinfo {author} {\bibfnamefont {M.~A.}\ \bibnamefont
  {{Morales}}}, \bibinfo {author} {\bibfnamefont {C.~E.}\ \bibnamefont
  {{Dreyer}}}, \bibinfo {author} {\bibfnamefont {A.~J.}\ \bibnamefont
  {{Millis}}}, \bibinfo {author} {\bibfnamefont {S.}~\bibnamefont {{Zhang}}},
  \bibinfo {author} {\bibfnamefont {T.~O.}\ \bibnamefont {{Wehling}}}, \bibinfo
  {author} {\bibfnamefont {L.~K.}\ \bibnamefont {{Wagner}}}, \ and\ \bibinfo
  {author} {\bibfnamefont {M.}~\bibnamefont {{R{\"o}sner}}},\ }\href@noop {}
  {\bibfield  {journal} {\bibinfo  {journal} {arXiv}\ } (\bibinfo {year}
  {2023})},\ \Eprint {http://arxiv.org/abs/2311.05987} {arXiv:2311.05987
  [cond-mat.str-el]} \BibitemShut {NoStop}%
\bibitem [{\citenamefont {Peters}\ \emph {et~al.}(2018)\citenamefont {Peters},
  \citenamefont {\ifmmode \mbox{\c{S}}\else \c{S}\fi{}a\ifmmode
  \mbox{\c{s}}\else \c{s}\fi{}\ifmmode \imath \else \i
  \fi{}o\ifmmode~\breve{g}\else \u{g}\fi{}lu}, \citenamefont {Mertig},\ and\
  \citenamefont {Katsnelson}}]{Peters2018abinitio}%
  \BibitemOpen
  \bibfield  {author} {\bibinfo {author} {\bibfnamefont {L.}~\bibnamefont
  {Peters}}, \bibinfo {author} {\bibfnamefont {E.}~\bibnamefont {\ifmmode
  \mbox{\c{S}}\else \c{S}\fi{}a\ifmmode \mbox{\c{s}}\else \c{s}\fi{}\ifmmode
  \imath \else \i \fi{}o\ifmmode~\breve{g}\else \u{g}\fi{}lu}}, \bibinfo
  {author} {\bibfnamefont {I.}~\bibnamefont {Mertig}}, \ and\ \bibinfo {author}
  {\bibfnamefont {M.~I.}\ \bibnamefont {Katsnelson}},\ }\href {\doibase
  10.1103/PhysRevB.97.045121} {\bibfield  {journal} {\bibinfo  {journal} {Phys.
  Rev. B}\ }\textbf {\bibinfo {volume} {97}},\ \bibinfo {pages} {045121}
  (\bibinfo {year} {2018})}\BibitemShut {NoStop}%
\bibitem [{\citenamefont {van Loon}\ \emph {et~al.}(2021)\citenamefont {van
  Loon}, \citenamefont {R\"osner}, \citenamefont {Katsnelson},\ and\
  \citenamefont {Wehling}}]{vanLoon2021random}%
  \BibitemOpen
  \bibfield  {author} {\bibinfo {author} {\bibfnamefont {E.~G. C.~P.}\
  \bibnamefont {van Loon}}, \bibinfo {author} {\bibfnamefont {M.}~\bibnamefont
  {R\"osner}}, \bibinfo {author} {\bibfnamefont {M.~I.}\ \bibnamefont
  {Katsnelson}}, \ and\ \bibinfo {author} {\bibfnamefont {T.~O.}\ \bibnamefont
  {Wehling}},\ }\href {\doibase 10.1103/PhysRevB.104.045134} {\bibfield
  {journal} {\bibinfo  {journal} {Phys. Rev. B}\ }\textbf {\bibinfo {volume}
  {104}},\ \bibinfo {pages} {045134} (\bibinfo {year} {2021})}\BibitemShut
  {NoStop}%
\bibitem [{\citenamefont {Scott}\ and\ \citenamefont
  {Booth}(2024)}]{Scott2024rigorous}%
  \BibitemOpen
  \bibfield  {author} {\bibinfo {author} {\bibfnamefont {C.~J.~C.}\
  \bibnamefont {Scott}}\ and\ \bibinfo {author} {\bibfnamefont {G.~H.}\
  \bibnamefont {Booth}},\ }\href {\doibase 10.1103/PhysRevLett.132.076401}
  {\bibfield  {journal} {\bibinfo  {journal} {Phys. Rev. Lett.}\ }\textbf
  {\bibinfo {volume} {132}},\ \bibinfo {pages} {076401} (\bibinfo {year}
  {2024})}\BibitemShut {NoStop}%
\bibitem [{\citenamefont {Dhawan}\ \emph {et~al.}(2021)\citenamefont {Dhawan},
  \citenamefont {Metcalf},\ and\ \citenamefont {Zgid}}]{Dhawan2021dynamical}%
  \BibitemOpen
  \bibfield  {author} {\bibinfo {author} {\bibfnamefont {D.}~\bibnamefont
  {Dhawan}}, \bibinfo {author} {\bibfnamefont {M.}~\bibnamefont {Metcalf}}, \
  and\ \bibinfo {author} {\bibfnamefont {D.}~\bibnamefont {Zgid}},\ }\href
  {\doibase 10.1021/acs.jctc.1c00931} {\bibfield  {journal} {\bibinfo
  {journal} {J. Chem. Theory Comput.}\ }\textbf {\bibinfo {volume} {17}},\
  \bibinfo {pages} {7622} (\bibinfo {year} {2021})}\BibitemShut {NoStop}%
\bibitem [{\citenamefont {Daniel}\ \emph {et~al.}(2021)\citenamefont {Daniel},
  \citenamefont {Dhawan}, \citenamefont {Zgid},\ and\ \citenamefont
  {Freericks}}]{Daniel2021sparse}%
  \BibitemOpen
  \bibfield  {author} {\bibinfo {author} {\bibfnamefont {C.}~\bibnamefont
  {Daniel}}, \bibinfo {author} {\bibfnamefont {D.}~\bibnamefont {Dhawan}},
  \bibinfo {author} {\bibfnamefont {D.}~\bibnamefont {Zgid}}, \ and\ \bibinfo
  {author} {\bibfnamefont {J.~K.}\ \bibnamefont {Freericks}},\ }\href {\doibase
  10.1140/epjs/s11734-021-00098-w} {\bibfield  {journal} {\bibinfo  {journal}
  {Eur. Phys. J. Spec. Top.}\ }\textbf {\bibinfo {volume} {230}},\ \bibinfo
  {pages} {1067} (\bibinfo {year} {2021})}\BibitemShut {NoStop}%
\bibitem [{\citenamefont {Koridon}\ \emph {et~al.}(2021)\citenamefont
  {Koridon}, \citenamefont {Yalouz}, \citenamefont {Senjean}, \citenamefont
  {Buda}, \citenamefont {O'Brien},\ and\ \citenamefont
  {Visscher}}]{Koridon2021}%
  \BibitemOpen
  \bibfield  {author} {\bibinfo {author} {\bibfnamefont {E.}~\bibnamefont
  {Koridon}}, \bibinfo {author} {\bibfnamefont {S.}~\bibnamefont {Yalouz}},
  \bibinfo {author} {\bibfnamefont {B.}~\bibnamefont {Senjean}}, \bibinfo
  {author} {\bibfnamefont {F.}~\bibnamefont {Buda}}, \bibinfo {author}
  {\bibfnamefont {T.~E.}\ \bibnamefont {O'Brien}}, \ and\ \bibinfo {author}
  {\bibfnamefont {L.}~\bibnamefont {Visscher}},\ }\href {\doibase
  10.1103/PhysRevResearch.3.033127} {\bibfield  {journal} {\bibinfo  {journal}
  {Phys. Rev. Res.}\ }\textbf {\bibinfo {volume} {3}},\ \bibinfo {pages}
  {033127} (\bibinfo {year} {2021})}\BibitemShut {NoStop}%
\bibitem [{\citenamefont {Campbell}(2019)}]{campbell2019random}%
  \BibitemOpen
  \bibfield  {author} {\bibinfo {author} {\bibfnamefont {E.}~\bibnamefont
  {Campbell}},\ }\href {\doibase 10.1103/PhysRevLett.123.070503} {\bibfield
  {journal} {\bibinfo  {journal} {Phys. Rev. Lett.}\ }\textbf {\bibinfo
  {volume} {123}},\ \bibinfo {pages} {070503} (\bibinfo {year}
  {2019})}\BibitemShut {NoStop}%
\bibitem [{\citenamefont {Peruzzo}\ \emph {et~al.}(2014)\citenamefont
  {Peruzzo}, \citenamefont {McClean}, \citenamefont {Shadbolt}, \citenamefont
  {Yung}, \citenamefont {Zhou}, \citenamefont {Love}, \citenamefont
  {Aspuru-Guzik},\ and\ \citenamefont {O'Brien}}]{Peruzzo2014variational}%
  \BibitemOpen
  \bibfield  {author} {\bibinfo {author} {\bibfnamefont {A.}~\bibnamefont
  {Peruzzo}}, \bibinfo {author} {\bibfnamefont {J.}~\bibnamefont {McClean}},
  \bibinfo {author} {\bibfnamefont {P.}~\bibnamefont {Shadbolt}}, \bibinfo
  {author} {\bibfnamefont {M.-H.}\ \bibnamefont {Yung}}, \bibinfo {author}
  {\bibfnamefont {X.-Q.}\ \bibnamefont {Zhou}}, \bibinfo {author}
  {\bibfnamefont {P.~J.}\ \bibnamefont {Love}}, \bibinfo {author}
  {\bibfnamefont {A.}~\bibnamefont {Aspuru-Guzik}}, \ and\ \bibinfo {author}
  {\bibfnamefont {J.~L.}\ \bibnamefont {O'Brien}},\ }\href {\doibase
  10.1038/ncomms5213} {\bibfield  {journal} {\bibinfo  {journal} {Nat.
  Commun.}\ }\textbf {\bibinfo {volume} {5}},\ \bibinfo {pages} {4213}
  (\bibinfo {year} {2014})}\BibitemShut {NoStop}%
\bibitem [{\citenamefont {Tilly}\ \emph {et~al.}(2022)\citenamefont {Tilly},
  \citenamefont {Chen}, \citenamefont {Cao}, \citenamefont {Picozzi},
  \citenamefont {Setia}, \citenamefont {Li}, \citenamefont {Grant},
  \citenamefont {Wossnig}, \citenamefont {Rungger}, \citenamefont {Booth},\
  and\ \citenamefont {Tennyson}}]{Tilly2022variational}%
  \BibitemOpen
  \bibfield  {author} {\bibinfo {author} {\bibfnamefont {J.}~\bibnamefont
  {Tilly}}, \bibinfo {author} {\bibfnamefont {H.}~\bibnamefont {Chen}},
  \bibinfo {author} {\bibfnamefont {S.}~\bibnamefont {Cao}}, \bibinfo {author}
  {\bibfnamefont {D.}~\bibnamefont {Picozzi}}, \bibinfo {author} {\bibfnamefont
  {K.}~\bibnamefont {Setia}}, \bibinfo {author} {\bibfnamefont
  {Y.}~\bibnamefont {Li}}, \bibinfo {author} {\bibfnamefont {E.}~\bibnamefont
  {Grant}}, \bibinfo {author} {\bibfnamefont {L.}~\bibnamefont {Wossnig}},
  \bibinfo {author} {\bibfnamefont {I.}~\bibnamefont {Rungger}}, \bibinfo
  {author} {\bibfnamefont {G.~H.}\ \bibnamefont {Booth}}, \ and\ \bibinfo
  {author} {\bibfnamefont {J.}~\bibnamefont {Tennyson}},\ }\href {\doibase
  https://doi.org/10.1016/j.physrep.2022.08.003} {\bibfield  {journal}
  {\bibinfo  {journal} {Phys. Rep.}\ }\textbf {\bibinfo {volume} {986}},\
  \bibinfo {pages} {1} (\bibinfo {year} {2022})}\BibitemShut {NoStop}%
\bibitem [{\citenamefont {Wecker}\ \emph {et~al.}(2015)\citenamefont {Wecker},
  \citenamefont {Hastings},\ and\ \citenamefont {Troyer}}]{wecker2015progress}%
  \BibitemOpen
  \bibfield  {author} {\bibinfo {author} {\bibfnamefont {D.}~\bibnamefont
  {Wecker}}, \bibinfo {author} {\bibfnamefont {M.~B.}\ \bibnamefont
  {Hastings}}, \ and\ \bibinfo {author} {\bibfnamefont {M.}~\bibnamefont
  {Troyer}},\ }\href {\doibase 10.1103/PhysRevA.92.042303} {\bibfield
  {journal} {\bibinfo  {journal} {Phys. Rev. A}\ }\textbf {\bibinfo {volume}
  {92}},\ \bibinfo {pages} {042303} (\bibinfo {year} {2015})}\BibitemShut
  {NoStop}%
\bibitem [{\citenamefont {Rubin}\ \emph {et~al.}(2018)\citenamefont {Rubin},
  \citenamefont {Babbush},\ and\ \citenamefont {McClean}}]{Rubin2018}%
  \BibitemOpen
  \bibfield  {author} {\bibinfo {author} {\bibfnamefont {N.~C.}\ \bibnamefont
  {Rubin}}, \bibinfo {author} {\bibfnamefont {R.}~\bibnamefont {Babbush}}, \
  and\ \bibinfo {author} {\bibfnamefont {J.}~\bibnamefont {McClean}},\ }\href
  {\doibase 10.1088/1367-2630/aab919} {\bibfield  {journal} {\bibinfo
  {journal} {New J. Phys.}\ }\textbf {\bibinfo {volume} {20}},\ \bibinfo
  {pages} {053020} (\bibinfo {year} {2018})}\BibitemShut {NoStop}%
\bibitem [{\citenamefont {Giannozzi}\ \emph {et~al.}(2009)\citenamefont
  {Giannozzi}, \citenamefont {Baroni}, \citenamefont {Bonini}, \citenamefont
  {Calandra}, \citenamefont {Car}, \citenamefont {Cavazzoni}, \citenamefont
  {Ceresoli}, \citenamefont {Chiarotti}, \citenamefont {Cococcioni},
  \citenamefont {Dabo}, \citenamefont {Corso}, \citenamefont {de~Gironcoli},
  \citenamefont {Fabris}, \citenamefont {Fratesi}, \citenamefont {Gebauer},
  \citenamefont {Gerstmann}, \citenamefont {Gougoussis}, \citenamefont
  {Kokalj}, \citenamefont {Lazzeri}, \citenamefont {Martin-Samos},
  \citenamefont {Marzari}, \citenamefont {Mauri}, \citenamefont {Mazzarello},
  \citenamefont {Paolini}, \citenamefont {Pasquarello}, \citenamefont
  {Paulatto}, \citenamefont {Sbraccia}, \citenamefont {Scandolo}, \citenamefont
  {Sclauzero}, \citenamefont {Seitsonen}, \citenamefont {Smogunov},
  \citenamefont {Umari},\ and\ \citenamefont {Wentzcovitch}}]{Espresso1}%
  \BibitemOpen
  \bibfield  {author} {\bibinfo {author} {\bibfnamefont {P.}~\bibnamefont
  {Giannozzi}}, \bibinfo {author} {\bibfnamefont {S.}~\bibnamefont {Baroni}},
  \bibinfo {author} {\bibfnamefont {N.}~\bibnamefont {Bonini}}, \bibinfo
  {author} {\bibfnamefont {M.}~\bibnamefont {Calandra}}, \bibinfo {author}
  {\bibfnamefont {R.}~\bibnamefont {Car}}, \bibinfo {author} {\bibfnamefont
  {C.}~\bibnamefont {Cavazzoni}}, \bibinfo {author} {\bibfnamefont
  {D.}~\bibnamefont {Ceresoli}}, \bibinfo {author} {\bibfnamefont {G.~L.}\
  \bibnamefont {Chiarotti}}, \bibinfo {author} {\bibfnamefont {M.}~\bibnamefont
  {Cococcioni}}, \bibinfo {author} {\bibfnamefont {I.}~\bibnamefont {Dabo}},
  \bibinfo {author} {\bibfnamefont {A.~D.}\ \bibnamefont {Corso}}, \bibinfo
  {author} {\bibfnamefont {S.}~\bibnamefont {de~Gironcoli}}, \bibinfo {author}
  {\bibfnamefont {S.}~\bibnamefont {Fabris}}, \bibinfo {author} {\bibfnamefont
  {G.}~\bibnamefont {Fratesi}}, \bibinfo {author} {\bibfnamefont
  {R.}~\bibnamefont {Gebauer}}, \bibinfo {author} {\bibfnamefont
  {U.}~\bibnamefont {Gerstmann}}, \bibinfo {author} {\bibfnamefont
  {C.}~\bibnamefont {Gougoussis}}, \bibinfo {author} {\bibfnamefont
  {A.}~\bibnamefont {Kokalj}}, \bibinfo {author} {\bibfnamefont
  {M.}~\bibnamefont {Lazzeri}}, \bibinfo {author} {\bibfnamefont
  {L.}~\bibnamefont {Martin-Samos}}, \bibinfo {author} {\bibfnamefont
  {N.}~\bibnamefont {Marzari}}, \bibinfo {author} {\bibfnamefont
  {F.}~\bibnamefont {Mauri}}, \bibinfo {author} {\bibfnamefont
  {R.}~\bibnamefont {Mazzarello}}, \bibinfo {author} {\bibfnamefont
  {S.}~\bibnamefont {Paolini}}, \bibinfo {author} {\bibfnamefont
  {A.}~\bibnamefont {Pasquarello}}, \bibinfo {author} {\bibfnamefont
  {L.}~\bibnamefont {Paulatto}}, \bibinfo {author} {\bibfnamefont
  {C.}~\bibnamefont {Sbraccia}}, \bibinfo {author} {\bibfnamefont
  {S.}~\bibnamefont {Scandolo}}, \bibinfo {author} {\bibfnamefont
  {G.}~\bibnamefont {Sclauzero}}, \bibinfo {author} {\bibfnamefont {A.~P.}\
  \bibnamefont {Seitsonen}}, \bibinfo {author} {\bibfnamefont {A.}~\bibnamefont
  {Smogunov}}, \bibinfo {author} {\bibfnamefont {P.}~\bibnamefont {Umari}}, \
  and\ \bibinfo {author} {\bibfnamefont {R.~M.}\ \bibnamefont {Wentzcovitch}},\
  }\href {\doibase 10.1088/0953-8984/21/39/395502} {\bibfield  {journal}
  {\bibinfo  {journal} {J. Phys.: Condens. Matter}\ }\textbf {\bibinfo {volume}
  {21}},\ \bibinfo {pages} {395502} (\bibinfo {year} {2009})}\BibitemShut
  {NoStop}%
\bibitem [{\citenamefont {Giannozzi}\ \emph {et~al.}(2017)\citenamefont
  {Giannozzi}, \citenamefont {Andreussi}, \citenamefont {Brumme}, \citenamefont
  {Bunau}, \citenamefont {Nardelli}, \citenamefont {Calandra}, \citenamefont
  {Car}, \citenamefont {Cavazzoni}, \citenamefont {Ceresoli}, \citenamefont
  {Cococcioni}, \citenamefont {Colonna}, \citenamefont {Carnimeo},
  \citenamefont {Corso}, \citenamefont {de~Gironcoli}, \citenamefont {Delugas},
  \citenamefont {DiStasio}, \citenamefont {Ferretti}, \citenamefont {Floris},
  \citenamefont {Fratesi}, \citenamefont {Fugallo}, \citenamefont {Gebauer},
  \citenamefont {Gerstmann}, \citenamefont {Giustino}, \citenamefont {Gorni},
  \citenamefont {Jia}, \citenamefont {Kawamura}, \citenamefont {Ko},
  \citenamefont {Kokalj}, \citenamefont {K\"u{\c{c}}\"ukbenli}, \citenamefont
  {Lazzeri}, \citenamefont {Marsili}, \citenamefont {Marzari}, \citenamefont
  {Mauri}, \citenamefont {Nguyen}, \citenamefont {Nguyen}, \citenamefont {de-la
  Roza}, \citenamefont {Paulatto}, \citenamefont {Ponc{\'{e}}}, \citenamefont
  {Rocca}, \citenamefont {Sabatini}, \citenamefont {Santra}, \citenamefont
  {Schlipf}, \citenamefont {Seitsonen}, \citenamefont {Smogunov}, \citenamefont
  {Timrov}, \citenamefont {Thonhauser}, \citenamefont {Umari}, \citenamefont
  {Vast}, \citenamefont {Wu},\ and\ \citenamefont {Baroni}}]{Espresso2}%
  \BibitemOpen
  \bibfield  {author} {\bibinfo {author} {\bibfnamefont {P.}~\bibnamefont
  {Giannozzi}}, \bibinfo {author} {\bibfnamefont {O.}~\bibnamefont
  {Andreussi}}, \bibinfo {author} {\bibfnamefont {T.}~\bibnamefont {Brumme}},
  \bibinfo {author} {\bibfnamefont {O.}~\bibnamefont {Bunau}}, \bibinfo
  {author} {\bibfnamefont {M.~B.}\ \bibnamefont {Nardelli}}, \bibinfo {author}
  {\bibfnamefont {M.}~\bibnamefont {Calandra}}, \bibinfo {author}
  {\bibfnamefont {R.}~\bibnamefont {Car}}, \bibinfo {author} {\bibfnamefont
  {C.}~\bibnamefont {Cavazzoni}}, \bibinfo {author} {\bibfnamefont
  {D.}~\bibnamefont {Ceresoli}}, \bibinfo {author} {\bibfnamefont
  {M.}~\bibnamefont {Cococcioni}}, \bibinfo {author} {\bibfnamefont
  {N.}~\bibnamefont {Colonna}}, \bibinfo {author} {\bibfnamefont
  {I.}~\bibnamefont {Carnimeo}}, \bibinfo {author} {\bibfnamefont {A.~D.}\
  \bibnamefont {Corso}}, \bibinfo {author} {\bibfnamefont {S.}~\bibnamefont
  {de~Gironcoli}}, \bibinfo {author} {\bibfnamefont {P.}~\bibnamefont
  {Delugas}}, \bibinfo {author} {\bibfnamefont {R.~A.}\ \bibnamefont
  {DiStasio}}, \bibinfo {author} {\bibfnamefont {A.}~\bibnamefont {Ferretti}},
  \bibinfo {author} {\bibfnamefont {A.}~\bibnamefont {Floris}}, \bibinfo
  {author} {\bibfnamefont {G.}~\bibnamefont {Fratesi}}, \bibinfo {author}
  {\bibfnamefont {G.}~\bibnamefont {Fugallo}}, \bibinfo {author} {\bibfnamefont
  {R.}~\bibnamefont {Gebauer}}, \bibinfo {author} {\bibfnamefont
  {U.}~\bibnamefont {Gerstmann}}, \bibinfo {author} {\bibfnamefont
  {F.}~\bibnamefont {Giustino}}, \bibinfo {author} {\bibfnamefont
  {T.}~\bibnamefont {Gorni}}, \bibinfo {author} {\bibfnamefont
  {J.}~\bibnamefont {Jia}}, \bibinfo {author} {\bibfnamefont {M.}~\bibnamefont
  {Kawamura}}, \bibinfo {author} {\bibfnamefont {H.-Y.}\ \bibnamefont {Ko}},
  \bibinfo {author} {\bibfnamefont {A.}~\bibnamefont {Kokalj}}, \bibinfo
  {author} {\bibfnamefont {E.}~\bibnamefont {K\"u{\c{c}}\"ukbenli}}, \bibinfo
  {author} {\bibfnamefont {M.}~\bibnamefont {Lazzeri}}, \bibinfo {author}
  {\bibfnamefont {M.}~\bibnamefont {Marsili}}, \bibinfo {author} {\bibfnamefont
  {N.}~\bibnamefont {Marzari}}, \bibinfo {author} {\bibfnamefont
  {F.}~\bibnamefont {Mauri}}, \bibinfo {author} {\bibfnamefont {N.~L.}\
  \bibnamefont {Nguyen}}, \bibinfo {author} {\bibfnamefont {H.-V.}\
  \bibnamefont {Nguyen}}, \bibinfo {author} {\bibfnamefont {A.~O.}\
  \bibnamefont {de-la Roza}}, \bibinfo {author} {\bibfnamefont
  {L.}~\bibnamefont {Paulatto}}, \bibinfo {author} {\bibfnamefont
  {S.}~\bibnamefont {Ponc{\'{e}}}}, \bibinfo {author} {\bibfnamefont
  {D.}~\bibnamefont {Rocca}}, \bibinfo {author} {\bibfnamefont
  {R.}~\bibnamefont {Sabatini}}, \bibinfo {author} {\bibfnamefont
  {B.}~\bibnamefont {Santra}}, \bibinfo {author} {\bibfnamefont
  {M.}~\bibnamefont {Schlipf}}, \bibinfo {author} {\bibfnamefont {A.~P.}\
  \bibnamefont {Seitsonen}}, \bibinfo {author} {\bibfnamefont {A.}~\bibnamefont
  {Smogunov}}, \bibinfo {author} {\bibfnamefont {I.}~\bibnamefont {Timrov}},
  \bibinfo {author} {\bibfnamefont {T.}~\bibnamefont {Thonhauser}}, \bibinfo
  {author} {\bibfnamefont {P.}~\bibnamefont {Umari}}, \bibinfo {author}
  {\bibfnamefont {N.}~\bibnamefont {Vast}}, \bibinfo {author} {\bibfnamefont
  {X.}~\bibnamefont {Wu}}, \ and\ \bibinfo {author} {\bibfnamefont
  {S.}~\bibnamefont {Baroni}},\ }\href {\doibase 10.1088/1361-648x/aa8f79}
  {\bibfield  {journal} {\bibinfo  {journal} {J. Phys.: Condens. Matter}\
  }\textbf {\bibinfo {volume} {29}},\ \bibinfo {pages} {465901} (\bibinfo
  {year} {2017})}\BibitemShut {NoStop}%
\bibitem [{\citenamefont {Giannozzi}\ \emph {et~al.}(2020)\citenamefont
  {Giannozzi}, \citenamefont {Baseggio}, \citenamefont {Bonfà}, \citenamefont
  {Brunato}, \citenamefont {Car}, \citenamefont {Carnimeo}, \citenamefont
  {Cavazzoni}, \citenamefont {de~Gironcoli}, \citenamefont {Delugas},
  \citenamefont {Ferrari~Ruffino}, \citenamefont {Ferretti}, \citenamefont
  {Marzari}, \citenamefont {Timrov}, \citenamefont {Urru},\ and\ \citenamefont
  {Baroni}}]{Espresso3}%
  \BibitemOpen
  \bibfield  {author} {\bibinfo {author} {\bibfnamefont {P.}~\bibnamefont
  {Giannozzi}}, \bibinfo {author} {\bibfnamefont {O.}~\bibnamefont {Baseggio}},
  \bibinfo {author} {\bibfnamefont {P.}~\bibnamefont {Bonfà}}, \bibinfo
  {author} {\bibfnamefont {D.}~\bibnamefont {Brunato}}, \bibinfo {author}
  {\bibfnamefont {R.}~\bibnamefont {Car}}, \bibinfo {author} {\bibfnamefont
  {I.}~\bibnamefont {Carnimeo}}, \bibinfo {author} {\bibfnamefont
  {C.}~\bibnamefont {Cavazzoni}}, \bibinfo {author} {\bibfnamefont
  {S.}~\bibnamefont {de~Gironcoli}}, \bibinfo {author} {\bibfnamefont
  {P.}~\bibnamefont {Delugas}}, \bibinfo {author} {\bibfnamefont
  {F.}~\bibnamefont {Ferrari~Ruffino}}, \bibinfo {author} {\bibfnamefont
  {A.}~\bibnamefont {Ferretti}}, \bibinfo {author} {\bibfnamefont
  {N.}~\bibnamefont {Marzari}}, \bibinfo {author} {\bibfnamefont
  {I.}~\bibnamefont {Timrov}}, \bibinfo {author} {\bibfnamefont
  {A.}~\bibnamefont {Urru}}, \ and\ \bibinfo {author} {\bibfnamefont
  {S.}~\bibnamefont {Baroni}},\ }\href {\doibase 10.1063/5.0005082} {\bibfield
  {journal} {\bibinfo  {journal} {J. Chem. Phys.}\ }\textbf {\bibinfo {volume}
  {152}},\ \bibinfo {pages} {154105} (\bibinfo {year} {2020})}\BibitemShut
  {NoStop}%
\bibitem [{PP()}]{PP}%
  \BibitemOpen
  \href@noop {} {}\bibinfo {howpublished}
  {\url{http://www.quantum-espresso.org}}\BibitemShut {NoStop}%
\bibitem [{\citenamefont {Hamann}(2013)}]{ONCV}%
  \BibitemOpen
  \bibfield  {author} {\bibinfo {author} {\bibfnamefont {D.~R.}\ \bibnamefont
  {Hamann}},\ }\href {\doibase 10.1103/PhysRevB.88.085117} {\bibfield
  {journal} {\bibinfo  {journal} {Phys. Rev. B}\ }\textbf {\bibinfo {volume}
  {88}},\ \bibinfo {pages} {085117} (\bibinfo {year} {2013})}\BibitemShut
  {NoStop}%
\bibitem [{\citenamefont {Becke}(1993)}]{B3LYP1}%
  \BibitemOpen
  \bibfield  {author} {\bibinfo {author} {\bibfnamefont {A.~D.}\ \bibnamefont
  {Becke}},\ }\href {\doibase 10.1063/1.464913} {\bibfield  {journal} {\bibinfo
   {journal} {J. Chem. Phys.}\ }\textbf {\bibinfo {volume} {98}},\ \bibinfo
  {pages} {5648} (\bibinfo {year} {1993})}\BibitemShut {NoStop}%
\bibitem [{\citenamefont {Lee}\ \emph {et~al.}(1988)\citenamefont {Lee},
  \citenamefont {Yang},\ and\ \citenamefont {Parr}}]{B3LYP2}%
  \BibitemOpen
  \bibfield  {author} {\bibinfo {author} {\bibfnamefont {C.}~\bibnamefont
  {Lee}}, \bibinfo {author} {\bibfnamefont {W.}~\bibnamefont {Yang}}, \ and\
  \bibinfo {author} {\bibfnamefont {R.~G.}\ \bibnamefont {Parr}},\ }\href
  {\doibase 10.1103/PhysRevB.37.785} {\bibfield  {journal} {\bibinfo  {journal}
  {Phys. Rev. B}\ }\textbf {\bibinfo {volume} {37}},\ \bibinfo {pages} {785}
  (\bibinfo {year} {1988})}\BibitemShut {NoStop}%
\bibitem [{\citenamefont {Frisch}\ \emph {et~al.}(2016)\citenamefont {Frisch},
  \citenamefont {Trucks}, \citenamefont {Schlegel}, \citenamefont {Scuseria},
  \citenamefont {Robb}, \citenamefont {Cheeseman}, \citenamefont {Scalmani},
  \citenamefont {Barone}, \citenamefont {Petersson}, \citenamefont {Nakatsuji},
  \citenamefont {Li}, \citenamefont {Caricato}, \citenamefont {Marenich},
  \citenamefont {Bloino}, \citenamefont {Janesko}, \citenamefont {Gomperts},
  \citenamefont {Mennucci}, \citenamefont {Hratchian}, \citenamefont {Ortiz},
  \citenamefont {Izmaylov}, \citenamefont {Sonnenberg}, \citenamefont
  {Williams-Young}, \citenamefont {Ding}, \citenamefont {Lipparini},
  \citenamefont {Egidi}, \citenamefont {Goings}, \citenamefont {Peng},
  \citenamefont {Petrone}, \citenamefont {Henderson}, \citenamefont
  {Ranasinghe}, \citenamefont {Zakrzewski}, \citenamefont {Gao}, \citenamefont
  {Rega}, \citenamefont {Zheng}, \citenamefont {Liang}, \citenamefont {Hada},
  \citenamefont {Ehara}, \citenamefont {Toyota}, \citenamefont {Fukuda},
  \citenamefont {Hasegawa}, \citenamefont {Ishida}, \citenamefont {Nakajima},
  \citenamefont {Honda}, \citenamefont {Kitao}, \citenamefont {Nakai},
  \citenamefont {Vreven}, \citenamefont {Throssell}, \citenamefont
  {Montgomery}, \citenamefont {Peralta}, \citenamefont {Ogliaro}, \citenamefont
  {Bearpark}, \citenamefont {Heyd}, \citenamefont {Brothers}, \citenamefont
  {Kudin}, \citenamefont {Staroverov}, \citenamefont {Keith}, \citenamefont
  {Kobayashi}, \citenamefont {Normand}, \citenamefont {Raghavachari},
  \citenamefont {Rendell}, \citenamefont {Burant}, \citenamefont {Iyengar},
  \citenamefont {Tomasi}, \citenamefont {Cossi}, \citenamefont {Millam},
  \citenamefont {Klene}, \citenamefont {Adamo}, \citenamefont {Cammi},
  \citenamefont {Ochterski}, \citenamefont {Martin}, \citenamefont {Morokuma},
  \citenamefont {Farkas}, \citenamefont {Foresman},\ and\ \citenamefont
  {Fox}}]{g16}%
  \BibitemOpen
  \bibfield  {author} {\bibinfo {author} {\bibfnamefont {M.~J.}\ \bibnamefont
  {Frisch}}, \bibinfo {author} {\bibfnamefont {G.~W.}\ \bibnamefont {Trucks}},
  \bibinfo {author} {\bibfnamefont {H.~B.}\ \bibnamefont {Schlegel}}, \bibinfo
  {author} {\bibfnamefont {G.~E.}\ \bibnamefont {Scuseria}}, \bibinfo {author}
  {\bibfnamefont {M.~A.}\ \bibnamefont {Robb}}, \bibinfo {author}
  {\bibfnamefont {J.~R.}\ \bibnamefont {Cheeseman}}, \bibinfo {author}
  {\bibfnamefont {G.}~\bibnamefont {Scalmani}}, \bibinfo {author}
  {\bibfnamefont {V.}~\bibnamefont {Barone}}, \bibinfo {author} {\bibfnamefont
  {G.~A.}\ \bibnamefont {Petersson}}, \bibinfo {author} {\bibfnamefont
  {H.}~\bibnamefont {Nakatsuji}}, \bibinfo {author} {\bibfnamefont
  {X.}~\bibnamefont {Li}}, \bibinfo {author} {\bibfnamefont {M.}~\bibnamefont
  {Caricato}}, \bibinfo {author} {\bibfnamefont {A.~V.}\ \bibnamefont
  {Marenich}}, \bibinfo {author} {\bibfnamefont {J.}~\bibnamefont {Bloino}},
  \bibinfo {author} {\bibfnamefont {B.~G.}\ \bibnamefont {Janesko}}, \bibinfo
  {author} {\bibfnamefont {R.}~\bibnamefont {Gomperts}}, \bibinfo {author}
  {\bibfnamefont {B.}~\bibnamefont {Mennucci}}, \bibinfo {author}
  {\bibfnamefont {H.~P.}\ \bibnamefont {Hratchian}}, \bibinfo {author}
  {\bibfnamefont {J.~V.}\ \bibnamefont {Ortiz}}, \bibinfo {author}
  {\bibfnamefont {A.~F.}\ \bibnamefont {Izmaylov}}, \bibinfo {author}
  {\bibfnamefont {J.~L.}\ \bibnamefont {Sonnenberg}}, \bibinfo {author}
  {\bibfnamefont {D.}~\bibnamefont {Williams-Young}}, \bibinfo {author}
  {\bibfnamefont {F.}~\bibnamefont {Ding}}, \bibinfo {author} {\bibfnamefont
  {F.}~\bibnamefont {Lipparini}}, \bibinfo {author} {\bibfnamefont
  {F.}~\bibnamefont {Egidi}}, \bibinfo {author} {\bibfnamefont
  {J.}~\bibnamefont {Goings}}, \bibinfo {author} {\bibfnamefont
  {B.}~\bibnamefont {Peng}}, \bibinfo {author} {\bibfnamefont {A.}~\bibnamefont
  {Petrone}}, \bibinfo {author} {\bibfnamefont {T.}~\bibnamefont {Henderson}},
  \bibinfo {author} {\bibfnamefont {D.}~\bibnamefont {Ranasinghe}}, \bibinfo
  {author} {\bibfnamefont {V.~G.}\ \bibnamefont {Zakrzewski}}, \bibinfo
  {author} {\bibfnamefont {J.}~\bibnamefont {Gao}}, \bibinfo {author}
  {\bibfnamefont {N.}~\bibnamefont {Rega}}, \bibinfo {author} {\bibfnamefont
  {G.}~\bibnamefont {Zheng}}, \bibinfo {author} {\bibfnamefont
  {W.}~\bibnamefont {Liang}}, \bibinfo {author} {\bibfnamefont
  {M.}~\bibnamefont {Hada}}, \bibinfo {author} {\bibfnamefont {M.}~\bibnamefont
  {Ehara}}, \bibinfo {author} {\bibfnamefont {K.}~\bibnamefont {Toyota}},
  \bibinfo {author} {\bibfnamefont {R.}~\bibnamefont {Fukuda}}, \bibinfo
  {author} {\bibfnamefont {J.}~\bibnamefont {Hasegawa}}, \bibinfo {author}
  {\bibfnamefont {M.}~\bibnamefont {Ishida}}, \bibinfo {author} {\bibfnamefont
  {T.}~\bibnamefont {Nakajima}}, \bibinfo {author} {\bibfnamefont
  {Y.}~\bibnamefont {Honda}}, \bibinfo {author} {\bibfnamefont
  {O.}~\bibnamefont {Kitao}}, \bibinfo {author} {\bibfnamefont
  {H.}~\bibnamefont {Nakai}}, \bibinfo {author} {\bibfnamefont
  {T.}~\bibnamefont {Vreven}}, \bibinfo {author} {\bibfnamefont
  {K.}~\bibnamefont {Throssell}}, \bibinfo {author} {\bibfnamefont {J.~A.}\
  \bibnamefont {Montgomery}, \bibfnamefont {{Jr.}}}, \bibinfo {author}
  {\bibfnamefont {J.~E.}\ \bibnamefont {Peralta}}, \bibinfo {author}
  {\bibfnamefont {F.}~\bibnamefont {Ogliaro}}, \bibinfo {author} {\bibfnamefont
  {M.~J.}\ \bibnamefont {Bearpark}}, \bibinfo {author} {\bibfnamefont {J.~J.}\
  \bibnamefont {Heyd}}, \bibinfo {author} {\bibfnamefont {E.~N.}\ \bibnamefont
  {Brothers}}, \bibinfo {author} {\bibfnamefont {K.~N.}\ \bibnamefont {Kudin}},
  \bibinfo {author} {\bibfnamefont {V.~N.}\ \bibnamefont {Staroverov}},
  \bibinfo {author} {\bibfnamefont {T.~A.}\ \bibnamefont {Keith}}, \bibinfo
  {author} {\bibfnamefont {R.}~\bibnamefont {Kobayashi}}, \bibinfo {author}
  {\bibfnamefont {J.}~\bibnamefont {Normand}}, \bibinfo {author} {\bibfnamefont
  {K.}~\bibnamefont {Raghavachari}}, \bibinfo {author} {\bibfnamefont {A.~P.}\
  \bibnamefont {Rendell}}, \bibinfo {author} {\bibfnamefont {J.~C.}\
  \bibnamefont {Burant}}, \bibinfo {author} {\bibfnamefont {S.~S.}\
  \bibnamefont {Iyengar}}, \bibinfo {author} {\bibfnamefont {J.}~\bibnamefont
  {Tomasi}}, \bibinfo {author} {\bibfnamefont {M.}~\bibnamefont {Cossi}},
  \bibinfo {author} {\bibfnamefont {J.~M.}\ \bibnamefont {Millam}}, \bibinfo
  {author} {\bibfnamefont {M.}~\bibnamefont {Klene}}, \bibinfo {author}
  {\bibfnamefont {C.}~\bibnamefont {Adamo}}, \bibinfo {author} {\bibfnamefont
  {R.}~\bibnamefont {Cammi}}, \bibinfo {author} {\bibfnamefont {J.~W.}\
  \bibnamefont {Ochterski}}, \bibinfo {author} {\bibfnamefont {R.~L.}\
  \bibnamefont {Martin}}, \bibinfo {author} {\bibfnamefont {K.}~\bibnamefont
  {Morokuma}}, \bibinfo {author} {\bibfnamefont {O.}~\bibnamefont {Farkas}},
  \bibinfo {author} {\bibfnamefont {J.~B.}\ \bibnamefont {Foresman}}, \ and\
  \bibinfo {author} {\bibfnamefont {D.~J.}\ \bibnamefont {Fox}},\ }\href@noop
  {} {\enquote {\bibinfo {title} {Gaussian~16 {R}evision {C}.01},}\ } (\bibinfo
  {year} {2016}),\ \bibinfo {note} {{Gaussian Inc. Wallingford CT}}\BibitemShut
  {NoStop}%
\bibitem [{\citenamefont {Fujiwara}\ \emph {et~al.}(2003)\citenamefont
  {Fujiwara}, \citenamefont {Yamamoto},\ and\ \citenamefont
  {Ishii}}]{Fujiwara2003generalization}%
  \BibitemOpen
  \bibfield  {author} {\bibinfo {author} {\bibfnamefont {T.}~\bibnamefont
  {Fujiwara}}, \bibinfo {author} {\bibfnamefont {S.}~\bibnamefont {Yamamoto}},
  \ and\ \bibinfo {author} {\bibfnamefont {Y.}~\bibnamefont {Ishii}},\ }\href
  {\doibase 10.1143/JPSJ.72.777} {\bibfield  {journal} {\bibinfo  {journal} {J.
  Phys. Soc. Jpn.}\ }\textbf {\bibinfo {volume} {72}},\ \bibinfo {pages} {777}
  (\bibinfo {year} {2003})}\BibitemShut {NoStop}%
\bibitem [{\citenamefont {Nohara}\ \emph {et~al.}(2009)\citenamefont {Nohara},
  \citenamefont {Yamamoto},\ and\ \citenamefont
  {Fujiwara}}]{Nohara2009electronic}%
  \BibitemOpen
  \bibfield  {author} {\bibinfo {author} {\bibfnamefont {Y.}~\bibnamefont
  {Nohara}}, \bibinfo {author} {\bibfnamefont {S.}~\bibnamefont {Yamamoto}}, \
  and\ \bibinfo {author} {\bibfnamefont {T.}~\bibnamefont {Fujiwara}},\ }\href
  {\doibase 10.1103/PhysRevB.79.195110} {\bibfield  {journal} {\bibinfo
  {journal} {Phys. Rev. B}\ }\textbf {\bibinfo {volume} {79}},\ \bibinfo
  {pages} {195110} (\bibinfo {year} {2009})}\BibitemShut {NoStop}%
\bibitem [{\citenamefont {Nakamura}\ \emph {et~al.}(2008)\citenamefont
  {Nakamura}, \citenamefont {Arita},\ and\ \citenamefont
  {Imada}}]{Nakamura2008abinitio}%
  \BibitemOpen
  \bibfield  {author} {\bibinfo {author} {\bibfnamefont {K.}~\bibnamefont
  {Nakamura}}, \bibinfo {author} {\bibfnamefont {R.}~\bibnamefont {Arita}}, \
  and\ \bibinfo {author} {\bibfnamefont {M.}~\bibnamefont {Imada}},\ }\href
  {\doibase 10.1143/JPSJ.77.093711} {\bibfield  {journal} {\bibinfo  {journal}
  {J. Phys. Soc. Jpn.}\ }\textbf {\bibinfo {volume} {77}},\ \bibinfo {pages}
  {093711} (\bibinfo {year} {2008})}\BibitemShut {NoStop}%
\bibitem [{\citenamefont {Nakamura}\ \emph {et~al.}(2009)\citenamefont
  {Nakamura}, \citenamefont {Yoshimoto}, \citenamefont {Kosugi}, \citenamefont
  {Arita},\ and\ \citenamefont {Imada}}]{Nakamura2009abinitio}%
  \BibitemOpen
  \bibfield  {author} {\bibinfo {author} {\bibfnamefont {K.}~\bibnamefont
  {Nakamura}}, \bibinfo {author} {\bibfnamefont {Y.}~\bibnamefont {Yoshimoto}},
  \bibinfo {author} {\bibfnamefont {T.}~\bibnamefont {Kosugi}}, \bibinfo
  {author} {\bibfnamefont {R.}~\bibnamefont {Arita}}, \ and\ \bibinfo {author}
  {\bibfnamefont {M.}~\bibnamefont {Imada}},\ }\href {\doibase
  10.1143/JPSJ.78.083710} {\bibfield  {journal} {\bibinfo  {journal} {J. Phys.
  Soc. Jpn.}\ }\textbf {\bibinfo {volume} {78}},\ \bibinfo {pages} {083710}
  (\bibinfo {year} {2009})}\BibitemShut {NoStop}%
\bibitem [{\citenamefont {Nakamura}\ \emph {et~al.}(2016)\citenamefont
  {Nakamura}, \citenamefont {Nohara}, \citenamefont {Yosimoto},\ and\
  \citenamefont {Nomura}}]{Nakamura2016abinitio}%
  \BibitemOpen
  \bibfield  {author} {\bibinfo {author} {\bibfnamefont {K.}~\bibnamefont
  {Nakamura}}, \bibinfo {author} {\bibfnamefont {Y.}~\bibnamefont {Nohara}},
  \bibinfo {author} {\bibfnamefont {Y.}~\bibnamefont {Yosimoto}}, \ and\
  \bibinfo {author} {\bibfnamefont {Y.}~\bibnamefont {Nomura}},\ }\href
  {\doibase 10.1103/PhysRevB.93.085124} {\bibfield  {journal} {\bibinfo
  {journal} {Phys. Rev. B}\ }\textbf {\bibinfo {volume} {93}},\ \bibinfo
  {pages} {085124} (\bibinfo {year} {2016})}\BibitemShut {NoStop}%
\bibitem [{\citenamefont {Momma}\ and\ \citenamefont {Izumi}(2011)}]{VESTA}%
  \BibitemOpen
  \bibfield  {author} {\bibinfo {author} {\bibfnamefont {K.}~\bibnamefont
  {Momma}}\ and\ \bibinfo {author} {\bibfnamefont {F.}~\bibnamefont {Izumi}},\
  }\href {\doibase 10.1107/S0021889811038970} {\bibfield  {journal} {\bibinfo
  {journal} {J. Appl. Cryst.}\ }\textbf {\bibinfo {volume} {44}},\ \bibinfo
  {pages} {1272} (\bibinfo {year} {2011})}\BibitemShut {NoStop}%
\bibitem [{\citenamefont {Sun}\ \emph {et~al.}(2020)\citenamefont {Sun},
  \citenamefont {Zhang}, \citenamefont {Banerjee}, \citenamefont {Bao},
  \citenamefont {Barbry}, \citenamefont {Blunt}, \citenamefont {Bogdanov},
  \citenamefont {Booth}, \citenamefont {Chen}, \citenamefont {Cui},
  \citenamefont {Eriksen}, \citenamefont {Gao}, \citenamefont {Guo},
  \citenamefont {Hermann}, \citenamefont {Hermes}, \citenamefont {Koh},
  \citenamefont {Koval}, \citenamefont {Lehtola}, \citenamefont {Li},
  \citenamefont {Liu}, \citenamefont {Mardirossian}, \citenamefont {McClain},
  \citenamefont {Motta}, \citenamefont {Mussard}, \citenamefont {Pham},
  \citenamefont {Pulkin}, \citenamefont {Purwanto}, \citenamefont {Robinson},
  \citenamefont {Ronca}, \citenamefont {Sayfutyarova}, \citenamefont
  {Scheurer}, \citenamefont {Schurkus}, \citenamefont {Smith}, \citenamefont
  {Sun}, \citenamefont {Sun}, \citenamefont {Upadhyay}, \citenamefont {Wagner},
  \citenamefont {Wang}, \citenamefont {White}, \citenamefont {Whitfield},
  \citenamefont {Williamson}, \citenamefont {Wouters}, \citenamefont {Yang},
  \citenamefont {Yu}, \citenamefont {Zhu}, \citenamefont {Berkelbach},
  \citenamefont {Sharma}, \citenamefont {Sokolov},\ and\ \citenamefont
  {Chan}}]{PySCF}%
  \BibitemOpen
  \bibfield  {author} {\bibinfo {author} {\bibfnamefont {Q.}~\bibnamefont
  {Sun}}, \bibinfo {author} {\bibfnamefont {X.}~\bibnamefont {Zhang}}, \bibinfo
  {author} {\bibfnamefont {S.}~\bibnamefont {Banerjee}}, \bibinfo {author}
  {\bibfnamefont {P.}~\bibnamefont {Bao}}, \bibinfo {author} {\bibfnamefont
  {M.}~\bibnamefont {Barbry}}, \bibinfo {author} {\bibfnamefont {N.~S.}\
  \bibnamefont {Blunt}}, \bibinfo {author} {\bibfnamefont {N.~A.}\ \bibnamefont
  {Bogdanov}}, \bibinfo {author} {\bibfnamefont {G.~H.}\ \bibnamefont {Booth}},
  \bibinfo {author} {\bibfnamefont {J.}~\bibnamefont {Chen}}, \bibinfo {author}
  {\bibfnamefont {Z.-H.}\ \bibnamefont {Cui}}, \bibinfo {author} {\bibfnamefont
  {J.~J.}\ \bibnamefont {Eriksen}}, \bibinfo {author} {\bibfnamefont
  {Y.}~\bibnamefont {Gao}}, \bibinfo {author} {\bibfnamefont {S.}~\bibnamefont
  {Guo}}, \bibinfo {author} {\bibfnamefont {J.}~\bibnamefont {Hermann}},
  \bibinfo {author} {\bibfnamefont {M.~R.}\ \bibnamefont {Hermes}}, \bibinfo
  {author} {\bibfnamefont {K.}~\bibnamefont {Koh}}, \bibinfo {author}
  {\bibfnamefont {P.}~\bibnamefont {Koval}}, \bibinfo {author} {\bibfnamefont
  {S.}~\bibnamefont {Lehtola}}, \bibinfo {author} {\bibfnamefont
  {Z.}~\bibnamefont {Li}}, \bibinfo {author} {\bibfnamefont {J.}~\bibnamefont
  {Liu}}, \bibinfo {author} {\bibfnamefont {N.}~\bibnamefont {Mardirossian}},
  \bibinfo {author} {\bibfnamefont {J.~D.}\ \bibnamefont {McClain}}, \bibinfo
  {author} {\bibfnamefont {M.}~\bibnamefont {Motta}}, \bibinfo {author}
  {\bibfnamefont {B.}~\bibnamefont {Mussard}}, \bibinfo {author} {\bibfnamefont
  {H.~Q.}\ \bibnamefont {Pham}}, \bibinfo {author} {\bibfnamefont
  {A.}~\bibnamefont {Pulkin}}, \bibinfo {author} {\bibfnamefont
  {W.}~\bibnamefont {Purwanto}}, \bibinfo {author} {\bibfnamefont {P.~J.}\
  \bibnamefont {Robinson}}, \bibinfo {author} {\bibfnamefont {E.}~\bibnamefont
  {Ronca}}, \bibinfo {author} {\bibfnamefont {E.~R.}\ \bibnamefont
  {Sayfutyarova}}, \bibinfo {author} {\bibfnamefont {M.}~\bibnamefont
  {Scheurer}}, \bibinfo {author} {\bibfnamefont {H.~F.}\ \bibnamefont
  {Schurkus}}, \bibinfo {author} {\bibfnamefont {J.~E.~T.}\ \bibnamefont
  {Smith}}, \bibinfo {author} {\bibfnamefont {C.}~\bibnamefont {Sun}}, \bibinfo
  {author} {\bibfnamefont {S.-N.}\ \bibnamefont {Sun}}, \bibinfo {author}
  {\bibfnamefont {S.}~\bibnamefont {Upadhyay}}, \bibinfo {author}
  {\bibfnamefont {L.~K.}\ \bibnamefont {Wagner}}, \bibinfo {author}
  {\bibfnamefont {X.}~\bibnamefont {Wang}}, \bibinfo {author} {\bibfnamefont
  {A.}~\bibnamefont {White}}, \bibinfo {author} {\bibfnamefont {J.~D.}\
  \bibnamefont {Whitfield}}, \bibinfo {author} {\bibfnamefont {M.~J.}\
  \bibnamefont {Williamson}}, \bibinfo {author} {\bibfnamefont
  {S.}~\bibnamefont {Wouters}}, \bibinfo {author} {\bibfnamefont
  {J.}~\bibnamefont {Yang}}, \bibinfo {author} {\bibfnamefont {J.~M.}\
  \bibnamefont {Yu}}, \bibinfo {author} {\bibfnamefont {T.}~\bibnamefont
  {Zhu}}, \bibinfo {author} {\bibfnamefont {T.~C.}\ \bibnamefont {Berkelbach}},
  \bibinfo {author} {\bibfnamefont {S.}~\bibnamefont {Sharma}}, \bibinfo
  {author} {\bibfnamefont {A.~Y.}\ \bibnamefont {Sokolov}}, \ and\ \bibinfo
  {author} {\bibfnamefont {G.~K.-L.}\ \bibnamefont {Chan}},\ }\href {\doibase
  10.1063/5.0006074} {\bibfield  {journal} {\bibinfo  {journal} {J. Chem.
  Phys.}\ }\textbf {\bibinfo {volume} {153}},\ \bibinfo {pages} {024109}
  (\bibinfo {year} {2020})}\BibitemShut {NoStop}%
\bibitem [{\citenamefont {Sun}\ \emph {et~al.}(2018)\citenamefont {Sun},
  \citenamefont {Berkelbach}, \citenamefont {Blunt}, \citenamefont {Booth},
  \citenamefont {Guo}, \citenamefont {Li}, \citenamefont {Liu}, \citenamefont
  {McClain}, \citenamefont {Sayfutyarova}, \citenamefont {Sharma},
  \citenamefont {Wouters},\ and\ \citenamefont {Chan}}]{PySCF2}%
  \BibitemOpen
  \bibfield  {author} {\bibinfo {author} {\bibfnamefont {Q.}~\bibnamefont
  {Sun}}, \bibinfo {author} {\bibfnamefont {T.~C.}\ \bibnamefont {Berkelbach}},
  \bibinfo {author} {\bibfnamefont {N.~S.}\ \bibnamefont {Blunt}}, \bibinfo
  {author} {\bibfnamefont {G.~H.}\ \bibnamefont {Booth}}, \bibinfo {author}
  {\bibfnamefont {S.}~\bibnamefont {Guo}}, \bibinfo {author} {\bibfnamefont
  {Z.}~\bibnamefont {Li}}, \bibinfo {author} {\bibfnamefont {J.}~\bibnamefont
  {Liu}}, \bibinfo {author} {\bibfnamefont {J.~D.}\ \bibnamefont {McClain}},
  \bibinfo {author} {\bibfnamefont {E.~R.}\ \bibnamefont {Sayfutyarova}},
  \bibinfo {author} {\bibfnamefont {S.}~\bibnamefont {Sharma}}, \bibinfo
  {author} {\bibfnamefont {S.}~\bibnamefont {Wouters}}, \ and\ \bibinfo
  {author} {\bibfnamefont {G.~K.-L.}\ \bibnamefont {Chan}},\ }\href {\doibase
  https://doi.org/10.1002/wcms.1340} {\bibfield  {journal} {\bibinfo  {journal}
  {WIREs Comput. Mol. Sci.}\ }\textbf {\bibinfo {volume} {8}},\ \bibinfo
  {pages} {e1340} (\bibinfo {year} {2018})}\BibitemShut {NoStop}%
\bibitem [{\citenamefont {McClean}\ \emph {et~al.}(2020)\citenamefont
  {McClean}, \citenamefont {Rubin}, \citenamefont {Sung}, \citenamefont
  {Kivlichan}, \citenamefont {Bonet-Monroig}, \citenamefont {Cao},
  \citenamefont {Dai}, \citenamefont {Fried}, \citenamefont {Gidney},
  \citenamefont {Gimby}, \citenamefont {Gokhale}, \citenamefont {Häner},
  \citenamefont {Hardikar}, \citenamefont {Havlíček}, \citenamefont
  {Higgott}, \citenamefont {Huang}, \citenamefont {Izaac}, \citenamefont
  {Jiang}, \citenamefont {Liu}, \citenamefont {McArdle}, \citenamefont
  {Neeley}, \citenamefont {O’Brien}, \citenamefont {O’Gorman},
  \citenamefont {Ozfidan}, \citenamefont {Radin}, \citenamefont {Romero},
  \citenamefont {Sawaya}, \citenamefont {Senjean}, \citenamefont {Setia},
  \citenamefont {Sim}, \citenamefont {Steiger}, \citenamefont {Steudtner},
  \citenamefont {Sun}, \citenamefont {Sun}, \citenamefont {Wang}, \citenamefont
  {Zhang},\ and\ \citenamefont {Babbush}}]{McClean2020}%
  \BibitemOpen
  \bibfield  {author} {\bibinfo {author} {\bibfnamefont {J.~R.}\ \bibnamefont
  {McClean}}, \bibinfo {author} {\bibfnamefont {N.~C.}\ \bibnamefont {Rubin}},
  \bibinfo {author} {\bibfnamefont {K.~J.}\ \bibnamefont {Sung}}, \bibinfo
  {author} {\bibfnamefont {I.~D.}\ \bibnamefont {Kivlichan}}, \bibinfo {author}
  {\bibfnamefont {X.}~\bibnamefont {Bonet-Monroig}}, \bibinfo {author}
  {\bibfnamefont {Y.}~\bibnamefont {Cao}}, \bibinfo {author} {\bibfnamefont
  {C.}~\bibnamefont {Dai}}, \bibinfo {author} {\bibfnamefont {E.~S.}\
  \bibnamefont {Fried}}, \bibinfo {author} {\bibfnamefont {C.}~\bibnamefont
  {Gidney}}, \bibinfo {author} {\bibfnamefont {B.}~\bibnamefont {Gimby}},
  \bibinfo {author} {\bibfnamefont {P.}~\bibnamefont {Gokhale}}, \bibinfo
  {author} {\bibfnamefont {T.}~\bibnamefont {Häner}}, \bibinfo {author}
  {\bibfnamefont {T.}~\bibnamefont {Hardikar}}, \bibinfo {author}
  {\bibfnamefont {V.}~\bibnamefont {Havlíček}}, \bibinfo {author}
  {\bibfnamefont {O.}~\bibnamefont {Higgott}}, \bibinfo {author} {\bibfnamefont
  {C.}~\bibnamefont {Huang}}, \bibinfo {author} {\bibfnamefont
  {J.}~\bibnamefont {Izaac}}, \bibinfo {author} {\bibfnamefont
  {Z.}~\bibnamefont {Jiang}}, \bibinfo {author} {\bibfnamefont
  {X.}~\bibnamefont {Liu}}, \bibinfo {author} {\bibfnamefont {S.}~\bibnamefont
  {McArdle}}, \bibinfo {author} {\bibfnamefont {M.}~\bibnamefont {Neeley}},
  \bibinfo {author} {\bibfnamefont {T.}~\bibnamefont {O’Brien}}, \bibinfo
  {author} {\bibfnamefont {B.}~\bibnamefont {O’Gorman}}, \bibinfo {author}
  {\bibfnamefont {I.}~\bibnamefont {Ozfidan}}, \bibinfo {author} {\bibfnamefont
  {M.~D.}\ \bibnamefont {Radin}}, \bibinfo {author} {\bibfnamefont
  {J.}~\bibnamefont {Romero}}, \bibinfo {author} {\bibfnamefont {N.~P.~D.}\
  \bibnamefont {Sawaya}}, \bibinfo {author} {\bibfnamefont {B.}~\bibnamefont
  {Senjean}}, \bibinfo {author} {\bibfnamefont {K.}~\bibnamefont {Setia}},
  \bibinfo {author} {\bibfnamefont {S.}~\bibnamefont {Sim}}, \bibinfo {author}
  {\bibfnamefont {D.~S.}\ \bibnamefont {Steiger}}, \bibinfo {author}
  {\bibfnamefont {M.}~\bibnamefont {Steudtner}}, \bibinfo {author}
  {\bibfnamefont {Q.}~\bibnamefont {Sun}}, \bibinfo {author} {\bibfnamefont
  {W.}~\bibnamefont {Sun}}, \bibinfo {author} {\bibfnamefont {D.}~\bibnamefont
  {Wang}}, \bibinfo {author} {\bibfnamefont {F.}~\bibnamefont {Zhang}}, \ and\
  \bibinfo {author} {\bibfnamefont {R.}~\bibnamefont {Babbush}},\ }\href
  {\doibase 10.1088/2058-9565/ab8ebc} {\bibfield  {journal} {\bibinfo
  {journal} {Quantum Sci. Technol.}\ }\textbf {\bibinfo {volume} {5}},\
  \bibinfo {pages} {034014} (\bibinfo {year} {2020})}\BibitemShut {NoStop}%
\bibitem [{\citenamefont {Folkestad}\ \emph {et~al.}(2020)\citenamefont
  {Folkestad}, \citenamefont {Kjønstad}, \citenamefont {Myhre}, \citenamefont
  {Andersen}, \citenamefont {Balbi}, \citenamefont {Coriani}, \citenamefont
  {Giovannini}, \citenamefont {Goletto}, \citenamefont {Haugland},
  \citenamefont {Hutcheson}, \citenamefont {Høyvik}, \citenamefont {Moitra},
  \citenamefont {Paul}, \citenamefont {Scavino}, \citenamefont {Skeidsvoll},
  \citenamefont {Tveten},\ and\ \citenamefont {Koch}}]{Folkestad2020eT}%
  \BibitemOpen
  \bibfield  {author} {\bibinfo {author} {\bibfnamefont {S.~D.}\ \bibnamefont
  {Folkestad}}, \bibinfo {author} {\bibfnamefont {E.~F.}\ \bibnamefont
  {Kjønstad}}, \bibinfo {author} {\bibfnamefont {R.~H.}\ \bibnamefont
  {Myhre}}, \bibinfo {author} {\bibfnamefont {J.~H.}\ \bibnamefont {Andersen}},
  \bibinfo {author} {\bibfnamefont {A.}~\bibnamefont {Balbi}}, \bibinfo
  {author} {\bibfnamefont {S.}~\bibnamefont {Coriani}}, \bibinfo {author}
  {\bibfnamefont {T.}~\bibnamefont {Giovannini}}, \bibinfo {author}
  {\bibfnamefont {L.}~\bibnamefont {Goletto}}, \bibinfo {author} {\bibfnamefont
  {T.~S.}\ \bibnamefont {Haugland}}, \bibinfo {author} {\bibfnamefont
  {A.}~\bibnamefont {Hutcheson}}, \bibinfo {author} {\bibfnamefont {I.-M.}\
  \bibnamefont {Høyvik}}, \bibinfo {author} {\bibfnamefont {T.}~\bibnamefont
  {Moitra}}, \bibinfo {author} {\bibfnamefont {A.~C.}\ \bibnamefont {Paul}},
  \bibinfo {author} {\bibfnamefont {M.}~\bibnamefont {Scavino}}, \bibinfo
  {author} {\bibfnamefont {A.~S.}\ \bibnamefont {Skeidsvoll}}, \bibinfo
  {author} {\bibfnamefont {{\AA}.~H.}\ \bibnamefont {Tveten}}, \ and\ \bibinfo
  {author} {\bibfnamefont {H.}~\bibnamefont {Koch}},\ }\href {\doibase
  10.1063/5.0004713} {\bibfield  {journal} {\bibinfo  {journal} {J. Chem.
  Phys.}\ }\textbf {\bibinfo {volume} {152}},\ \bibinfo {pages} {184103}
  (\bibinfo {year} {2020})}\BibitemShut {NoStop}%
\bibitem [{\citenamefont {Paul}\ \emph {et~al.}(2021)\citenamefont {Paul},
  \citenamefont {Myhre},\ and\ \citenamefont {Koch}}]{Paul2021new}%
  \BibitemOpen
  \bibfield  {author} {\bibinfo {author} {\bibfnamefont {A.~C.}\ \bibnamefont
  {Paul}}, \bibinfo {author} {\bibfnamefont {R.~H.}\ \bibnamefont {Myhre}}, \
  and\ \bibinfo {author} {\bibfnamefont {H.}~\bibnamefont {Koch}},\ }\href
  {\doibase 10.1021/acs.jctc.0c00686} {\bibfield  {journal} {\bibinfo
  {journal} {J. Chem. Theory Comput.}\ }\textbf {\bibinfo {volume} {17}},\
  \bibinfo {pages} {117} (\bibinfo {year} {2021})}\BibitemShut {NoStop}%
\bibitem [{\citenamefont {Li~Manni}\ \emph {et~al.}(2023)\citenamefont
  {Li~Manni}, \citenamefont {Fdez.~Galván}, \citenamefont {Alavi},
  \citenamefont {Aleotti}, \citenamefont {Aquilante}, \citenamefont
  {Autschbach}, \citenamefont {Avagliano}, \citenamefont {Baiardi},
  \citenamefont {Bao}, \citenamefont {Battaglia}, \citenamefont {Birnoschi},
  \citenamefont {Blanco-González}, \citenamefont {Bokarev}, \citenamefont
  {Broer}, \citenamefont {Cacciari}, \citenamefont {Calio}, \citenamefont
  {Carlson}, \citenamefont {Carvalho~Couto}, \citenamefont {Cerdán},
  \citenamefont {Chibotaru}, \citenamefont {Chilton}, \citenamefont {Church},
  \citenamefont {Conti}, \citenamefont {Coriani}, \citenamefont
  {Cuéllar-Zuquin}, \citenamefont {Daoud}, \citenamefont {Dattani},
  \citenamefont {Decleva}, \citenamefont {de~Graaf}, \citenamefont {Delcey},
  \citenamefont {De~Vico}, \citenamefont {Dobrautz}, \citenamefont {Dong},
  \citenamefont {Feng}, \citenamefont {Ferré}, \citenamefont {Filatov(Gulak)},
  \citenamefont {Gagliardi}, \citenamefont {Garavelli}, \citenamefont
  {González}, \citenamefont {Guan}, \citenamefont {Guo}, \citenamefont
  {Hennefarth}, \citenamefont {Hermes}, \citenamefont {Hoyer}, \citenamefont
  {Huix-Rotllant}, \citenamefont {Jaiswal}, \citenamefont {Kaiser},
  \citenamefont {Kaliakin}, \citenamefont {Khamesian}, \citenamefont {King},
  \citenamefont {Kochetov}, \citenamefont {Krośnicki}, \citenamefont {Kumaar},
  \citenamefont {Larsson}, \citenamefont {Lehtola}, \citenamefont {Lepetit},
  \citenamefont {Lischka}, \citenamefont {López~Ríos}, \citenamefont
  {Lundberg}, \citenamefont {Ma}, \citenamefont {Mai}, \citenamefont
  {Marquetand}, \citenamefont {Merritt}, \citenamefont {Montorsi},
  \citenamefont {Mörchen}, \citenamefont {Nenov}, \citenamefont {Nguyen},
  \citenamefont {Nishimoto}, \citenamefont {Oakley}, \citenamefont {Olivucci},
  \citenamefont {Oppel}, \citenamefont {Padula}, \citenamefont {Pandharkar},
  \citenamefont {Phung}, \citenamefont {Plasser}, \citenamefont {Raggi},
  \citenamefont {Rebolini}, \citenamefont {Reiher}, \citenamefont {Rivalta},
  \citenamefont {Roca-Sanjuán}, \citenamefont {Romig}, \citenamefont {Safari},
  \citenamefont {Sánchez-Mansilla}, \citenamefont {Sand}, \citenamefont
  {Schapiro}, \citenamefont {Scott}, \citenamefont {Segarra-Martí},
  \citenamefont {Segatta}, \citenamefont {Sergentu}, \citenamefont {Sharma},
  \citenamefont {Shepard}, \citenamefont {Shu}, \citenamefont {Staab},
  \citenamefont {Straatsma}, \citenamefont {S{\o}rensen}, \citenamefont
  {Tenorio}, \citenamefont {Truhlar}, \citenamefont {Ungur}, \citenamefont
  {Vacher}, \citenamefont {Veryazov}, \citenamefont {Vo{\ss}}, \citenamefont
  {Weser}, \citenamefont {Wu}, \citenamefont {Yang}, \citenamefont {Yarkony},
  \citenamefont {Zhou}, \citenamefont {Zobel},\ and\ \citenamefont
  {Lindh}}]{OpenMolcas1}%
  \BibitemOpen
  \bibfield  {author} {\bibinfo {author} {\bibfnamefont {G.}~\bibnamefont
  {Li~Manni}}, \bibinfo {author} {\bibfnamefont {I.}~\bibnamefont
  {Fdez.~Galván}}, \bibinfo {author} {\bibfnamefont {A.}~\bibnamefont
  {Alavi}}, \bibinfo {author} {\bibfnamefont {F.}~\bibnamefont {Aleotti}},
  \bibinfo {author} {\bibfnamefont {F.}~\bibnamefont {Aquilante}}, \bibinfo
  {author} {\bibfnamefont {J.}~\bibnamefont {Autschbach}}, \bibinfo {author}
  {\bibfnamefont {D.}~\bibnamefont {Avagliano}}, \bibinfo {author}
  {\bibfnamefont {A.}~\bibnamefont {Baiardi}}, \bibinfo {author} {\bibfnamefont
  {J.~J.}\ \bibnamefont {Bao}}, \bibinfo {author} {\bibfnamefont
  {S.}~\bibnamefont {Battaglia}}, \bibinfo {author} {\bibfnamefont
  {L.}~\bibnamefont {Birnoschi}}, \bibinfo {author} {\bibfnamefont
  {A.}~\bibnamefont {Blanco-González}}, \bibinfo {author} {\bibfnamefont
  {S.~I.}\ \bibnamefont {Bokarev}}, \bibinfo {author} {\bibfnamefont
  {R.}~\bibnamefont {Broer}}, \bibinfo {author} {\bibfnamefont
  {R.}~\bibnamefont {Cacciari}}, \bibinfo {author} {\bibfnamefont {P.~B.}\
  \bibnamefont {Calio}}, \bibinfo {author} {\bibfnamefont {R.~K.}\ \bibnamefont
  {Carlson}}, \bibinfo {author} {\bibfnamefont {R.}~\bibnamefont
  {Carvalho~Couto}}, \bibinfo {author} {\bibfnamefont {L.}~\bibnamefont
  {Cerdán}}, \bibinfo {author} {\bibfnamefont {L.~F.}\ \bibnamefont
  {Chibotaru}}, \bibinfo {author} {\bibfnamefont {N.~F.}\ \bibnamefont
  {Chilton}}, \bibinfo {author} {\bibfnamefont {J.~R.}\ \bibnamefont {Church}},
  \bibinfo {author} {\bibfnamefont {I.}~\bibnamefont {Conti}}, \bibinfo
  {author} {\bibfnamefont {S.}~\bibnamefont {Coriani}}, \bibinfo {author}
  {\bibfnamefont {J.}~\bibnamefont {Cuéllar-Zuquin}}, \bibinfo {author}
  {\bibfnamefont {R.~E.}\ \bibnamefont {Daoud}}, \bibinfo {author}
  {\bibfnamefont {N.}~\bibnamefont {Dattani}}, \bibinfo {author} {\bibfnamefont
  {P.}~\bibnamefont {Decleva}}, \bibinfo {author} {\bibfnamefont
  {C.}~\bibnamefont {de~Graaf}}, \bibinfo {author} {\bibfnamefont {M.~G.}\
  \bibnamefont {Delcey}}, \bibinfo {author} {\bibfnamefont {L.}~\bibnamefont
  {De~Vico}}, \bibinfo {author} {\bibfnamefont {W.}~\bibnamefont {Dobrautz}},
  \bibinfo {author} {\bibfnamefont {S.~S.}\ \bibnamefont {Dong}}, \bibinfo
  {author} {\bibfnamefont {R.}~\bibnamefont {Feng}}, \bibinfo {author}
  {\bibfnamefont {N.}~\bibnamefont {Ferré}}, \bibinfo {author} {\bibfnamefont
  {M.}~\bibnamefont {Filatov(Gulak)}}, \bibinfo {author} {\bibfnamefont
  {L.}~\bibnamefont {Gagliardi}}, \bibinfo {author} {\bibfnamefont
  {M.}~\bibnamefont {Garavelli}}, \bibinfo {author} {\bibfnamefont
  {L.}~\bibnamefont {González}}, \bibinfo {author} {\bibfnamefont
  {Y.}~\bibnamefont {Guan}}, \bibinfo {author} {\bibfnamefont {M.}~\bibnamefont
  {Guo}}, \bibinfo {author} {\bibfnamefont {M.~R.}\ \bibnamefont {Hennefarth}},
  \bibinfo {author} {\bibfnamefont {M.~R.}\ \bibnamefont {Hermes}}, \bibinfo
  {author} {\bibfnamefont {C.~E.}\ \bibnamefont {Hoyer}}, \bibinfo {author}
  {\bibfnamefont {M.}~\bibnamefont {Huix-Rotllant}}, \bibinfo {author}
  {\bibfnamefont {V.~K.}\ \bibnamefont {Jaiswal}}, \bibinfo {author}
  {\bibfnamefont {A.}~\bibnamefont {Kaiser}}, \bibinfo {author} {\bibfnamefont
  {D.~S.}\ \bibnamefont {Kaliakin}}, \bibinfo {author} {\bibfnamefont
  {M.}~\bibnamefont {Khamesian}}, \bibinfo {author} {\bibfnamefont {D.~S.}\
  \bibnamefont {King}}, \bibinfo {author} {\bibfnamefont {V.}~\bibnamefont
  {Kochetov}}, \bibinfo {author} {\bibfnamefont {M.}~\bibnamefont
  {Krośnicki}}, \bibinfo {author} {\bibfnamefont {A.~A.}\ \bibnamefont
  {Kumaar}}, \bibinfo {author} {\bibfnamefont {E.~D.}\ \bibnamefont {Larsson}},
  \bibinfo {author} {\bibfnamefont {S.}~\bibnamefont {Lehtola}}, \bibinfo
  {author} {\bibfnamefont {M.-B.}\ \bibnamefont {Lepetit}}, \bibinfo {author}
  {\bibfnamefont {H.}~\bibnamefont {Lischka}}, \bibinfo {author} {\bibfnamefont
  {P.}~\bibnamefont {López~Ríos}}, \bibinfo {author} {\bibfnamefont
  {M.}~\bibnamefont {Lundberg}}, \bibinfo {author} {\bibfnamefont
  {D.}~\bibnamefont {Ma}}, \bibinfo {author} {\bibfnamefont {S.}~\bibnamefont
  {Mai}}, \bibinfo {author} {\bibfnamefont {P.}~\bibnamefont {Marquetand}},
  \bibinfo {author} {\bibfnamefont {I.~C.~D.}\ \bibnamefont {Merritt}},
  \bibinfo {author} {\bibfnamefont {F.}~\bibnamefont {Montorsi}}, \bibinfo
  {author} {\bibfnamefont {M.}~\bibnamefont {Mörchen}}, \bibinfo {author}
  {\bibfnamefont {A.}~\bibnamefont {Nenov}}, \bibinfo {author} {\bibfnamefont
  {V.~H.~A.}\ \bibnamefont {Nguyen}}, \bibinfo {author} {\bibfnamefont
  {Y.}~\bibnamefont {Nishimoto}}, \bibinfo {author} {\bibfnamefont {M.~S.}\
  \bibnamefont {Oakley}}, \bibinfo {author} {\bibfnamefont {M.}~\bibnamefont
  {Olivucci}}, \bibinfo {author} {\bibfnamefont {M.}~\bibnamefont {Oppel}},
  \bibinfo {author} {\bibfnamefont {D.}~\bibnamefont {Padula}}, \bibinfo
  {author} {\bibfnamefont {R.}~\bibnamefont {Pandharkar}}, \bibinfo {author}
  {\bibfnamefont {Q.~M.}\ \bibnamefont {Phung}}, \bibinfo {author}
  {\bibfnamefont {F.}~\bibnamefont {Plasser}}, \bibinfo {author} {\bibfnamefont
  {G.}~\bibnamefont {Raggi}}, \bibinfo {author} {\bibfnamefont
  {E.}~\bibnamefont {Rebolini}}, \bibinfo {author} {\bibfnamefont
  {M.}~\bibnamefont {Reiher}}, \bibinfo {author} {\bibfnamefont
  {I.}~\bibnamefont {Rivalta}}, \bibinfo {author} {\bibfnamefont
  {D.}~\bibnamefont {Roca-Sanjuán}}, \bibinfo {author} {\bibfnamefont
  {T.}~\bibnamefont {Romig}}, \bibinfo {author} {\bibfnamefont {A.~A.}\
  \bibnamefont {Safari}}, \bibinfo {author} {\bibfnamefont {A.}~\bibnamefont
  {Sánchez-Mansilla}}, \bibinfo {author} {\bibfnamefont {A.~M.}\ \bibnamefont
  {Sand}}, \bibinfo {author} {\bibfnamefont {I.}~\bibnamefont {Schapiro}},
  \bibinfo {author} {\bibfnamefont {T.~R.}\ \bibnamefont {Scott}}, \bibinfo
  {author} {\bibfnamefont {J.}~\bibnamefont {Segarra-Martí}}, \bibinfo
  {author} {\bibfnamefont {F.}~\bibnamefont {Segatta}}, \bibinfo {author}
  {\bibfnamefont {D.-C.}\ \bibnamefont {Sergentu}}, \bibinfo {author}
  {\bibfnamefont {P.}~\bibnamefont {Sharma}}, \bibinfo {author} {\bibfnamefont
  {R.}~\bibnamefont {Shepard}}, \bibinfo {author} {\bibfnamefont
  {Y.}~\bibnamefont {Shu}}, \bibinfo {author} {\bibfnamefont {J.~K.}\
  \bibnamefont {Staab}}, \bibinfo {author} {\bibfnamefont {T.~P.}\ \bibnamefont
  {Straatsma}}, \bibinfo {author} {\bibfnamefont {L.~K.}\ \bibnamefont
  {S{\o}rensen}}, \bibinfo {author} {\bibfnamefont {B.~N.~C.}\ \bibnamefont
  {Tenorio}}, \bibinfo {author} {\bibfnamefont {D.~G.}\ \bibnamefont
  {Truhlar}}, \bibinfo {author} {\bibfnamefont {L.}~\bibnamefont {Ungur}},
  \bibinfo {author} {\bibfnamefont {M.}~\bibnamefont {Vacher}}, \bibinfo
  {author} {\bibfnamefont {V.}~\bibnamefont {Veryazov}}, \bibinfo {author}
  {\bibfnamefont {T.~A.}\ \bibnamefont {Vo{\ss}}}, \bibinfo {author}
  {\bibfnamefont {O.}~\bibnamefont {Weser}}, \bibinfo {author} {\bibfnamefont
  {D.}~\bibnamefont {Wu}}, \bibinfo {author} {\bibfnamefont {X.}~\bibnamefont
  {Yang}}, \bibinfo {author} {\bibfnamefont {D.}~\bibnamefont {Yarkony}},
  \bibinfo {author} {\bibfnamefont {C.}~\bibnamefont {Zhou}}, \bibinfo {author}
  {\bibfnamefont {J.~P.}\ \bibnamefont {Zobel}}, \ and\ \bibinfo {author}
  {\bibfnamefont {R.}~\bibnamefont {Lindh}},\ }\href {\doibase
  10.1021/acs.jctc.3c00182} {\bibfield  {journal} {\bibinfo  {journal} {J.
  Chem. Theory Comput.}\ }\textbf {\bibinfo {volume} {19}},\ \bibinfo {pages}
  {6933} (\bibinfo {year} {2023})}\BibitemShut {NoStop}%
\bibitem [{\citenamefont {Aquilante}\ \emph {et~al.}(2020)\citenamefont
  {Aquilante}, \citenamefont {Autschbach}, \citenamefont {Baiardi},
  \citenamefont {Battaglia}, \citenamefont {Borin}, \citenamefont {Chibotaru},
  \citenamefont {Conti}, \citenamefont {De~Vico}, \citenamefont {Delcey},
  \citenamefont {Fdez.~Galv\'{a}n}, \citenamefont {Ferr\'{e}}, \citenamefont
  {Freitag}, \citenamefont {Garavelli}, \citenamefont {Gong}, \citenamefont
  {Knecht}, \citenamefont {Larsson}, \citenamefont {Lindh}, \citenamefont
  {Lundberg}, \citenamefont {Malmqvist}, \citenamefont {Nenov}, \citenamefont
  {Norell}, \citenamefont {Odelius}, \citenamefont {Olivucci}, \citenamefont
  {Pedersen}, \citenamefont {Pedraza-Gonz\'{a}lez}, \citenamefont {Phung},
  \citenamefont {Pierloot}, \citenamefont {Reiher}, \citenamefont {Schapiro},
  \citenamefont {Segarra-Mart\'{i}}, \citenamefont {Segatta}, \citenamefont
  {Seijo}, \citenamefont {Sen}, \citenamefont {Sergentu}, \citenamefont
  {Stein}, \citenamefont {Ungur}, \citenamefont {Vacher}, \citenamefont
  {Valentini},\ and\ \citenamefont {Veryazov}}]{OpenMolcas2}%
  \BibitemOpen
  \bibfield  {author} {\bibinfo {author} {\bibfnamefont {F.}~\bibnamefont
  {Aquilante}}, \bibinfo {author} {\bibfnamefont {J.}~\bibnamefont
  {Autschbach}}, \bibinfo {author} {\bibfnamefont {A.}~\bibnamefont {Baiardi}},
  \bibinfo {author} {\bibfnamefont {S.}~\bibnamefont {Battaglia}}, \bibinfo
  {author} {\bibfnamefont {V.~A.}\ \bibnamefont {Borin}}, \bibinfo {author}
  {\bibfnamefont {L.~F.}\ \bibnamefont {Chibotaru}}, \bibinfo {author}
  {\bibfnamefont {I.}~\bibnamefont {Conti}}, \bibinfo {author} {\bibfnamefont
  {L.}~\bibnamefont {De~Vico}}, \bibinfo {author} {\bibfnamefont
  {M.}~\bibnamefont {Delcey}}, \bibinfo {author} {\bibfnamefont
  {I.}~\bibnamefont {Fdez.~Galv\'{a}n}}, \bibinfo {author} {\bibfnamefont
  {N.}~\bibnamefont {Ferr\'{e}}}, \bibinfo {author} {\bibfnamefont
  {L.}~\bibnamefont {Freitag}}, \bibinfo {author} {\bibfnamefont
  {M.}~\bibnamefont {Garavelli}}, \bibinfo {author} {\bibfnamefont
  {X.}~\bibnamefont {Gong}}, \bibinfo {author} {\bibfnamefont {S.}~\bibnamefont
  {Knecht}}, \bibinfo {author} {\bibfnamefont {E.~D.}\ \bibnamefont {Larsson}},
  \bibinfo {author} {\bibfnamefont {R.}~\bibnamefont {Lindh}}, \bibinfo
  {author} {\bibfnamefont {M.}~\bibnamefont {Lundberg}}, \bibinfo {author}
  {\bibfnamefont {P.~{\AA}.}\ \bibnamefont {Malmqvist}}, \bibinfo {author}
  {\bibfnamefont {A.}~\bibnamefont {Nenov}}, \bibinfo {author} {\bibfnamefont
  {J.}~\bibnamefont {Norell}}, \bibinfo {author} {\bibfnamefont
  {M.}~\bibnamefont {Odelius}}, \bibinfo {author} {\bibfnamefont
  {M.}~\bibnamefont {Olivucci}}, \bibinfo {author} {\bibfnamefont {T.~B.}\
  \bibnamefont {Pedersen}}, \bibinfo {author} {\bibfnamefont {L.}~\bibnamefont
  {Pedraza-Gonz\'{a}lez}}, \bibinfo {author} {\bibfnamefont {Q.~M.}\
  \bibnamefont {Phung}}, \bibinfo {author} {\bibfnamefont {K.}~\bibnamefont
  {Pierloot}}, \bibinfo {author} {\bibfnamefont {M.}~\bibnamefont {Reiher}},
  \bibinfo {author} {\bibfnamefont {I.}~\bibnamefont {Schapiro}}, \bibinfo
  {author} {\bibfnamefont {J.}~\bibnamefont {Segarra-Mart\'{i}}}, \bibinfo
  {author} {\bibfnamefont {F.}~\bibnamefont {Segatta}}, \bibinfo {author}
  {\bibfnamefont {L.}~\bibnamefont {Seijo}}, \bibinfo {author} {\bibfnamefont
  {S.}~\bibnamefont {Sen}}, \bibinfo {author} {\bibfnamefont {D.-C.}\
  \bibnamefont {Sergentu}}, \bibinfo {author} {\bibfnamefont {C.~J.}\
  \bibnamefont {Stein}}, \bibinfo {author} {\bibfnamefont {L.}~\bibnamefont
  {Ungur}}, \bibinfo {author} {\bibfnamefont {M.}~\bibnamefont {Vacher}},
  \bibinfo {author} {\bibfnamefont {A.}~\bibnamefont {Valentini}}, \ and\
  \bibinfo {author} {\bibfnamefont {V.}~\bibnamefont {Veryazov}},\ }\href
  {\doibase 10.1063/5.0004835} {\bibfield  {journal} {\bibinfo  {journal} {J.
  Chem. Phys.}\ }\textbf {\bibinfo {volume} {152}},\ \bibinfo {pages} {214117}
  (\bibinfo {year} {2020})}\BibitemShut {NoStop}%
\bibitem [{\citenamefont {Fdez.~Galv\'{a}n}\ \emph {et~al.}(2019)\citenamefont
  {Fdez.~Galv\'{a}n}, \citenamefont {Vacher}, \citenamefont {Alavi},
  \citenamefont {Angeli}, \citenamefont {Aquilante}, \citenamefont
  {Autschbach}, \citenamefont {Bao}, \citenamefont {Bokarev}, \citenamefont
  {Bogdanov}, \citenamefont {Carlson}, \citenamefont {Chibotaru}, \citenamefont
  {Creutzberg}, \citenamefont {Dattani}, \citenamefont {Delcey}, \citenamefont
  {Dong}, \citenamefont {Dreuw}, \citenamefont {Freitag}, \citenamefont
  {Frutos}, \citenamefont {Gagliardi}, \citenamefont {Gendron}, \citenamefont
  {Giussani}, \citenamefont {González}, \citenamefont {Grell}, \citenamefont
  {Guo}, \citenamefont {Hoyer}, \citenamefont {Johansson}, \citenamefont
  {Keller}, \citenamefont {Knecht}, \citenamefont {Kovačević}, \citenamefont
  {Källman}, \citenamefont {Li~Manni}, \citenamefont {Lundberg}, \citenamefont
  {Ma}, \citenamefont {Mai}, \citenamefont {Malhado}, \citenamefont
  {Malmqvist}, \citenamefont {Marquetand}, \citenamefont {Mewes}, \citenamefont
  {Norell}, \citenamefont {Olivucci}, \citenamefont {Oppel}, \citenamefont
  {Phung}, \citenamefont {Pierloot}, \citenamefont {Plasser}, \citenamefont
  {Reiher}, \citenamefont {Sand}, \citenamefont {Schapiro}, \citenamefont
  {Sharma}, \citenamefont {Stein}, \citenamefont {S{\o}rensen}, \citenamefont
  {Truhlar}, \citenamefont {Ugandi}, \citenamefont {Ungur}, \citenamefont
  {Valentini}, \citenamefont {Vancoillie}, \citenamefont {Veryazov},
  \citenamefont {Weser}, \citenamefont {Wesołowski}, \citenamefont {Widmark},
  \citenamefont {Wouters}, \citenamefont {Zech}, \citenamefont {Zobel},\ and\
  \citenamefont {Lindh}}]{OpenMolcas3}%
  \BibitemOpen
  \bibfield  {author} {\bibinfo {author} {\bibfnamefont {I.}~\bibnamefont
  {Fdez.~Galv\'{a}n}}, \bibinfo {author} {\bibfnamefont {M.}~\bibnamefont
  {Vacher}}, \bibinfo {author} {\bibfnamefont {A.}~\bibnamefont {Alavi}},
  \bibinfo {author} {\bibfnamefont {C.}~\bibnamefont {Angeli}}, \bibinfo
  {author} {\bibfnamefont {F.}~\bibnamefont {Aquilante}}, \bibinfo {author}
  {\bibfnamefont {J.}~\bibnamefont {Autschbach}}, \bibinfo {author}
  {\bibfnamefont {J.~J.}\ \bibnamefont {Bao}}, \bibinfo {author} {\bibfnamefont
  {S.~I.}\ \bibnamefont {Bokarev}}, \bibinfo {author} {\bibfnamefont {N.~A.}\
  \bibnamefont {Bogdanov}}, \bibinfo {author} {\bibfnamefont {R.~K.}\
  \bibnamefont {Carlson}}, \bibinfo {author} {\bibfnamefont {L.~F.}\
  \bibnamefont {Chibotaru}}, \bibinfo {author} {\bibfnamefont {J.}~\bibnamefont
  {Creutzberg}}, \bibinfo {author} {\bibfnamefont {N.}~\bibnamefont {Dattani}},
  \bibinfo {author} {\bibfnamefont {M.~G.}\ \bibnamefont {Delcey}}, \bibinfo
  {author} {\bibfnamefont {S.~S.}\ \bibnamefont {Dong}}, \bibinfo {author}
  {\bibfnamefont {A.}~\bibnamefont {Dreuw}}, \bibinfo {author} {\bibfnamefont
  {L.}~\bibnamefont {Freitag}}, \bibinfo {author} {\bibfnamefont {L.~M.}\
  \bibnamefont {Frutos}}, \bibinfo {author} {\bibfnamefont {L.}~\bibnamefont
  {Gagliardi}}, \bibinfo {author} {\bibfnamefont {F.}~\bibnamefont {Gendron}},
  \bibinfo {author} {\bibfnamefont {A.}~\bibnamefont {Giussani}}, \bibinfo
  {author} {\bibfnamefont {L.}~\bibnamefont {González}}, \bibinfo {author}
  {\bibfnamefont {G.}~\bibnamefont {Grell}}, \bibinfo {author} {\bibfnamefont
  {M.}~\bibnamefont {Guo}}, \bibinfo {author} {\bibfnamefont {C.~E.}\
  \bibnamefont {Hoyer}}, \bibinfo {author} {\bibfnamefont {M.}~\bibnamefont
  {Johansson}}, \bibinfo {author} {\bibfnamefont {S.}~\bibnamefont {Keller}},
  \bibinfo {author} {\bibfnamefont {S.}~\bibnamefont {Knecht}}, \bibinfo
  {author} {\bibfnamefont {G.}~\bibnamefont {Kovačević}}, \bibinfo {author}
  {\bibfnamefont {E.}~\bibnamefont {Källman}}, \bibinfo {author}
  {\bibfnamefont {G.}~\bibnamefont {Li~Manni}}, \bibinfo {author}
  {\bibfnamefont {M.}~\bibnamefont {Lundberg}}, \bibinfo {author}
  {\bibfnamefont {Y.}~\bibnamefont {Ma}}, \bibinfo {author} {\bibfnamefont
  {S.}~\bibnamefont {Mai}}, \bibinfo {author} {\bibfnamefont {J.~P.}\
  \bibnamefont {Malhado}}, \bibinfo {author} {\bibfnamefont {P.~{\AA}.}\
  \bibnamefont {Malmqvist}}, \bibinfo {author} {\bibfnamefont {P.}~\bibnamefont
  {Marquetand}}, \bibinfo {author} {\bibfnamefont {S.~A.}\ \bibnamefont
  {Mewes}}, \bibinfo {author} {\bibfnamefont {J.}~\bibnamefont {Norell}},
  \bibinfo {author} {\bibfnamefont {M.}~\bibnamefont {Olivucci}}, \bibinfo
  {author} {\bibfnamefont {M.}~\bibnamefont {Oppel}}, \bibinfo {author}
  {\bibfnamefont {Q.~M.}\ \bibnamefont {Phung}}, \bibinfo {author}
  {\bibfnamefont {K.}~\bibnamefont {Pierloot}}, \bibinfo {author}
  {\bibfnamefont {F.}~\bibnamefont {Plasser}}, \bibinfo {author} {\bibfnamefont
  {M.}~\bibnamefont {Reiher}}, \bibinfo {author} {\bibfnamefont {A.~M.}\
  \bibnamefont {Sand}}, \bibinfo {author} {\bibfnamefont {I.}~\bibnamefont
  {Schapiro}}, \bibinfo {author} {\bibfnamefont {P.}~\bibnamefont {Sharma}},
  \bibinfo {author} {\bibfnamefont {C.~J.}\ \bibnamefont {Stein}}, \bibinfo
  {author} {\bibfnamefont {L.~K.}\ \bibnamefont {S{\o}rensen}}, \bibinfo
  {author} {\bibfnamefont {D.~G.}\ \bibnamefont {Truhlar}}, \bibinfo {author}
  {\bibfnamefont {M.}~\bibnamefont {Ugandi}}, \bibinfo {author} {\bibfnamefont
  {L.}~\bibnamefont {Ungur}}, \bibinfo {author} {\bibfnamefont
  {A.}~\bibnamefont {Valentini}}, \bibinfo {author} {\bibfnamefont
  {S.}~\bibnamefont {Vancoillie}}, \bibinfo {author} {\bibfnamefont
  {V.}~\bibnamefont {Veryazov}}, \bibinfo {author} {\bibfnamefont
  {O.}~\bibnamefont {Weser}}, \bibinfo {author} {\bibfnamefont {T.~A.}\
  \bibnamefont {Wesołowski}}, \bibinfo {author} {\bibfnamefont {P.-O.}\
  \bibnamefont {Widmark}}, \bibinfo {author} {\bibfnamefont {S.}~\bibnamefont
  {Wouters}}, \bibinfo {author} {\bibfnamefont {A.}~\bibnamefont {Zech}},
  \bibinfo {author} {\bibfnamefont {J.~P.}\ \bibnamefont {Zobel}}, \ and\
  \bibinfo {author} {\bibfnamefont {R.}~\bibnamefont {Lindh}},\ }\href
  {\doibase 10.1021/acs.jctc.9b00532} {\bibfield  {journal} {\bibinfo
  {journal} {J. Chem. Theory Comput.}\ }\textbf {\bibinfo {volume} {15}},\
  \bibinfo {pages} {5925} (\bibinfo {year} {2019})}\BibitemShut {NoStop}%
\bibitem [{\citenamefont {Guther}\ \emph {et~al.}(2020)\citenamefont {Guther},
  \citenamefont {Anderson}, \citenamefont {Blunt}, \citenamefont {Bogdanov},
  \citenamefont {Cleland}, \citenamefont {Dattani}, \citenamefont {Dobrautz},
  \citenamefont {Ghanem}, \citenamefont {Jeszenszki}, \citenamefont
  {Liebermann}, \citenamefont {Li~Manni}, \citenamefont {Lozovoi},
  \citenamefont {Luo}, \citenamefont {Ma}, \citenamefont {Merz}, \citenamefont
  {Overy}, \citenamefont {Rampp}, \citenamefont {Samanta}, \citenamefont
  {Schwarz}, \citenamefont {Shepherd}, \citenamefont {Smart}, \citenamefont
  {Vitale}, \citenamefont {Weser}, \citenamefont {Booth},\ and\ \citenamefont
  {Alavi}}]{guther2020neci}%
  \BibitemOpen
  \bibfield  {author} {\bibinfo {author} {\bibfnamefont {K.}~\bibnamefont
  {Guther}}, \bibinfo {author} {\bibfnamefont {R.~J.}\ \bibnamefont
  {Anderson}}, \bibinfo {author} {\bibfnamefont {N.~S.}\ \bibnamefont {Blunt}},
  \bibinfo {author} {\bibfnamefont {N.~A.}\ \bibnamefont {Bogdanov}}, \bibinfo
  {author} {\bibfnamefont {D.}~\bibnamefont {Cleland}}, \bibinfo {author}
  {\bibfnamefont {N.}~\bibnamefont {Dattani}}, \bibinfo {author} {\bibfnamefont
  {W.}~\bibnamefont {Dobrautz}}, \bibinfo {author} {\bibfnamefont
  {K.}~\bibnamefont {Ghanem}}, \bibinfo {author} {\bibfnamefont
  {P.}~\bibnamefont {Jeszenszki}}, \bibinfo {author} {\bibfnamefont
  {N.}~\bibnamefont {Liebermann}}, \bibinfo {author} {\bibfnamefont
  {G.}~\bibnamefont {Li~Manni}}, \bibinfo {author} {\bibfnamefont {A.~Y.}\
  \bibnamefont {Lozovoi}}, \bibinfo {author} {\bibfnamefont {H.}~\bibnamefont
  {Luo}}, \bibinfo {author} {\bibfnamefont {D.}~\bibnamefont {Ma}}, \bibinfo
  {author} {\bibfnamefont {F.}~\bibnamefont {Merz}}, \bibinfo {author}
  {\bibfnamefont {C.}~\bibnamefont {Overy}}, \bibinfo {author} {\bibfnamefont
  {M.}~\bibnamefont {Rampp}}, \bibinfo {author} {\bibfnamefont {P.~K.}\
  \bibnamefont {Samanta}}, \bibinfo {author} {\bibfnamefont {L.~R.}\
  \bibnamefont {Schwarz}}, \bibinfo {author} {\bibfnamefont {J.~J.}\
  \bibnamefont {Shepherd}}, \bibinfo {author} {\bibfnamefont {S.~D.}\
  \bibnamefont {Smart}}, \bibinfo {author} {\bibfnamefont {E.}~\bibnamefont
  {Vitale}}, \bibinfo {author} {\bibfnamefont {O.}~\bibnamefont {Weser}},
  \bibinfo {author} {\bibfnamefont {G.~H.}\ \bibnamefont {Booth}}, \ and\
  \bibinfo {author} {\bibfnamefont {A.}~\bibnamefont {Alavi}},\ }\href
  {\doibase 10.1063/5.0005754} {\bibfield  {journal} {\bibinfo  {journal} {J.
  Chem. Phys.}\ }\textbf {\bibinfo {volume} {153}},\ \bibinfo {pages} {034107}
  (\bibinfo {year} {2020})}\BibitemShut {NoStop}%
\bibitem [{\citenamefont {{Qiskit contributors}}(2023)}]{Qiskit}%
  \BibitemOpen
  \bibfield  {author} {\bibinfo {author} {\bibnamefont {{Qiskit
  contributors}}},\ }\href {\doibase 10.5281/zenodo.2573505} {\enquote
  {\bibinfo {title} {Qiskit: An open-source framework for quantum computing},}\
  } (\bibinfo {year} {2023})\BibitemShut {NoStop}%
\bibitem [{\citenamefont {{The Qiskit Nature developers and
  contributors}}(2023)}]{the-qiskit-nature-developers-and-contrib-2023-7828768}%
  \BibitemOpen
  \bibfield  {author} {\bibinfo {author} {\bibnamefont {{The Qiskit Nature
  developers and contributors}}},\ }\href {\doibase 10.5281/zenodo.7828768}
  {\enquote {\bibinfo {title} {Qiskit nature 0.6.0},}\ } (\bibinfo {year}
  {2023}),\ \bibinfo {note} {{Qiskit Nature has some code that is included
  under other licensing. These files have been removed from the zip repository
  provided here and are only available via Github. See
  https://github.com/Qiskit/qiskit-nature\#license for more
  details.}}\BibitemShut {Stop}%
\bibitem [{\citenamefont {Gard}\ \emph {et~al.}(2020)\citenamefont {Gard},
  \citenamefont {Zhu}, \citenamefont {Barron}, \citenamefont {Mayhall},
  \citenamefont {Economou},\ and\ \citenamefont {Barnes}}]{Gard2020}%
  \BibitemOpen
  \bibfield  {author} {\bibinfo {author} {\bibfnamefont {B.~T.}\ \bibnamefont
  {Gard}}, \bibinfo {author} {\bibfnamefont {L.}~\bibnamefont {Zhu}}, \bibinfo
  {author} {\bibfnamefont {G.~S.}\ \bibnamefont {Barron}}, \bibinfo {author}
  {\bibfnamefont {N.~J.}\ \bibnamefont {Mayhall}}, \bibinfo {author}
  {\bibfnamefont {S.~E.}\ \bibnamefont {Economou}}, \ and\ \bibinfo {author}
  {\bibfnamefont {E.}~\bibnamefont {Barnes}},\ }\href {\doibase
  10.1038/s41534-019-0240-1} {\bibfield  {journal} {\bibinfo  {journal} {npj
  Quantum Inf.}\ }\textbf {\bibinfo {volume} {6}},\ \bibinfo {pages} {10}
  (\bibinfo {year} {2020})}\BibitemShut {NoStop}%
\bibitem [{\citenamefont {Schulten}\ and\ \citenamefont
  {Karplus}(1972)}]{Schulten1972origin}%
  \BibitemOpen
  \bibfield  {author} {\bibinfo {author} {\bibfnamefont {K.}~\bibnamefont
  {Schulten}}\ and\ \bibinfo {author} {\bibfnamefont {M.}~\bibnamefont
  {Karplus}},\ }\href {\doibase https://doi.org/10.1016/0009-2614(72)80120-9}
  {\bibfield  {journal} {\bibinfo  {journal} {Chem. Phys. Lett.}\ }\textbf
  {\bibinfo {volume} {14}},\ \bibinfo {pages} {305} (\bibinfo {year}
  {1972})}\BibitemShut {NoStop}%
\bibitem [{\citenamefont {Tavan}\ and\ \citenamefont
  {Schulten}(1986)}]{Tavan1986low}%
  \BibitemOpen
  \bibfield  {author} {\bibinfo {author} {\bibfnamefont {P.}~\bibnamefont
  {Tavan}}\ and\ \bibinfo {author} {\bibfnamefont {K.}~\bibnamefont
  {Schulten}},\ }\href {\doibase 10.1063/1.451442} {\bibfield  {journal}
  {\bibinfo  {journal} {J. Chem. Phys.}\ }\textbf {\bibinfo {volume} {85}},\
  \bibinfo {pages} {6602} (\bibinfo {year} {1986})}\BibitemShut {NoStop}%
\bibitem [{\citenamefont {Pariser}\ and\ \citenamefont
  {Parr}(1953{\natexlab{a}})}]{PPP1}%
  \BibitemOpen
  \bibfield  {author} {\bibinfo {author} {\bibfnamefont {R.}~\bibnamefont
  {Pariser}}\ and\ \bibinfo {author} {\bibfnamefont {R.~G.}\ \bibnamefont
  {Parr}},\ }\href {\doibase 10.1063/1.1698929} {\bibfield  {journal} {\bibinfo
   {journal} {J. Chem. Phys.}\ }\textbf {\bibinfo {volume} {21}},\ \bibinfo
  {pages} {466} (\bibinfo {year} {1953}{\natexlab{a}})}\BibitemShut {NoStop}%
\bibitem [{\citenamefont {Pariser}\ and\ \citenamefont
  {Parr}(1953{\natexlab{b}})}]{PPP2}%
  \BibitemOpen
  \bibfield  {author} {\bibinfo {author} {\bibfnamefont {R.}~\bibnamefont
  {Pariser}}\ and\ \bibinfo {author} {\bibfnamefont {R.~G.}\ \bibnamefont
  {Parr}},\ }\href {\doibase 10.1063/1.1699030} {\bibfield  {journal} {\bibinfo
   {journal} {J. Chem. Phys.}\ }\textbf {\bibinfo {volume} {21}},\ \bibinfo
  {pages} {767} (\bibinfo {year} {1953}{\natexlab{b}})}\BibitemShut {NoStop}%
\bibitem [{\citenamefont {Pople}(1953)}]{PPP3}%
  \BibitemOpen
  \bibfield  {author} {\bibinfo {author} {\bibfnamefont {J.~A.}\ \bibnamefont
  {Pople}},\ }\href {\doibase 10.1039/TF9534901375} {\bibfield  {journal}
  {\bibinfo  {journal} {Trans. Faraday Soc.}\ }\textbf {\bibinfo {volume}
  {49}},\ \bibinfo {pages} {1375} (\bibinfo {year} {1953})}\BibitemShut
  {NoStop}%
\bibitem [{\citenamefont {Hudson}\ and\ \citenamefont
  {Kohler}(1972)}]{Hudson1972}%
  \BibitemOpen
  \bibfield  {author} {\bibinfo {author} {\bibfnamefont {B.}~\bibnamefont
  {Hudson}}\ and\ \bibinfo {author} {\bibfnamefont {B.}~\bibnamefont
  {Kohler}},\ }\href {\doibase https://doi.org/10.1016/0009-2614(72)80119-2}
  {\bibfield  {journal} {\bibinfo  {journal} {Chem. Phys. Lett.}\ }\textbf
  {\bibinfo {volume} {14}},\ \bibinfo {pages} {299} (\bibinfo {year}
  {1972})}\BibitemShut {NoStop}%
\bibitem [{\citenamefont {Hudson}\ and\ \citenamefont
  {Kohler}(1984)}]{Hudson1984}%
  \BibitemOpen
  \bibfield  {author} {\bibinfo {author} {\bibfnamefont {B.}~\bibnamefont
  {Hudson}}\ and\ \bibinfo {author} {\bibfnamefont {B.}~\bibnamefont
  {Kohler}},\ }\href {\doibase https://doi.org/10.1016/0379-6779(84)90062-6}
  {\bibfield  {journal} {\bibinfo  {journal} {Synth. Met.}\ }\textbf {\bibinfo
  {volume} {9}},\ \bibinfo {pages} {241} (\bibinfo {year} {1984})}\BibitemShut
  {NoStop}%
\bibitem [{\citenamefont {Ido}\ \emph {et~al.}(2023)\citenamefont {Ido},
  \citenamefont {Kawamura}, \citenamefont {Motoyama}, \citenamefont {Yoshimi},
  \citenamefont {Yamaji}, \citenamefont {Todo}, \citenamefont {Kawashima},\
  and\ \citenamefont {Misawa}}]{HPhi2}%
  \BibitemOpen
  \bibfield  {author} {\bibinfo {author} {\bibfnamefont {K.}~\bibnamefont
  {Ido}}, \bibinfo {author} {\bibfnamefont {M.}~\bibnamefont {Kawamura}},
  \bibinfo {author} {\bibfnamefont {Y.}~\bibnamefont {Motoyama}}, \bibinfo
  {author} {\bibfnamefont {K.}~\bibnamefont {Yoshimi}}, \bibinfo {author}
  {\bibfnamefont {Y.}~\bibnamefont {Yamaji}}, \bibinfo {author} {\bibfnamefont
  {S.}~\bibnamefont {Todo}}, \bibinfo {author} {\bibfnamefont {N.}~\bibnamefont
  {Kawashima}}, \ and\ \bibinfo {author} {\bibfnamefont {T.}~\bibnamefont
  {Misawa}},\ }\href@noop {} {\bibfield  {journal} {\bibinfo  {journal}
  {arXiv}\ } (\bibinfo {year} {2023})},\ \Eprint
  {http://arxiv.org/abs/2307.13222} {arXiv:2307.13222 [cond-mat.str-el]}
  \BibitemShut {NoStop}%
\bibitem [{\citenamefont {Rohatgi}(2022)}]{Rohatgi2022}%
  \BibitemOpen
  \bibfield  {author} {\bibinfo {author} {\bibfnamefont {A.}~\bibnamefont
  {Rohatgi}},\ }\href {https://automeris.io/WebPlotDigitizer} {\enquote
  {\bibinfo {title} {Webplotdigitizer: Version 4.6},}\ } (\bibinfo {year}
  {2022})\BibitemShut {NoStop}%
\bibitem [{\citenamefont {Cave}\ \emph {et~al.}(2004)\citenamefont {Cave},
  \citenamefont {Zhang}, \citenamefont {Maitra},\ and\ \citenamefont
  {Burke}}]{CAVE2004}%
  \BibitemOpen
  \bibfield  {author} {\bibinfo {author} {\bibfnamefont {R.~J.}\ \bibnamefont
  {Cave}}, \bibinfo {author} {\bibfnamefont {F.}~\bibnamefont {Zhang}},
  \bibinfo {author} {\bibfnamefont {N.~T.}\ \bibnamefont {Maitra}}, \ and\
  \bibinfo {author} {\bibfnamefont {K.}~\bibnamefont {Burke}},\ }\href
  {\doibase https://doi.org/10.1016/j.cplett.2004.03.051} {\bibfield  {journal}
  {\bibinfo  {journal} {Chem. Phys. Lett.}\ }\textbf {\bibinfo {volume}
  {389}},\ \bibinfo {pages} {39} (\bibinfo {year} {2004})}\BibitemShut
  {NoStop}%
\bibitem [{Emi()}]{Emieeel}%
  \BibitemOpen
  \href@noop {} {}\bibinfo {howpublished}
  {\url{https://github.com/Emieeel/1-Norm_calculations.git}}\BibitemShut
  {NoStop}%
\bibitem [{\citenamefont {Preskill}(2018)}]{Preskill2018quantumcomputingin}%
  \BibitemOpen
  \bibfield  {author} {\bibinfo {author} {\bibfnamefont {J.}~\bibnamefont
  {Preskill}},\ }\href {\doibase 10.22331/q-2018-08-06-79} {\bibfield
  {journal} {\bibinfo  {journal} {{Quantum}}\ }\textbf {\bibinfo {volume}
  {2}},\ \bibinfo {pages} {79} (\bibinfo {year} {2018})}\BibitemShut {NoStop}%
\bibitem [{\citenamefont {Yoshioka}\ \emph {et~al.}(2022)\citenamefont
  {Yoshioka}, \citenamefont {Okubo}, \citenamefont {Suzuki}, \citenamefont
  {Koizumi},\ and\ \citenamefont {Mizukami}}]{yoshioka2022hunting}%
  \BibitemOpen
  \bibfield  {author} {\bibinfo {author} {\bibfnamefont {N.}~\bibnamefont
  {Yoshioka}}, \bibinfo {author} {\bibfnamefont {T.}~\bibnamefont {Okubo}},
  \bibinfo {author} {\bibfnamefont {Y.}~\bibnamefont {Suzuki}}, \bibinfo
  {author} {\bibfnamefont {Y.}~\bibnamefont {Koizumi}}, \ and\ \bibinfo
  {author} {\bibfnamefont {W.}~\bibnamefont {Mizukami}},\ }\href@noop {} {}
  (\bibinfo {year} {2022}),\ \Eprint {http://arxiv.org/abs/2210.14109}
  {arXiv:2210.14109 [quant-ph]} \BibitemShut {NoStop}%
\bibitem [{\citenamefont {Ichikawa}\ \emph {et~al.}(2023)\citenamefont
  {Ichikawa}, \citenamefont {Hakoshima}, \citenamefont {Inui}, \citenamefont
  {Ito}, \citenamefont {Matsuda}, \citenamefont {Mitarai}, \citenamefont
  {Miyamoto}, \citenamefont {Mizukami}, \citenamefont {Mizuta}, \citenamefont
  {Mori}, \citenamefont {Nakano}, \citenamefont {Nakayama}, \citenamefont
  {Okada}, \citenamefont {Sugimoto}, \citenamefont {Takahira}, \citenamefont
  {Takemori}, \citenamefont {Tsukano}, \citenamefont {Ueda}, \citenamefont
  {Watanabe}, \citenamefont {Yoshida},\ and\ \citenamefont
  {Fujii}}]{ichikawa2023comprehensive}%
  \BibitemOpen
  \bibfield  {author} {\bibinfo {author} {\bibfnamefont {T.}~\bibnamefont
  {Ichikawa}}, \bibinfo {author} {\bibfnamefont {H.}~\bibnamefont {Hakoshima}},
  \bibinfo {author} {\bibfnamefont {K.}~\bibnamefont {Inui}}, \bibinfo {author}
  {\bibfnamefont {K.}~\bibnamefont {Ito}}, \bibinfo {author} {\bibfnamefont
  {R.}~\bibnamefont {Matsuda}}, \bibinfo {author} {\bibfnamefont
  {K.}~\bibnamefont {Mitarai}}, \bibinfo {author} {\bibfnamefont
  {K.}~\bibnamefont {Miyamoto}}, \bibinfo {author} {\bibfnamefont
  {W.}~\bibnamefont {Mizukami}}, \bibinfo {author} {\bibfnamefont
  {K.}~\bibnamefont {Mizuta}}, \bibinfo {author} {\bibfnamefont
  {T.}~\bibnamefont {Mori}}, \bibinfo {author} {\bibfnamefont {Y.}~\bibnamefont
  {Nakano}}, \bibinfo {author} {\bibfnamefont {A.}~\bibnamefont {Nakayama}},
  \bibinfo {author} {\bibfnamefont {K.~N.}\ \bibnamefont {Okada}}, \bibinfo
  {author} {\bibfnamefont {T.}~\bibnamefont {Sugimoto}}, \bibinfo {author}
  {\bibfnamefont {S.}~\bibnamefont {Takahira}}, \bibinfo {author}
  {\bibfnamefont {N.}~\bibnamefont {Takemori}}, \bibinfo {author}
  {\bibfnamefont {S.}~\bibnamefont {Tsukano}}, \bibinfo {author} {\bibfnamefont
  {H.}~\bibnamefont {Ueda}}, \bibinfo {author} {\bibfnamefont {R.}~\bibnamefont
  {Watanabe}}, \bibinfo {author} {\bibfnamefont {Y.}~\bibnamefont {Yoshida}}, \
  and\ \bibinfo {author} {\bibfnamefont {K.}~\bibnamefont {Fujii}},\
  }\href@noop {} {} (\bibinfo {year} {2023}),\ \Eprint
  {http://arxiv.org/abs/2307.16130} {arXiv:2307.16130 [quant-ph]} \BibitemShut
  {NoStop}%
\bibitem [{\citenamefont {Shinaoka}\ \emph {et~al.}(2015)\citenamefont
  {Shinaoka}, \citenamefont {Troyer},\ and\ \citenamefont
  {Werner}}]{Shinaoka2015accuracy}%
  \BibitemOpen
  \bibfield  {author} {\bibinfo {author} {\bibfnamefont {H.}~\bibnamefont
  {Shinaoka}}, \bibinfo {author} {\bibfnamefont {M.}~\bibnamefont {Troyer}}, \
  and\ \bibinfo {author} {\bibfnamefont {P.}~\bibnamefont {Werner}},\ }\href
  {\doibase 10.1103/PhysRevB.91.245156} {\bibfield  {journal} {\bibinfo
  {journal} {Phys. Rev. B}\ }\textbf {\bibinfo {volume} {91}},\ \bibinfo
  {pages} {245156} (\bibinfo {year} {2015})}\BibitemShut {NoStop}%
\bibitem [{\citenamefont {Honerkamp}\ \emph {et~al.}(2018)\citenamefont
  {Honerkamp}, \citenamefont {Shinaoka}, \citenamefont {Assaad},\ and\
  \citenamefont {Werner}}]{Carsten2018limitations}%
  \BibitemOpen
  \bibfield  {author} {\bibinfo {author} {\bibfnamefont {C.}~\bibnamefont
  {Honerkamp}}, \bibinfo {author} {\bibfnamefont {H.}~\bibnamefont {Shinaoka}},
  \bibinfo {author} {\bibfnamefont {F.~F.}\ \bibnamefont {Assaad}}, \ and\
  \bibinfo {author} {\bibfnamefont {P.}~\bibnamefont {Werner}},\ }\href
  {\doibase 10.1103/PhysRevB.98.235151} {\bibfield  {journal} {\bibinfo
  {journal} {Phys. Rev. B}\ }\textbf {\bibinfo {volume} {98}},\ \bibinfo
  {pages} {235151} (\bibinfo {year} {2018})}\BibitemShut {NoStop}%
\end{thebibliography}%

\appendix
\section{State characterization \label{sec:appendix}}

We describe the characterization of the 1${}^1{\rm A_g}$ ground state, the singly excited 1${}^1{\rm B_u}$ state, and the doubly excited 2${}^1{\rm A_g}$ state of our models.
In this Appendix, the model parameters specified in Table~\ref{tab:params} are used.

First, we perform a basis transformation by SCF calculation. The resulting canonical molecular orbital $\tilde{\phi}_j$ is represented by the linear combination of the Wannier functions $\phi_i$ as
\begin{align}
    \tilde{\phi}_j = \sum_{i=1}^K \phi_i C_{ij}, \label{eq:linear}
\end{align}
where $C_{ij}$ is the molecular orbital coefficient and $K$ is the number of the Wannier functions.

Next, the model Hamiltonian in the transformed basis is diagonalized.
The eigenstate of the model Hamiltonian $\ket{\Psi^{({I})}}$ is represented by the superposition of the computational basis states $\ket{\widetilde{\Phi}_k}$: 
\begin{align}
    \ket{\Psi^{({I})}} = \sum_k \tilde{d}_k^{({I})} \ket{\widetilde{\Phi}_k},
\end{align}
where $\tilde{d}_k^{({I})}$ is the CI coefficient of the $k$-th basis state in the state $I$. The tilde represents the use of the canonical orbital basis.

The computational basis state $\ket{\widetilde{\Phi}_k}$ can be written as the occupation number vector defined as follows:
\begin{align}
    \ket{\widetilde{\Phi}_k} = \ket{\tilde{n}_{1\alpha,k}\tilde{n}_{1\beta,k}\tilde{n}_{2\alpha,k}\tilde{n}_{2\beta,k}\cdots \tilde{n}_{K\alpha,k}\tilde{n}_{K\beta,k}},
\end{align}
where $\tilde{n}_{i\sigma,k}$ is the occupation number of the spin-orbital consisting of the $i$-th canonical orbital $\tilde{\phi}_i$ with spin $\sigma = \{\alpha, \beta\}$ in the $k$-th basis state, and thus $\tilde{n}_{i\sigma,k} = \{0, 1\}$. For example, the Hartree-Fock (HF) state in the (4e, 4o) models corresponds to $\ket{11110000}$.

Table~\ref{tab:chara} shows the CI coefficients and the computational basis states of the eigenstates of our models. 
The definitions of Model~1 and Model~2 are the same as that defined in Sec.~\ref{sec:vert}, employing $\mathcal{H}_{\rm int}^{(1)}$ and $\mathcal{H}_{\rm int}^{(2)}$ as the electron-electron interaction part, respectively.
Note that this table only shows the coefficients of the large absolute values required for the characterization. For explanation, we use $\Ket{\widetilde{\Phi}_k^{({I})}}$ for the $k$-th computational basis state for each state $I$ instead of $\ket{\widetilde{\Phi}_k}$. This means that the different basis states $\ket{\widetilde{\Phi}_k}$ are defined for each state $I$.
\begin{table*}[ht]
    \centering
    \caption{The basis state definition and the CI coefficients of the eigenstates of our models. Only the leading terms required for the state characterization are displayed.}
    \resizebox{2.0 \columnwidth}{!}{
    \begin{tabular}{llcrlrlrlrlrl} \hline \hline
    Molecule, & \multirow{2}{*}{Model}   & \multirow{2}{*}{$I$} & \multirow{2}{*}{$\tilde{d}_1^{({I})}$} & \multirow{2}{*}{$\Ket{\widetilde{\Phi}_1^{({I})}}$} & \multirow{2}{*}{$\tilde{d}_2^{({I})}$} & \multirow{2}{*}{$\Ket{\widetilde{\Phi}_2^{({I})}}$} & \multirow{2}{*}{$\tilde{d}_3^{({I})}$} & \multirow{2}{*}{$\Ket{\widetilde{\Phi}_3^{({I})}}$} & \multirow{2}{*}{$\tilde{d}_4^{({I})}$} & \multirow{2}{*}{$\Ket{\widetilde{\Phi}_4^{({I})}}$} & \multirow{2}{*}{$\tilde{d}_5^{({I})}$} & \multirow{2}{*}{$\Ket{\widetilde{\Phi}_5^{({I})}}$} \\ 
    Space &&&&&&&&&&&& \\ \hline \hline
    Ethylene,              & Model 1 & 1${}^1{\rm A_g}$ & $-$0.986 & $\ket{1100}$         &   +0.169 & $\ket{0011}$         &       &       \\ 
    (2e, 2o)               &         & 1${}^1{\rm B_u}$ & $-$0.707 & $\ket{0110}$         &   +0.707 & $\ket{1001}$         &       &       \\ 
                           &         & 2${}^1{\rm A_g}$ &   +0.986 & $\ket{0011}$         &   +0.169 & $\ket{1100}$         &       &       \\ \cline{2-13}
                           & Model 2 & 1${}^1{\rm A_g}$ & $-$0.986 & $\ket{1100}$         &   +0.169 & $\ket{0011}$         &       &       \\ 
                           &         & 1${}^1{\rm B_u}$ & $-$0.707 & $\ket{0110}$         &   +0.707 & $\ket{1001}$         &       &       \\ 
                           &         & 2${}^1{\rm A_g}$ &   +0.986 & $\ket{0011}$         &   +0.169 & $\ket{1100}$         &       &       \\ \hline
    Butadiene,             & Model 1 & 1${}^1{\rm A_g}$ & $-$0.982 & $\ket{11110000}$     &   +0.139 & $\ket{11001100}$     &       &       \\ 
    (4e, 4o)               &         & 1${}^1{\rm B_u}$ & $-$0.695 & $\ket{11100100}$     &   +0.695 & $\ket{11011000}$     & $-$0.110 & $\ket{01101100}$ &   +0.110 & $\ket{10011100}$ \\ 
                           &         & 2${}^1{\rm A_g}$ &   +0.478 & $\ket{11001100}$     &   +0.410 & $\ket{11010010}$     & $-$0.410 & $\ket{11100001}$ &   +0.403 & $\ket{10110100}$ & $-$0.403 & $\ket{01111000}$ \\ \cline{2-13}
                           & Model 2 & 1${}^1{\rm A_g}$ & $-$0.982 & $\ket{11110000}$     &   +0.135 & $\ket{11001100}$       \\ 
                           &         & 1${}^1{\rm B_u}$ & $-$0.697 & $\ket{11100100}$     &   +0.697 & $\ket{11011000}$     & $-$0.105 & $\ket{01101100}$ &   +0.105 & $\ket{10011100}$      \\ 
                           &         & 2${}^1{\rm A_g}$ &   +0.489 & $\ket{11001100}$     &   +0.410 & $\ket{10110100}$     & $-$0.410 & $\ket{01111000}$ &   +0.399 & $\ket{11010010}$ & $-$0.399 & $\ket{11100001}$ \\ \hline
    Hexatriene,              & Model 1 & 1${}^1{\rm A_g}$ &   +0.982 & $\ket{11110000}$     & $-$0.144 & $\ket{11001100}$       \\ 
     (4e, 4o)                &         & 1${}^1{\rm B_u}$ & $-$0.694 & $\ket{11100100}$     &   +0.694 & $\ket{11011000}$     &   +0.092 & $\ket{01101100}$ & $-$0.092 & $\ket{10011100}$ &    \\ 
                             &         & 2${}^1{\rm A_g}$ & $-$0.561 & $\ket{11001100}$     &   +0.490 & $\ket{10110100}$     & $-$0.490 & $\ket{01111000}$ &   +0.224 & $\ket{11010010}$ & $-$0.224 & $\ket{11100001}$     \\ \cline{2-13}
                             & Model 2 & 1${}^1{\rm A_g}$ &   +0.984 & $\ket{11110000}$     & $-$0.128 & $\ket{11001100}$     &        \\ 
                             &         & 1${}^1{\rm B_u}$ & $-$0.695 & $\ket{11100100}$     &   +0.695 & $\ket{11011000}$     &   +0.093 & $\ket{01101100}$ & $-$0.093 & $\ket{10011100}$        \\ 
                             &         & 2${}^1{\rm A_g}$ & $-$0.565 & $\ket{11001100}$     &   +0.475 & $\ket{10110100}$     & $-$0.475 & $\ket{01111000}$ &   +0.237 & $\ket{11010010}$ & $-$0.237 & $\ket{11100001}$      \\ \hline
    Hexatriene,              & Model 1 & 1${}^1{\rm A_g}$ &   +0.974 & $\ket{111111000000}$ & $-$0.134 & $\ket{111100110000}$ &        \\ 
    (6e, 6o)                 &         & 1${}^1{\rm B_u}$ &   +0.682 & $\ket{111110010000}$ & $-$0.682 & $\ket{111101100000}$ &   +0.128 & $\ket{111100011000}$ & $-$0.128 & $\ket{111100100100}$        \\ 
                             &         & 2${}^1{\rm A_g}$ & $-$0.557 & $\ket{111100110000}$ & $-$0.388 & $\ket{111101001000}$ &   +0.388 & $\ket{111110000100}$ & $-$0.324 & $\ket{111011010000}$ &   +0.324 & $\ket{110111100000}$      \\ \cline{2-13}
                             & Model 2 & 1${}^1{\rm A_g}$ & $-$0.974 & $\ket{111111000000}$ &   +0.127 & $\ket{111100110000}$ &        \\ 
                             &         & 1${}^1{\rm B_u}$ &   +0.685 & $\ket{111110010000}$ & $-$0.685 & $\ket{111101100000}$ &   +0.111 & $\ket{111100011000}$ & $-$0.111 & $\ket{111100100100}$        \\ 
                             &         & 2${}^1{\rm A_g}$ &   +0.564 & $\ket{111100110000}$ &   +0.389 & $\ket{111101001000}$ & $-$0.389 & $\ket{111110000100}$ & +0.319 & $\ket{111011010000}$ & $-$0.319 & $\ket{110111100000}$       \\ \hline \hline
    \end{tabular}}
    \label{tab:chara}
\end{table*}

Table~\ref{tab:chara} shows that the eigenstates are represented as the linear combination of the spin-adapted configurations.
In all cases, the HF state is dominant in the 1${}^1{\rm A_g}$ ground state, and the 1${}^1{\rm B_u}$ and 2${}^1{\rm A_g}$ states are reasonably characterized as the excited states of the one- and two-electron excitations from the highest occupied to the lowest unoccupied molecular orbitals.

\end{document}